\documentclass[fleqn,usenatbib]{mnras}

\usepackage{newtxtext,newtxmath}
\usepackage{enumitem}
\usepackage{hyperref}

\usepackage[T1]{fontenc}
\usepackage{upgreek}

\DeclareRobustCommand{\VAN}[3]{#2}
\let\VANthebibliography\thebibliography
\def\thebibliography{\DeclareRobustCommand{\VAN}[3]{##3}\VANthebibliography}

\usepackage{graphicx}	
\usepackage{amsmath}	
\newcommand{\sarc}{$^{\prime\prime}$}

\title[The ILT view of the Bo\"{o}tes Deep Field]{The sub-arcsecond ILT view of the Bo\"{o}tes Deep Field: A link between low-frequency kiloparsec radio morphology and AGN driven ionised outflows}

\author[Emmy L. Escott]{\parbox{\textwidth}{
Emmy L. Escott$^{1,2}$\thanks{E-mail:Emmy.Escott@csiro.au},
Leah K. Morabito$^{1,3}$,
Frits Sweijen$^{1}$,
Chris M. Harrison$^{4}$,
James Petley$^{5}$,
Jurjen M. G. H. J. de Jong$^{5,6}$,
Reinout J. van Weeren$^{5}$,
Thomas S. Higginson$^{7}$,
Isabella Prandoni$^{8}$,
George Miley$^{5}$,
Huub J. A. Röttgering$^{5}$} \\
\\
$^{1}$Centre for Extragalactic Astronomy, Department of Physics, Durham
University, Durham DH1 3LE, UK\\
$^{2}$Australia Telescope National Facility, CSIRO, Space and Astronomy, PO Box 1130, Bentley, WA 6102, Australia\\
$^{3}$Institute for Computational Cosmology, Department of Physics, University of Durham, South Road, Durham DH1 3LE, UK\\
$^{4}$School of Mathematics, Statistics and Physics, Newcastle University, NE1 7RU, UK\\
$^{5}$Leiden Observatory, Leiden University, PO Box 9513, 2300 RA
Leiden, The Netherlands\\
$^{6}$ASTRON, The Netherlands Institute for Radio Astronomy, Postbus 2, 7990 AA Dwingeloo, The Netherlands \\
$^{7}$School of Physics, University of Bristol, HH Wills Physics Laboratory, Tyndall Avenue, Bristol BS8 1TL, UK \\
$^{8}$INAF-IRA, Via P. Gobetti 101, 40129 Bologna, Italy \\ 
}

\date{Accepted 2026 February 03. Received 2026 January 29; in original form 2025 August 29}

\pubyear{2026}

\begin{document}
\label{firstpage}
\pagerange{\pageref{firstpage}--\pageref{lastpage}}
\maketitle

\begin{abstract}
\noindent
Active Galactic Nuclei (AGN) outflows can regulate host galaxy evolution via AGN feedback. Ionised gas outflows have been linked to enhanced radio emission. In the first paper of this series, AGN detected with the International LOFAR Telescope (ILT) at 6\sarc\ were more likely to host an [O~{\sc iii}] $\lambdaup$5007 \AA\ outflow than AGN not detected, although only high-powered jets were ruled out as the origin of radio emission. New wide-field, sub-arcsecond resolution imaging at 144~MHz with the ILT now enables a resolved morphological study of this sample. We present the first wide-field, sub-arcsecond images of the Boötes Deep Field at 144 MHz, detecting 4074 sources in the $\sim$0.3\sarc\ image with a central sensitivity of 33.8~$\muup$Jy~$\mathrm{beam^{-1}}$. For 47 AGN matched in AGN luminosity, we probe radio emission on kiloparsec-scales to investigate correlations with [O~{\sc iii}] outflows. This sample spans $z<0.83$, $10^{40}<L_{\mathrm{[OIII]}}<10^{43}~\mathrm{erg~s^{-1}}$, and $10^{21}<L_{\mathrm{144MHz}}<10^{24.5}~\mathrm{W~Hz^{-1}}$. We find that if we detect an AGN on both large-scales (6\sarc) and small-scales (0.3\sarc), 90$\pm$7 per cent have an [O~{\sc iii}] outflow, compared to 63$\pm$9 per cent of sources detected on large-scales, but undetected on small-scales. Furthermore, 17$\pm$6 per cent of sources without an [O~{\sc iii}] outflow are detected on kiloparsec-scales, compared to 51$\pm$12 per cent of sources with an [O~{\sc iii}] outflow. This implies a connection between [O~{\sc iii}] outflows and kiloparsec-scale radio emission, which is likely AGN-driven. In contrast, AGN without an [O~{\sc iii}] outflow are dominated by diffuse radio emission, likely to be associated with star formation.

\end{abstract}

\begin{keywords}
galaxies: active -  quasars: emission lines - galaxies: kinematics and dynamics - techniques: interferometric - techniques: image processing - ISM: jets and outflows 
\end{keywords}



\section{Introduction}

How feedback from Active Galactic Nuclei (AGN) operates is one of the major open questions in modern day astronomy. These accreting objects lie in the central regions of galaxies and are so powerful that they can alter the evolution of their host galaxy. This is clearly seen in both observations and simulations. Observations show that the mass of the Super Massive Black Hole \citep[SMBH;][]{kormendy_evidence_1992, kormendy_inward_1995, magorrian_demography_1998} is correlated with the velocity dispersion of their galaxy \citep{gebhardt_relationship_2000,merritt_mbh-sigma_2001}. Additionally, in cosmological simulations, AGN feedback is required to reproduce the observable Universe. For example, these AGN provide crucial heating processes in these models \citep{bower_breaking_2006,croton_erratum_2006}. This feedback mechanism can be understood using a variety of observational techniques \citep[see,][for a review]{harrison_observational_2024}. Although we can see the effects of AGN feedback occurring in our Universe, it still remains unclear how this feedback operates. One research area which can bring us closer to the answer is the study of AGN outflows. \par

AGN outflows can propagate into the interstellar medium (ISM) on kiloparsec scales and insert energy into the surroundings affecting a galaxies evolution \citep[e.g.,][]{fiore_agn_2017, harrison_agn_2018, veilleux_cool_2020}. [O~{\sc iii}] $\lambdaup$5007 \AA\ is a common tracer of ionised gas outflows and we can locate [O~{\sc iii}] outflows by seeing if a second, broad, asymmetric, blueshifted, component is present. [O~{\sc iii}] outflows have been linked to radio emission and investigating this connection helps us understand AGN feedback \citep[e.g.,][]{rawlings_relations_1989, mullaney_narrow-line_2013, nesvadba_gas_2017, zakamska_discovery_2016, alban_mapping_2024}. The narrow line region, where [O~{\sc iii}] originates, typically spans a few kiloparsec, which is on a comparable scale to the radio emission we observe from steep spectrum radio cores \cite[see,][for a review]{miley_structure_1980}, further demonstrating the link between [O~{\sc iii}] and radio emission.

Previous works have begun to investigate the link between radio morphologies and [O~{\sc iii}] emission, however these have been limited by the resolution of large radio surveys. \cite{molyneux_extreme_2019} obtained a sample of 2922 spectroscopically confirmed AGN below a redshift of 0.2, selected to be jet dominated. The authors used this sample to study the relationship between 1.4~GHz radio morphologies with [O~{\sc iii}] kinematics. For their morphological classifications the authors utilised the radio survey Faint Images of the Radio Sky at Twenty-centimeters \citep[FIRST;][]{becker_first_1995} at an angular resolution of $\sim$5\sarc. Using a combination of machine learning and size measurements to classify radio morphology they found that compact sources have the most extreme [O~{\sc iii}] gas kinematics. Another notable work connecting [O~{\sc iii}] to radio emission is  \cite{kukreti_connecting_2024}. The authors build on the work of \cite{kukreti_ionised_2023} using the LOFAR Two-metre Sky Survey \citep[LoTSS;][]{shimwell_lofar_2017, shimwell_lofar_2019, shimwell_lofar_2022} alongside FIRST and the Very Large Array Sky Survey \citep[VLASS;][]{lacy_karl_2020} to derive the spectral slopes for a sample of 5720 radio AGN and they study the [O~{\sc iii}] line profiles using stacking analysis. They discover that peaked spectrum sources show more disturbed gas than non-peaked spectrum AGN, demonstrating that young jets have the strongest impact on [O~{\sc iii}] kinematics. The authors extend this into a morphological study by defining compact radio emission where the FIRST deconvolved (DC) major axis size is $< 3$ \sarc\ and extended radio emission where the DC major axis is $> 3$\sarc. The authors find more distributed gas when a compact morphology is present compared to extended radio emission. 

The studies above investigate radio emission at high frequencies with moderate angular resolution. In contrast, this paper is part of a series studying the connection between the properties of the [O~{\sc iii}] $\lambdaup$5007 \AA\ emission line and radio emission in a novel regime combining low-frequency with high angular resolution down to 0.3\sarc\ allowing us to probe kpc-scale emission.

In our first paper, \cite{escott_unveiling_2025} discovered that AGN detected in the LoTSS Deep Fields DR1 \citep{sabater_lofar_2021, tasse_lofar_2021} have a higher [O~{\sc iii}] outflow detection rate (67.2$\pm$3 per cent), compared to AGN without a detection (44.6$\pm$3 per cent). This confirmed a clear link between low-frequency radio emission at 144~MHz and these ionised gas outflows as traced by [O~{\sc iii}] $\lambdaup$5007.


Using the standard 6\sarc\ resolution, \cite{escott_unveiling_2025} found that the majority of the radio detected AGN did not show a radio excess \citep{delvecchio_vla-cosmos_2017, best_lofar_2023} and therefore the origin of radio emission is not due to high-powered jets. We were unable to refine the origin of radio emission physically driving the increased [O~{\sc iii}] outflow detection rate further because around 90 per cent of the radio detected sample were unresolved. Other possibilities for the emission could be low-powered jets \citep[e.g.][]{maini_compact_2016,jarvis_quasar_2021, njeri_quasar_2025}, shocks produced from wide angled disk winds which are AGN-driven \citep[e.g.][]{zakamska_quasar_2014,hwang_winds_2018,petley_connecting_2022}, or star formation \citep[e.g.][]{vries_star-formation_2007,bonzini_star_2015, padovani_faint_2016, delvecchio_vla-cosmos_2017}. Therefore, the question still remains: Which physical mechanism is driving the link between [O~{\sc iii}] outflows and low-frequency radio emission?

Very long baseline interferometry (VLBI) allows us to produce high-resolution images which provides us with vital information on the radio structures of sources. For example, the LeMMINGs survey conducted by e-MERLIN \citep{baldi_lemmings_2018,baldi_lemmings_2020} provides $\leqslant$0.2\sarc\ resolution images of 280 nearby galaxies at 1.5~GHz while the Very Long Baseline Array (VLBA) can probe parsec-scale jet structures at 15~GHz such as in the MOJAVE survey \citep[Monitoring of Jets in AGN with VLBA Experiments;][]{lister_mojave_2005}. At similar frequencies as the VLBA the European VLBI Network (EVN) is capable of $\sim$0.025\sarc\ \citep[e.g.][]{garrett_agn_2001, panessa_sub-parsec_2013, krezinger_revealing_2024}. Such high-resolution VLBI images give us invaluable information about the structures of radio sources to help entangle the origin of radio emission. While VLBI at high frequencies is relatively common practice, VLBI at low frequencies is particularly challenging due to the complex calibration strategies required to remove ionospheric directional dependent effects from visibilities.

The International LOFAR Telescope (ILT) is a powerful instrument to perform both wide area and deep high-resolution VLBI surveys \citep{morabito_decade_2025}. Three Deep Fields were released alongside LoTSS at standard 6\sarc\ resolution: Lockman Hole, European Large Area Infrared Space Observatory Survey North 1 (ELAIS-N1), and Bo\"{o}tes \citep{sabater_lofar_2021, tasse_lofar_2021}. ELAIS-N1 is now processed at its full depth of 10.7 $\muup$Jy~$\mathrm{beam^{-1}}$ using 505 hours of observations and is, to date, the deepest low-frequency radio image produced \citep{shimwell_lofar_2025}. Creating sub-arcsecond images at 144~MHz is challenging due to ionospheric effects and hence, to date, only two wide-field images have been published at the highest resolution at 144~MHz. \cite{sweijen_deep_2022} published a sub-arcsecond resolution image of the Lockman Hole using a 8-h integration time reaching a central sensitivity of 25~$\muup$Jy~$\mathrm{beam^{-1}}$ with 2483 sources detected. A sub-arcsecond resolution image of ELAIS-N1 with an integration time of 32-hours is published in \cite{de_jong_into_2024} with a depth of 14 $\muup$Jy~$\mathrm{beam^{-1}}$ and 9203 source detections. This paper presents the first high-resolution images of the third Deep Field, Bo\"{o}tes. Therefore, we now have access to a combined area of $\sim$19~$\mathrm{deg^{2}}$ of sub-arcsecond resolution images. A sub-arcsecond image of Euclid Deep Field North \citep[EDFN;][]{bondi_lofar_2024} will soon be available (Bondi et al. in prep) alongside a $\sim$2~$\mathrm{deg^{2}}$ of Abell 2255 \citep[]{rubeis_revealing_2025, rubeis_revealing_2026}.

Now that we have access to three of the LoTSS Deep Fields at sub-arcsecond resolution, covering $\sim$19~$\mathrm{deg^{2}}$, we can revisit the well-defined sample from \cite{escott_unveiling_2025} which lies within the ILT's FoV to obtain their sub-arcsecond radio morphologies. In this paper, we use a combination of the low-frequency, sub-arcsecond morphology alongside brightness temperature measurements \citep{morabito_identifying_2022, morabito_hidden_2025} to identify the physical mechanism driving the low-frequency radio emission and its relation to the presence of [O~{\sc iii}] outflows using our previous measurements from SDSS spectroscopy.

This paper is organised in the following sections: Section \ref{samp} describes the sample selection and the data including a summary of the [O~{\sc iii}] fitting procedure from \cite{escott_unveiling_2025}. Section \ref{data} describes the ILT data as well as presenting the first images of the Bo\"{o}tes field at sub-arcsecond resolution. Section \ref{morph} describes the morphological results. Section \ref{temp} discusses brightness temperature results. A discussion then follows in Section \ref{disscussion} and the conclusions are in Section \ref{conclusion}. In this work we assume a WMAP9 cosmology \citep{hinshaw_nine-year_2013} with $H_{0}$ = 69.32~km~$\mathrm{s^{-1}}$ $\mathrm{Mpc^{-1}}$, $\Omega_{m}$ = 0.287, and $\Omega_{\Lambda}$ = 0.713.

\section{Sample Selection and Definitions} \label{samp}

Here, we briefly summarise the sample selection of the radio detected AGN from \cite{escott_unveiling_2025}. We take optical spectroscopic measurements from SDSS using two catalogues: the SDSS DR16 Quasar catalogue \citep{lyke_sloan_2020} and a broad line AGN catalogue \citep{liu_comprehensive_2019}. We locate sources which lie within the three Deep Fields from LoTSS Deep DR1 and remove AGN with z~>~0.83 to ensure [O~{\sc iii}] is visible in the SDSS spectra. We also remove spectra with a signal to noise (SNR) ratio below 5 and remove a further 31 AGN after visual inspection revealed poor quality spectra, (e.g., missing [O~{\sc iii}] spectral information), leaving a sample of 198 AGN. We crossmatch these sources to the LoTSS Deep Fields DR1 using the catalogue of \cite{kondapally_lofar_2021}. We find that 115 AGN have a detection in LoTSS, which in \cite{escott_unveiling_2025} we coin the "radio detected AGN". In this paper we will only focus on the radio detected population as we are interested in linking the [O~{\sc iii}] outflow detections to sub-arcsecond 144~MHz morphologies. To obtain these sub-arcsecond resolution images we use wide-field images produced using the ILT. The FoV for these images is smaller in comparison to the FoV of the Dutch stations due to the larger size of the international stations. Consequently, 50 radio detected AGN fall out of the international station FoV and the radio detected population decreases to 76. Therefore, for these 76 AGN we already have access to [O~{\sc iii}] fitting results using SDSS, and do not need to conduct further spectroscopic decomposition. 

\subsection{[O~{\sc iii}] Spectral Fitting and Outflow Diagnostics}

We take the [O~{\sc iii}] spectroscopic fits of the 76 ILT detected AGN as described in \cite{escott_unveiling_2025}. To summarise the fitting procedure, we use the fitting module, \texttt{QUBESPEC} \footnote{Available at \url{https://github.com/honzascholtz/Qubespec}} \citep{scholtz_impact_2021}, which incorporates the Markov Chain Monte Carlo approach \citep[MCMC;][]{goodman_ensemble_2010}. We fit various models to the spectra, for example one model will fit [O~{\sc iii}] with a single Gaussian, and another will fit [O~{\sc iii}] with two Gaussians. We also implement Fe-II templates in several of the models, for full details, see \cite{escott_unveiling_2025}. We firstly use the Bayesian Information Criterion (BIC) to select the model with the best fit to the data for each spectrum and confirm the selection via visual inspection. 

In this paper, we are interested in whether the AGN hosts an [O~{\sc iii}] outflow or not. We use the same three diagnostics to determine if an [O~{\sc iii}] outflow is present as \cite{escott_unveiling_2025}. If two Gaussians produce the best fit to the emission line, we class this as an [O~{\sc iii}] fitted outflow. When a single component has the best fit and the width of [O~{\sc iii}] at 80 per cent of the flux ($W_{80}$) is over 800~km~$\mathrm{s^{-1}}$, we class this as an $W_{80}$ outflow, and we class AGN as hosting a $W_{80}$ likely outflow if one component is fitted and 600~km~$\mathrm{s^{-1}}$~<~$W_{80}$~<~800~km~$\mathrm{s^{-1}}$. This leaves AGN which have one Gaussian fitted with $W_{80}$~<~600~km~$\mathrm{s^{-1}}$ to be defined as not hosting an [O~{\sc iii}] outflow.



We acknowledge that due to the complexity of the kinematics of [O~{\sc iii}], the classifications above could lead us to missing outflowing structure in this emission line, for example AGN within our "no outflow" category could also show signs of outflows at smaller scales. For example, in \cite{ward_agn-driven_2024} the authors discuss that models demonstrate that physically slow outflows may not manifest themselves in optical spectroscopy. Therefore these weak outflows would be missed in our AGN within our "no outflow" population, meaning the [O~{\sc iii}] outflows which we trace in this paper and \cite{escott_unveiling_2025} are notably strong [O~{\sc iii}] outflows. 

\begin{figure*}
    \centering
    \includegraphics[width=\linewidth]{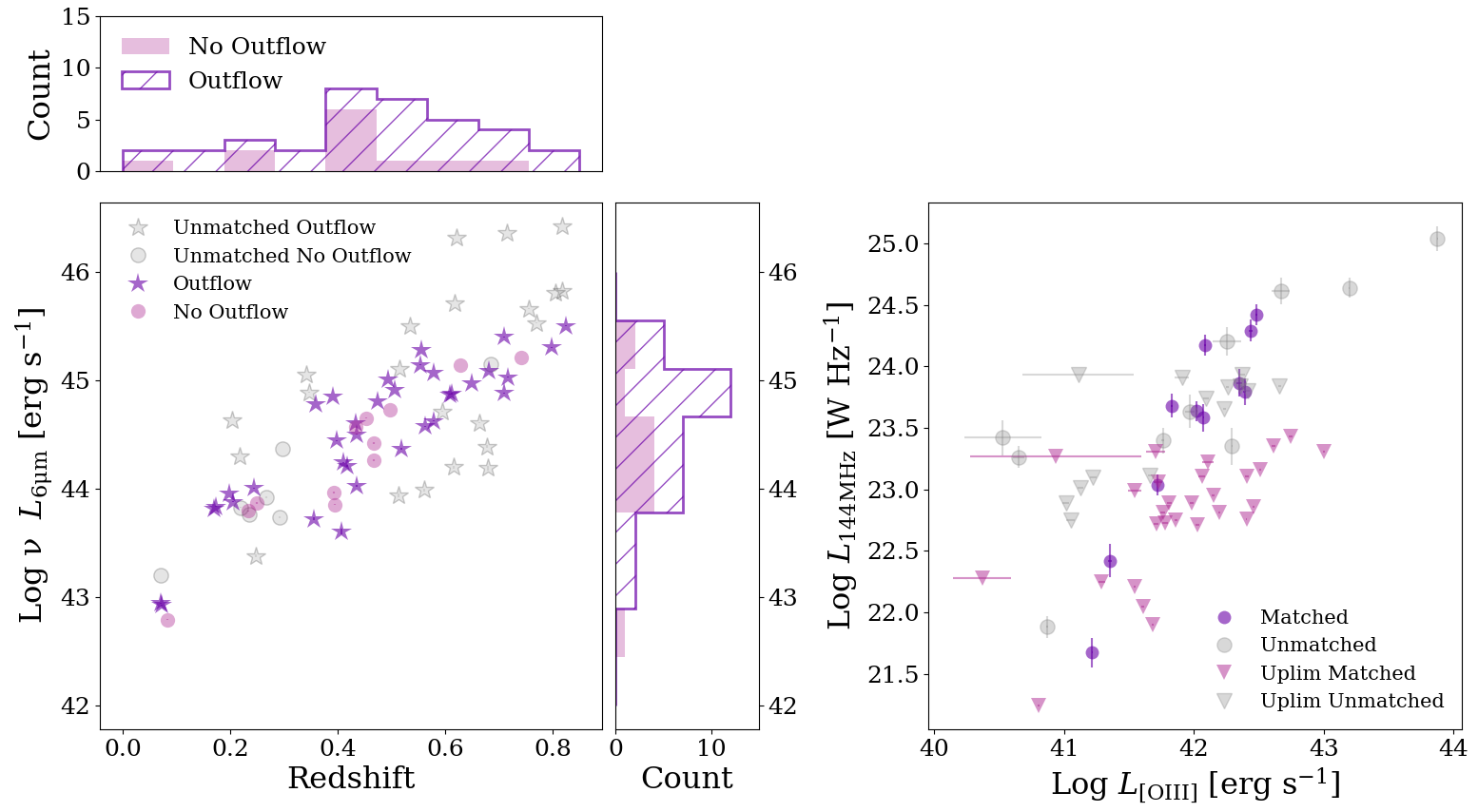}
    \caption{$\textit{Left:}$ The $L_{\mathrm{6\muup\\ m}}$ and redshift relation of the AGN with and without an [O~{\sc iii}] outflow. The coloured markers show the matched populations in $L_{\mathrm{6\muup\\ m}}$ and redshift, with an [O~{\sc iii}] outflow (purple stars) and without an [O~{\sc iii}] outflow (pink circles). The grey points represent the AGN which are removed as these are unmatched, with diamonds portraying the AGN with [O~{\sc iii}] outflows, and circles for the AGN without an [O~{\sc iii}] outflow. We show uncertainties but these are minimal. The top histogram conveys the redshift distribution and the right histogram is the distribution of $L_{\mathrm{6\muup\\ m}}$. The hashed purple histograms are the AGN with [O~{\sc iii}] outflows and the pink solid histograms are the AGN which do not have an [O~{\sc iii}] outflow. $\textit{Right:}$ $L_{\mathrm{144MHz}}$ as a function of $L_{\mathrm{[OIII]}}$, the coloured markers are from the $L_{\mathrm{6\muup\\ m}}$ and redshift matched population we use in this analysis, and the grey are unmatched. $L_{\mathrm{144MHz}}$ in this figure is calculated at 0.3\sarc. Circles represent the detected AGN and downward triangles are the upper radio luminosity limits.}
    \label{fig:lum}
\end{figure*}

To ensure that any morphological differences which appear between the outflowing and non-outflowing populations are not driven by a luminosity bias, we match the two populations in $L_{\mathrm{6\muup\\ m}}$ and redshift, where $L_{\mathrm{6\muup\\ m}}$ is a tracer for AGN bolometric luminosity \citep{richards_spectral_2006}. We calculate this luminosity using the flux densities at 5.8$\muup$m and 8$\muup$m of the $Spitzer$ Deep, Wide-Field Survey \citep[SDWFS;][]{ashby_Spitzer_2009} for AGN in Bo\"{o}tes, and $Spitzer$ Wide-Area Infrared Extragalactic Survey \citep[SWIRE;][]{lonsdale_swire_2003} fluxes for AGN in Lockman Hole or ELAIS-N1. We adopt the same tolerances used in \cite{escott_unveiling_2025} of $\Delta$ $z=$ 0.06 and $\Delta$ log $L_{\mathrm{6\muup\\ m}}$ = 0.3, as these successfully match our data as well as not substantially reducing our sample size. This reduces the overall AGN population from 76 AGN to 47, where we remove outliers in the $L_{\mathrm{6\muup\\ m}}$ and redshift distributions between the [O~{\sc iii}] outflow and non-outflow populations, where 35 exhibit an [O~{\sc iii}] outflow, while the remaining 12 show no evidence of such an [O~{\sc iii}] outflow. We perform a 2D Kolmogorov-Smirnov (KS) test on these matched populations using the public code \texttt{ndtest}\footnote{Written by Zhaozhou Li, \url{https://github.com/syrte/ndtest}} and obtain a p-value of 0.192, confirming that the two populations are statistically indistinguishable. We present our matched $L_{\mathrm{6\muup\\ m}}$ and redshift population as coloured markers in the left subplot in Figure \ref{fig:lum} and the unmatched AGN as grey markers, with AGN with [O~{\sc iii}] outflows presented as stars, and AGN without an [O~{\sc iii}] outflow as circles. The histograms represent the distribution in redshift (top) and $L_{\mathrm{6\muup\\ m}}$ (right), with the purple dashed histogram showing the outflowing population, and solid pink being the no outflow population. We show how many AGN are within each category and population in Table \ref{sample}, including the morphological classes which we discuss in Section {\ref{morph}}.

\begin{table}
    \centering
    \begin{tabular}{ccccc}
    \hline
         \shortstack{[O~{\sc iii}] Outflow Type} & \shortstack{Undetected}  &  \shortstack{Compact} & \shortstack{Extended} & \shortstack{Total} \\
         \hline
         \hline
        [O~{\sc iii}] Fitted Outflow  & 21 (13) & 21 (12) & 5 (4) &  47 (29)\\
        $W_{80}$ Outflow & 2 (1) & 2 (1) & 0 (0) & 4 (2)\\ 
        $W_{80}$ Likely Outflow & 4 (3) & 2 (1) & 0 (0) & 6 (4) \\
        \hline
        All Outflows & 27 (17) & 25 (14) & 5 (4) & 57 (35)\\
        No Outflow & 14 (10) & 5 (2) & 0 (0) & 19 (12)\\
        \hline
        Total & 41 (27) & 30 (16) & 5 (4) & 76 (47) \\
    \end{tabular}
    \caption{Table summarising the number of AGN in each category of [O~{\sc iii}] outflow and morphology type. The numbers in parentheses represent the $L_{\mathrm{6\muup\\ m}}$ and $\textit{z}$ matched population.}
    \label{sample}
\end{table}

We present the relationship between $L_{\mathrm{144MHz}}$ and $L_{\mathrm{[OIII]}}$ on the right-hand side of Figure \ref{fig:lum}, with the $L_{\mathrm{6\muup\\ m}}$ and redshift matched population as coloured markers and circular markers as detected sources and downward triangles as the upper limit for the undetected sources. We calculate the $k$-corrected $L_{\mathrm{144MHz}}$ for the detected AGN using the 0.3\sarc\ flux density, the spectral index, $\alpha$ (where $S_{\nu} \propto \nu^{-\alpha}$), assuming a typical synchrotron spectral index of $\alpha=0.7$ \citep{klein_radio_2018}, and the SDSS spectroscopic redshift. To calculate the associated uncertainties, we use the reported scatter in the flux scaling correction in each field, 25 per cent for Bo\"{o}tes (see Section \ref{postpro}), 18 per cent for Lockman Hole, and 15 per cent for ELAIS-N1 and sum these uncertainties in quadrature with the flux density uncertainty from the source finding for each source. To calculate the upper limit of $L_{\mathrm{144MHz}}$ for the undetected sources at 0.3\sarc, we take the relevant wide-field image and extract the $5\sigmaup$ noise level within a 3\sarc\ radius of the source's location and use this as the associated flux density. We do not report uncertainties for the upper limits, as they are not direct measurements but rather thresholds defined at a specific confidence level. We utilise the $L_{\mathrm{[OIII]}}$ presented in \cite{escott_unveiling_2025} for these sources which are calculated by integrating the continuum-subtracted region between 4975 \AA\ and 5030 \AA\ using spectroscopic redshifts and the distance modulus, and then this is converted to a luminosity. The inverse square root of the inverse variance is taken to compute the associated uncertainties and then the sum of the uncertainties is taken in quadrature. We see that our $L_{\mathrm{6\muup\\ m}}$ and redshift matched population follows a positive correlation between $L_{\mathrm{144MHz}}$ and $L_{\mathrm{[OIII]}}$, as expected.

\section{Data} \label{data}

\subsection{LoTSS Deep Fields}

For this investigation, we use the derived properties from LoTSS Deep Fields DR1 in \cite{kondapally_lofar_2021}. \cite{best_lofar_2023} provides us with star formation rates for sources within the Deep Fields derived from Spectral Energy Distribution (SED) fitting.

We take the sub-arcsecond resolution morphologies and flux densities from the 0.3\sarc\ resolution images at 144~MHz of Lockman Hole \citep{sweijen_deep_2022}, ELAIS-N1 \citep{de_jong_into_2024}, and Bo\"{o}tes. For our sample of AGN with $z<0.83$, these sub-arcsecond resolution images allow us to probe low-frequency radio emission down to scales of $\sim$2~kpc, which now allows us probe down to sub-kiloparsec emission which is not possible at 6\sarc. Across all fields, these 0.3\sarc\ resolution images contain $\sim$14,000 sources at sub-arcsecond resolution.

\subsection{Bo\"{o}tes High-Resolution Images}

\begin{table}
    \centering
    \begin{tabular}{cccc}
    \hline
         \shortstack{Image} & \shortstack{Resolution\\($\mathrm{arcsec^{2}}$)}  &  \shortstack{Sensitivity\\($\muup$Jy~$\mathrm{beam^{-1}}$)} & \shortstack{Source\\Count} \\
         \hline
         \hline
        0.3\sarc\ & 0.50\sarc\ $\times$ 0.34\sarc\ & 33.8  & 4074\\
        0.6\sarc\ & 0.67\sarc\ $\times$ 0.60\sarc\ & 44.2 & 4455 \\
        1.2\sarc\ & 2.23\sarc\ $\times$ 1.03\sarc\ & 81.3 & 2480 \\
        \hline
    \end{tabular}
    \caption{Summary of the three high-resolution Bo\"{o}tes images presented in this work: angular resolution of each image, sensitivity and source counts.}
    \label{tab:image_details}
\end{table}

Here, we release the first images of the Bo\"{o}tes Deep Field at 0.3\sarc, 0.6\sarc, and 1.2\sarc at 144~MHz. The 0.6\sarc\ and 1.2\sarc\ images are important to understand the flux density distribution at different spatial scales. These high-resolution images and catalogues are available on the LOFAR surveys website \footnote{The Bo\"{o}tes high-resolution images and associated catalogues are publicly available at: \url{https://lofar-surveys.org/hd-bootes.html}}.

\begin{figure*}
    \centering
    \includegraphics[width=0.7\textwidth]{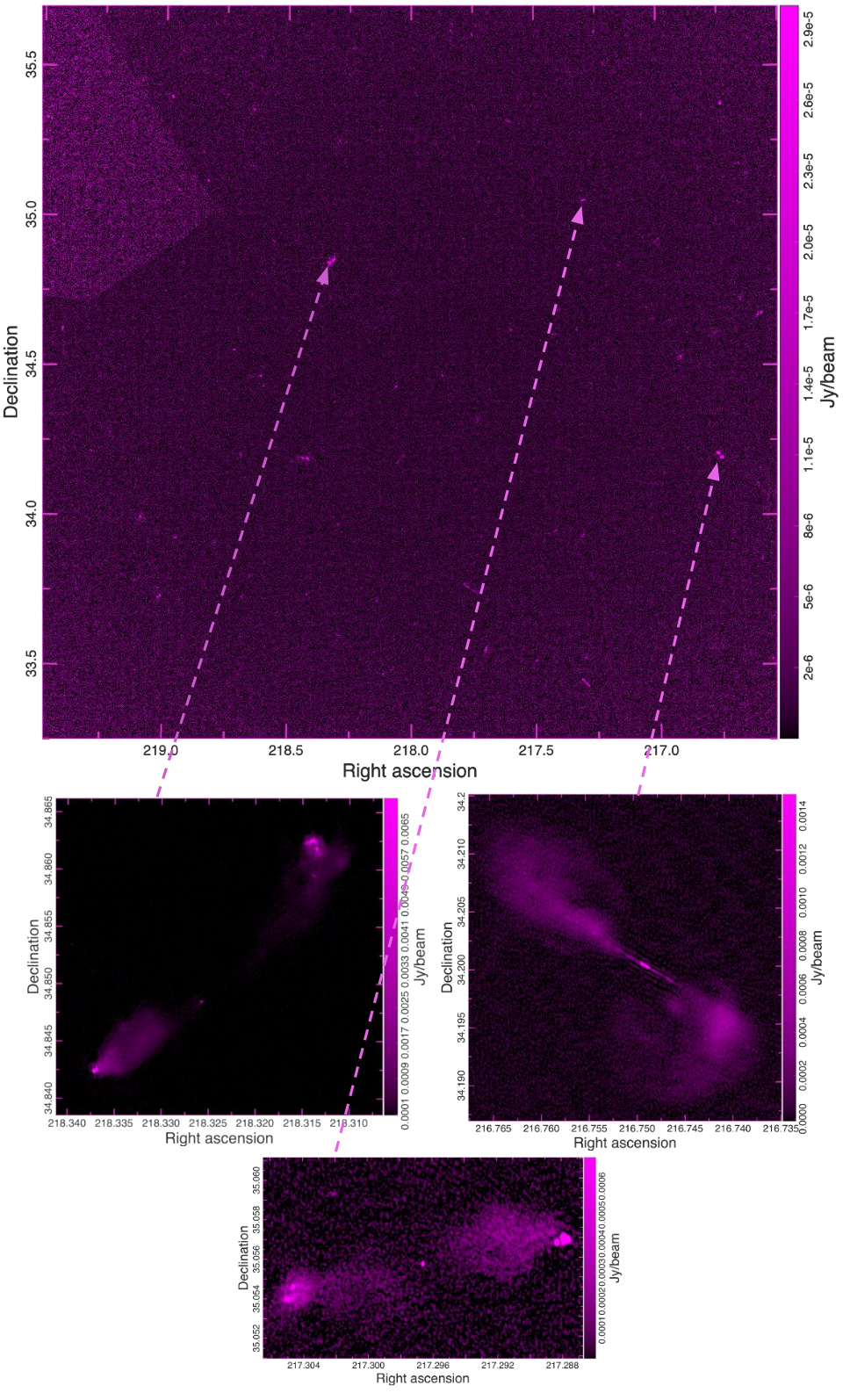}
    \caption{The first $\sim$0.3\sarc\ resolution image of the Bo\"{o}tes Deep Field. This image consists of $\sim$8~billion pixels, contains over 4,000 sources, and has a central sensitivity of 33.8~$\muup$Jy~$\mathrm{beam^{-1}}$. The image is 2.5~$\times$~2.5~$\mathrm{deg^{2}}$ with a restoring beam of 0.50\sarc~$\times$~0.34\sarc.  We highlight three interesting extended sources within this field beneath with the respective locations indicated by arrows. We can now probe radio emission of these sources to a new level thanks to the sub-arcsecond resolution.}
    \label{fig:0.3}
\end{figure*}

We present the 0.3\sarc\ resolution image of the Bo\"{o}tes Deep Field in Figure \ref{fig:0.3}. This image consists of 8.09 billion pixels, has a central sensitivity of 33.8~$\muup$Jy~$\mathrm{beam^{-1}}$, a resolution of 0.50\sarc~$\times$~0.34\sarc, contains 4074 sources with detections $\geq$ 5$\sigmaup$ (where $\sigmaup$ is the local RMS noise), and has an 8-h integration time.

Our 0.6\sarc\ resolution image consists of 2.02 billion pixels with 4455 sources and has a central sensitivity of 44.2~$\muup$Jy~$\mathrm{beam^{-1}}$ with a resolution of 0.67\sarc\ $\times$ 0.60\sarc. At 1.2\sarc\ the pixel number reduces to 0.506 billion with a source count of 2480 and a central sensitivity of 81.3~$\muup$Jy~$\mathrm{beam^{-1}}$ with a resolution of 2.23\sarc\ $\times$ 1.03\sarc.

\subsection{High-resolution Wide-Field Imaging Considerations} \label{high res}

This section briefly outlines the method we use to create the Bo\"{o}tes wide-field images using the ILT. For full details about the publicly available pipeline to produce VLBI images using the ILT, consult \cite{morabito_sub-arcsecond_2022} and for further information on the wide-field aspect of imaging using the ILT at sub-arcsecond, consult \cite{sweijen_deep_2022} and \cite{de_jong_into_2024}. We summarise the details of the data reduction techniques we implement in Appendix \ref{data reduction appen} as it differs slightly from the above papers, due the fact that data processing pipelines were still in development at this time. \par

We retrieve the raw data of a 8-h observation to produce the high-resolution images of Bo\"{o}tes from the Long Term Archive (LTA \footnote{\url{https://lta.lofar.eu/}{}}). The initial flux density calibrator observation has an observation ID of 726520 and the Bo\"{o}tes target field is 726524 with the project ID of LT10\_10. The observation date is 2019-10-20 with a central position of right ascension (RA) $\mathrm{218.0^{\circ}}$ and declination (Dec) $\mathrm{34.5^{\circ}}$. We use a frequency range of 121-166~MHz with 50 stations included in the final images: 24 core stations, 13 remote stations, and 13 international stations without the Swedish station (SE607), which we removed due to poor calibration solution for this station.

\subsubsection{Systematic Effects} \label{sys}

Radio data reduction relies heavily on calibration techniques that remove systematic effects embedded within the data visibilities. These systematic effects are removed by calibrating against a known in-field source, which we call a delay calibrator. An ideal calibrator source is a bright, compact, object which calibrates well through self-calibration cycles. Obtaining corrections using these calibrators allow us to understand how to calibrate for systematic effects across our FoV. We discuss in-field calibration in full using delay calibrators in Section \ref{delay calibration}.

There are two types of systematic effects present in the raw data. These are direction independent effects (DIEs) and direction dependent effects (DDEs). We briefly provide an overview of these effects; consult \cite{de_gasperin_systematic_2019} for full details. \par

\subsubsection{Direction Independent Effects} \label{DIE}

For ILT data reduction, the DIEs requiring correction are the polarisation offsets, the bandpass, and the clock offsets. LOFAR has two data streams, one for each of the X and Y polarisations. During observations these polarisations can deviate from each other, and therefore we must align the polarisations by applying a delay offset between the two streams. We calibrate these polarisations independently, allowing us to probe and correct for the offset between them. In this step, we assume that the calibrator sources are unpolarised. This polarisation offset is only present in the phases and is time independent. We therefore take one station as reference and apply a phase matrix which describes this delay offset causing the XX and YY streams to align. \par

Another DIE which we must correct for are systematic effects present across the bandpass. When we take observations with interferometers such as the ILT, we measure visibilities, and these visibilities have non-physical units. By correcting for the bandpass, we convert these visibility units to the physical units of Janskys (Jy) as well as correcting for any variation in sensitivity across the frequency band. The bandpass affects both the XX and YY polarisation in the same way, however the corrections we require are different for each polarisation stream. These bandpass effects are time independent. \par

The final DIE we must correct for is the offset between different station clocks. Each remote and international station has its own independent clock (all core stations use the same clock) and over time these clocks drift from one another. When correcting for the clock offsets for the remote and international stations, we use the clock of the core stations as reference. All these clocks need to be synchronised to enable coherent signal processing across the vast network of stations which make up the ILT. \par

\subsubsection{Direction Dependent Effects} \label{DDE}

The wide FoV of the ILT provides both advantages and challenges. The primary benefit is the increase in the number of sources that lie within one pointing and hence, in the wide-field mode of LOFAR-VLBI, we can produce wide-field images with thousands of sources within them. However, DDEs pose a significant challenge when reducing ILT data to sub-arcsecond resolution images. The DDEs are the ionospheric effects and the beam. These effects become more prominent at a wide-field scale, as ionospheric conditions exhibit variability on arc-minute scales, and can fluctuate across multiple degrees of the sky. It is therefore essential to correct for these effects when reducing ILT data.

The ILT's dipoles are fixed in place however, because the sources of interest move through the stations beam, any beam corrections we apply during calibration are time variable and the location of the source of interest within the FoV is directional dependent. When the dipole observations are correlated, the time and frequency of the observed phase varies between the stations and therefore this variation needs to be accounted for and thus a beam correction is applied. \par

The ionosphere poses the largest obstacle in LOFAR data reduction. The ionosphere is a layer of charged particles within the Earth's atmosphere. These particles interact with waves propagating through the atmosphere and as a result, these incoming waves become distorted in both phase and amplitude. These distortions have a frequency dependency of $\nu^{-2}$, and hence have a larger effect at low-frequency \citep{intema_ionospheric_2009, intema_deep_2011}. Therefore, in LOFAR observations, the ionosphere is the prevailing cause of DDEs. The effects are dependent on both the magnetic field strength along the line of sight and the Total Electron Content (TEC). The ionosphere poses a greater challenge for wide-field observations, as its conditions vary on degree scales. For VLBI, an even more serious concern is that ionospheric conditions can differ substantially between stations across Europe. \par

\section{Radio Morphological Classification} \label{morph}

\begin{figure}
    \centering
    \includegraphics[width=\linewidth]{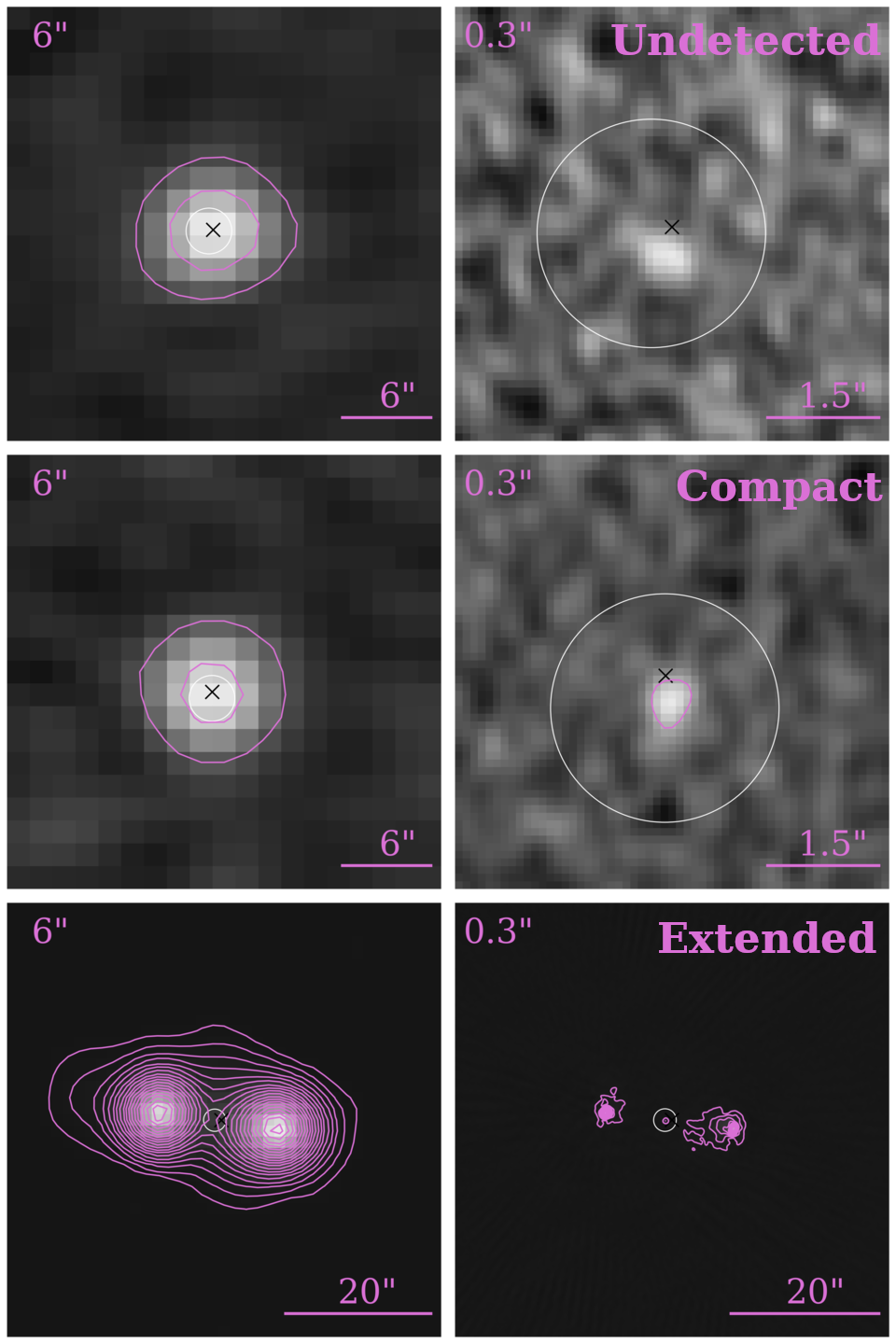}
    \caption{Demonstration of the morphology definitions in this work. \textit{Left:} 6\sarc\ resolution, \textit{right:} 0.3\sarc\ resolution. \textit{Top:} Example of an "undetected" source, which does not have a $5\sigmaup$ detected at 0.3\sarc\, but is detected at 6\sarc. \textit{Middle:} A "compact" morphology. \textit{Bottom:} "Extended" source. The black cross represents the LoTSS position, the white circle is the SDSS fibre and the contours are at a $5\sigmaup$ level.}
    \label{fig:morphs}
\end{figure}

To study the link between the presence of [O~{\sc iii}] outflows and the high-resolution morphologies we must first place each of the 47 of our luminosity matched AGN within our sample into a morphology category. The image we use for this is at the highest resolution: 0.3\sarc. We define three categories: undetected, compact, and extended. AGN are classed as undetected if the source is not within the catalogues published alongside the relevant field's image. We note that for these AGN which are undetected at 0.3\sarc\ resolution, the radio emission detected in the 6\sarc\ resolution image must be below the surface brightness sensitivity in the higher resolution image.

We define compact AGN in two ways. Images can suffer from ionospheric smearing, meaning an unresolved source maybe mistaken for an extended source via visual inspection. In an ideally calibrated image the ratio between the integrated flux density, $S_{I}$, and peak flux density, $S_{P}$, for a unresolved source will be equal to one. However, in reality this is not the case, but the natural logarithm of ($S_{I}$/$S_{P}$) follows a Gaussian distribution \citep{franzen_atlas_2015} and we can utilise this alongside its standard deviation to determine whether a source is resolved. We follow the procedure performed in \cite{shimwell_lofar_2019, shimwell_lofar_2022} to determine if a source is unresolved. We first locate the best candidates for real point sources by finding sources classified as S\_code == $\text{\lq S\rq}$ (pyBSDF has classified this source as a single compact component), and remove sources which are over three times (for Bo\"{o}tes and Lockman Hole) or four times (for ELAIS-N1) larger than the beam's major axis, as these sources could be resolved even if they are a single Gaussian component. For ELAIS-N1 we use four times the major axis of the beam as this provides a better sigmoid fit because ELAIS-N1 uses four observation nights meaning this image suffers from greater ionospheric smearing then the other fields. Also two of the four ELAIS-N1 pointings are averaged to 2 seconds instead of 1, hence contributing to time smearing. We perform a sigmoid curve fit and if a source is below this fit, then the source is unresolved, and hence we classify the source as compact. If a source is above this divide then this is a resolved source and we perform visual inspection on these to classify their morphology as these sources may still be compact. If a source shows spatially resolved structure, we define this AGN as extended, and if the source shows visually compact morphology we define these as compact. To summarise, we class the sources which are determined to be unresolved using \cite{shimwell_lofar_2019, shimwell_lofar_2022} as compact and perform visual inspection on all other sources. We present three sources, one of each morphological class in Figure \ref{fig:morphs}, where the left hand panels show the sources morphology at 6\sarc\ and on the right is the respective 0.3 \sarc\ morphology.

\begin{figure*}
    \centering
    \includegraphics[width=\textwidth]{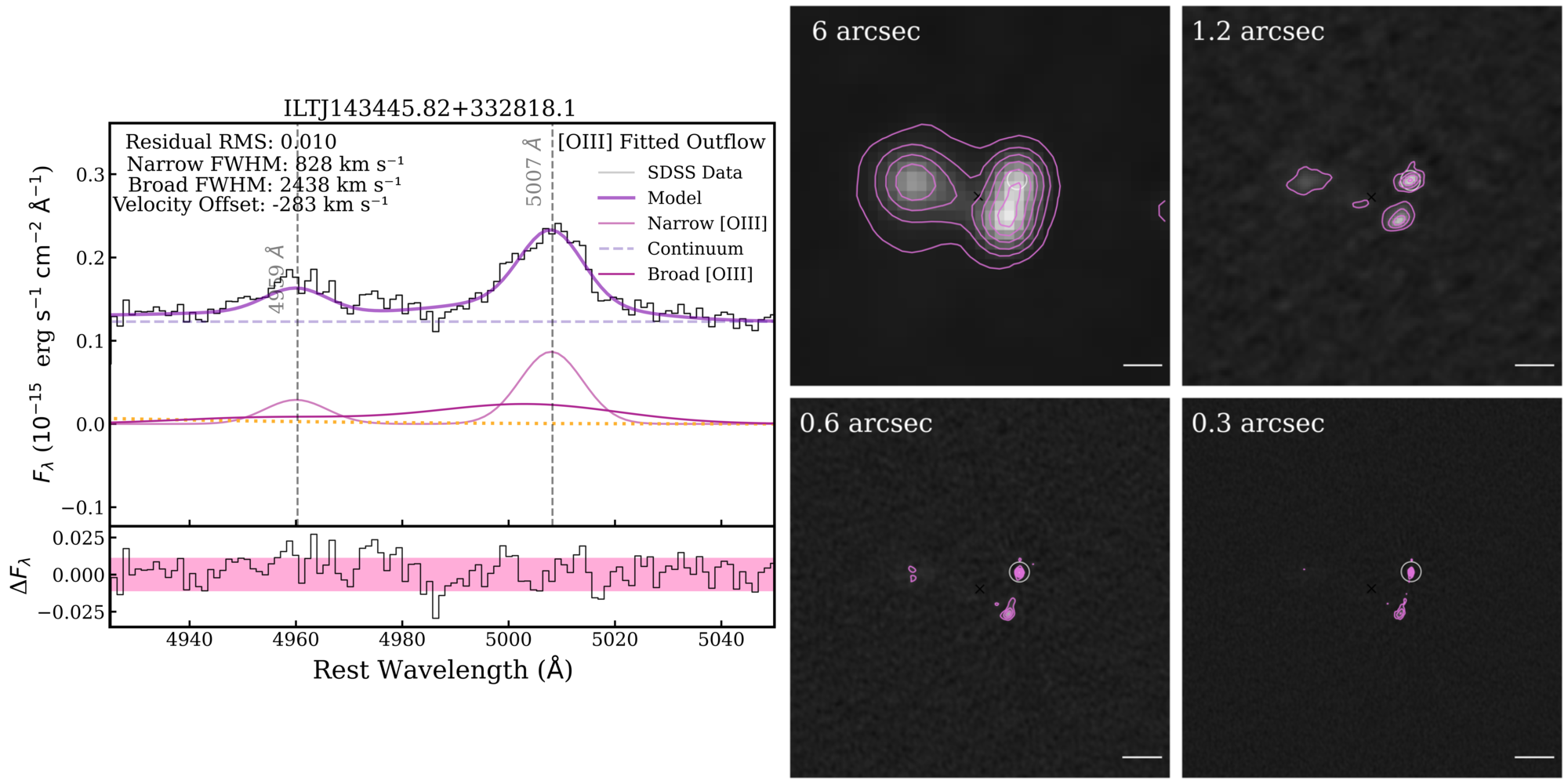}
    \caption{Montage showing the SDSS spectra and various morphologies of ILTJ143445.81+332818.1 ($z=0.197$) which lie within the Bo\"{o}tes field. \textit{Left:} SDSS spectra data (black) with the MCMC fitting results overlaid as a dark purple solid line. We display both the $\lambdaup$4959 \AA\ and $\lambdaup$5007 \AA\ [O~{\sc iii}] emission lines. The pink Gaussians show the narrow component of [O~{\sc iii}], the Gaussians in magenta are the broad component of [O~{\sc iii}] which implies an [O~{\sc iii}] outflow is occurring. The yellow dotted line shows the continuum. The lower panel of this subplot displays the residuals between the fitted model and SDSS data and the pink shaded region corresponds to the 1$\sigmaup$ RMS region which we calculate over the full spectral range of the model. This AGN hosts an [O~{\sc iii}] fitted outflow. \textit{Right:} Cutouts of ILTJ143445.81+332818.1 at multiple resolutions. Starting top left and proceeding clockwise: 6\sarc, 1.2\sarc, 0.3\sarc, and 0.6\sarc\ resolution. The background and the pink 5$\sigmaup$ contours are from the respective resolution image. The scale bar in the bottom right corner demonstrates 6\sarc\ and the white circle near the centre of each cutout is the location of the SDSS fibre. At each resolution this source has been classified as extended, and we believe the radio core is located in the northern components due to its compact structure.}
    \label{fig:spec_montage} 
\end{figure*}

We present an example source from the $L_{\mathrm{6\muup\\ m}}$ and redshift matched population which lies within the Bo\"{o}tes field in Figure \ref{fig:spec_montage}. On the left we see the SDSS spectra for this AGN with the [O~{\sc iii}] MCMC fit \citep[from][]{escott_unveiling_2025} overlaid and from this fit we see that this AGN hosts an [O~{\sc iii}] outflow, specifically a [O~{\sc iii}] fitted outflow as a second, broad, blueshifted Gaussian is present. In the right panels of this figure, we present four cutouts of this source where we show purple 5$\sigmaup$ contours. We start in the top left with the 6\sarc\ LoTSS cutout and proceeding clockwise with the 1.2\sarc, 0.3\sarc, and 0.6\sarc\ images. At all resolutions this source is extended.

\subsection{Connecting Kiloparsec-Scale Radio Morphologies to Ionised Outflows}

To help us understand whether there is a link between ionised outflows and physical processes which produce low-frequency radio emission, we first investigate the relationship between [O~{\sc iii}] outflow detection rates and the sub-arcsecond resolution morphology of the AGN within our sample. In Figure \ref{morph_outflow} we present the link between the various [O~{\sc iii}] outflow detection rates and morphology via two different methods. The detection rate of AGN with [O~{\sc iii}] outflows is shown with purple stars and the detection rate for the population without an [O~{\sc iii}] outflow is shown with pink circles. To estimate uncertainties, we assume a binomial distribution. Which is appropriate given the binary classification of sources as either outflowing or not. If the detection rate is equivalent to either 1 or 0, we do not show uncertainties. The grey dashed line separates the undetected AGN from the detected AGN. The results for Figure \ref{morph_outflow} are also shown quantitatively in Table \ref{tab:morph}.

On the left panel of Figure \ref{morph_outflow} we present the distribution of morphology class for sources with an [O~{\sc iii}] outflow and those without an [O~{\sc iii}] outflow, normalised by each [O~{\sc iii}] outflow category i.e. for the AGN with [O~{\sc iii}] outflows, summing the detection rates across the various morphological classes is equivalent to unity. We see that among AGN without [O~{\sc iii}] outflows, the majority of these AGN, 83.3$\pm$ 11 per cent, are undetected at sub-arcsecond resolution, compared to 48.6$\pm$8 of the outflowing population showing an undetected morphology. This is consistent with the results in \cite{escott_unveiling_2025}. Sources with a compact morphology are more likely to host an [O~{\sc iii}] outflow than not and all AGN which show extended radio structure at 0.3\sarc\ host an [O~{\sc iii}] outflow.

The right panel of Figure \ref{morph_outflow} shows the fraction of the population with/without [O~{\sc iii}] outflows per morphology category. Therefore, the sum of with and without [O~{\sc iii}] outflows within each morphology category is equivalent to unity. We see that in all morphological categories, there are more AGN which host an [O~{\sc iii}] outflow than to those which do not. This panel also reiterates that all four AGN with an extended structure host an [O~{\sc iii}] outflowing structure, and hence there are no AGN without an [O~{\sc iii}] outflow which have an extended morphology. In fact, even if we consider AGN which are not within our AGN luminosity and redshift matched population, all five extended sources have an [O~{\sc iii}] outflow. We also note that, for our extended population, not all of the radio flux density is confined within the SDSS fibre, and therefore we could be missing an [O~{\sc iii}] outflow from the spectroscopy. For example, see Figure \ref{fig:spec_montage}, where the SDSS fibre only covers the the radio core of this extended source at 0.3\sarc\ resolution. However every extended AGN is hosting an [O~{\sc iii}] outflow so we are not missing [O~{\sc iii}] outflows connected to the extended radio emission, but we could be missing additional [O~{\sc iii}] emission associated with the radio emission which does not lie within the SDSS fibre. 

To summarise, for AGN without [O~{\sc iii}] outflows, most are undetected at 0.3\sarc, but are detected at 6\sarc. This indicates that the radio emission has low surface brightness which is not detectable at high-resolution. Synchrotron emission in star-forming galaxies arises from supernovae and their remnants. As the supernova rate is correlated to the star formation rate, synchrotron emission provides a tracer of recent star formation \citep{condon_radio_1992}. We assume that this synchrotron star formation is a widespread phenomena spread across the galaxy occurring on kpc scales \citep[e.g.][]{walter_things_2008, tabatabaei_detailed_2013, heesen_calibrating_2019}, so the radio emission associated with star formation would be of a low surface brightness nature. It is consequently likely that if a source is detected at 6\sarc\ but not 0.3\sarc\ that the dominant radio emission mechanism is synchrotron star formation. If a galaxy has relatively intense star formation or is nearby, the low surface brightness emission could be above the detection limit and would manifest in a compact morphology. In such scenarios we can distinguish whether the radio emission is dominated by star formation or is AGN-driven using brightness temperature measurements (see Section \ref{Brightness}). These assumptions are consistent with \cite{morabito_identifying_2022}.   

In contrast, detected sources at high-resolution, are more likely to have [O~{\sc iii}] outflows than not. To be detected at both large and small scales, these sources have high surface brightness. To produce such high surface brightness the source is likely to be dominated by AGN-driven processes such as wide angle disk winds or radio jets \cite[e.g.][]{homan_intrinsic_2006, potzl_probing_2021, kravchenko_mojave_2025}. To distinguish the radio emission from winds and low-powered jets, simulations suggest we require milli-arcsecond resolution observations \citep{meenakshi_comparative_2024}. Therefore, we discuss wide-angle winds and low-powered jets together as AGN-driven phenomena. As discussed above, it is also possible that star formation can be above the detection limit and therefore we require brightness temperature measurements to determine whether the radio emission's origin is AGN-driven or due to star formation.


\subsubsection{Estimating Physical Sizes} \label{phy_size}

\begin{figure}
    \centering
    \includegraphics[width=\linewidth]{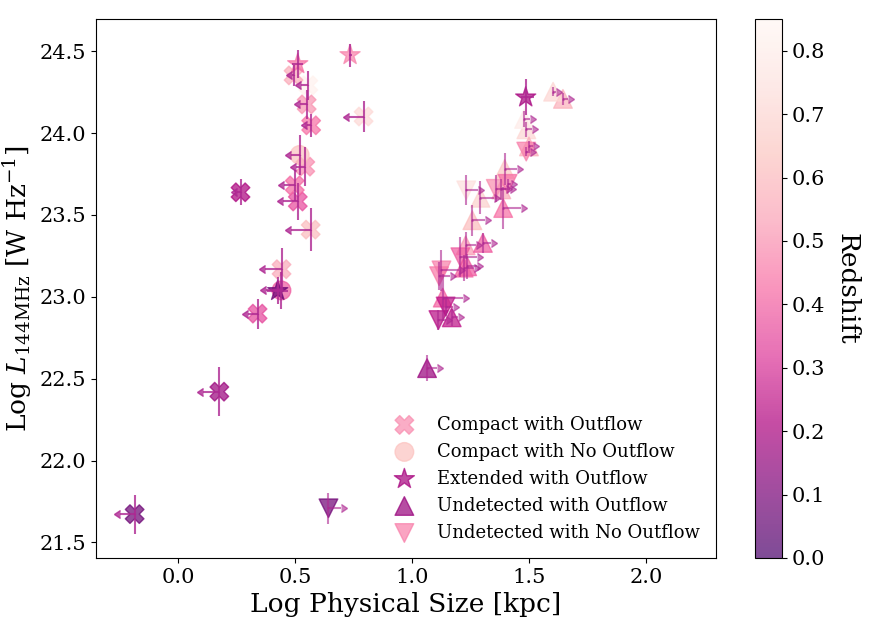}
    \caption{The relationship between the logarithm of the physical size, log $L_{144\mathrm{MHz}}$ and redshift (as traced by a colour bar). We show compact sources with an [O~{\sc iii}] outflow as crosses, compact source without an [O~{\sc iii}] outflow as circles, extended AGN with [O~{\sc iii}] outflows as stars, undetected AGN with an [O~{\sc iii}] outflow as upward-pointing triangles, and undetected AGN without an [O~{\sc iii}] outflow as downward-pointing triangles. The sizes of compact sources are upper limits, while the sizes for undetected AGN are lower limits.}
    \label{fig:size_lum}
\end{figure}

To investigate the structure of our sources at sub-arcsecond resolution, in Figure \ref{fig:size_lum} we examine the relationship between the luminosity at 144~MHz, their physical size, and redshift indicated by the colour bar. For detected sources, we adopt the major axis as a proxy for physical size for sources within Lockman Hole and ELAIS-N1 as well as compact AGN within Bo\"{o}tes, but if a source is extended in Bo\"{o}tes we use the LAS (see Section \ref{catalog}). To estimate the lower limit angular sizes for the undetected sources, we extract the 5$\sigma$ RMS noise from the 0.3\sarc\ image for the relevant field, and using the flux density at 6\sarc, we calculate the major axis required for a source to be detected at 0.3\sarc, assuming a circular morphology. For the associated asymmetric uncertainties, we use the upper and lower bounds of the flux density and then proceed to calculate the major axis. We convert these angular sizes to physical sizes using WMAP9 cosmology and spectroscopic redshift from SDSS. We note for the undetected sources, the $L_{144\mathrm{MHz}}$, and its uncertainties, are measured at 6\sarc\ \citep[presented in][]{escott_unveiling_2025}, whereas for the detected sources, we use 0.3\sarc\ measurements to calculate $L_{144\mathrm{MHz}}$. The use of radio maps with different angular resolutions in calculating $L_{144\mathrm{MHz}}$ may contribute to the apparent bimodality present between the detected and undetected sources.

In Figure \ref{fig:size_lum}, we depict compact sources with an [O~{\sc iii}] outflow as crosses, compact sources without an [O~{\sc iii}] outflow as circles, extended sources with a [O~{\sc iii}] outflow as stars, and the undetected sources as triangles, with upward triangles indicating undetected sources with an [O~{\sc iii}] outflow, and downward triangles as undetected sources without an [O~{\sc iii}] outflow. The sizes of compact objects are upper limits, as they remain unresolved at sub-arcsecond resolution. We observe a positive correlation between the physical size of sources and the luminosity at 144~MHz in all populations, along with a general increase in redshift with both variables. The largest source in our detected population is an extended source in the Bo\"{o}tes field. The undetected sources follow a tight trend with redshift and radio luminosity because their sizes depend on both variables. This population exhibits larger physical sizes than the detected sources, with values between 4~kpc and 39~kpc.

\section{Brightness Temperature as an AGN Diagnostic} \label{temp}

LOFAR, with its sub-arcsecond resolution at low frequencies and high sensitivity, offers surface brightness sensitivity comparable to VLBI observations at higher frequencies. This is therefore an ideal instrument to identify radio emission from AGN \citep{morabito_identifying_2022, morabito_decade_2025}. Star formation is expected to have a maximum surface brightness and anything above this must be produced by an AGN. Using the assumptions in \cite{condon_radio_1992}, at an observed frequency of 144~MHz, the maximum value of brightness temperature produced by star formation is $\gtrsim$$10^{6}$~K. Here we do not explicitly calculate brightness temperature for our sources, but rather the surface brightness. 

Following \cite{morabito_identifying_2022}, we calculate the surface brightness using the total flux density and solid angle of the source, where the solid angle is,

\begin{equation}
    \Omega = \frac{\pi\theta_{1}\theta_{2}}{4\ln2},
\end{equation}

\noindent
where $\theta_{1}$ and $\theta_{2}$ are the major and minor deconvolved axes. To determine the maximum surface brightness for star formation at 144~MHz, we use the same assumptions as \cite{morabito_identifying_2022}~\footnote{Electron temperature of $10^{4}$~K, a synchrotron spectral index of -0.8, and two key frequencies: the observed frequency of 144 MHz, as well as the assumed frequency of 3~GHz where synchrotron emission becomes optically thick.}. We find that 9 of our 16 compact sources are inconsistent with star formation, and are therefore due to an AGN (56.25 per cent), with 7 (43.75 per cent) sources falling below the threshold for AGN activity. We note that the brightness temperature measurements are lower limits as these measurements are conducted on compact sources with upper size limits, therefore the 7 compact sources not classified as AGN at the current resolution could still have low-luminosity AGN cores, for example higher resolution VLBI data can reveal resolved, jet structures where sources are compact on kilo-parsec scales \citep[e.g.][]{gabuzda_survey_1992,giroletti_low-power_2006, kharb_nature_2021}. Therefore, we cannot determine if these sources are consistent with star formation or AGN activity.

\begin{figure}
    \centering
    \includegraphics[width=0.5\textwidth]{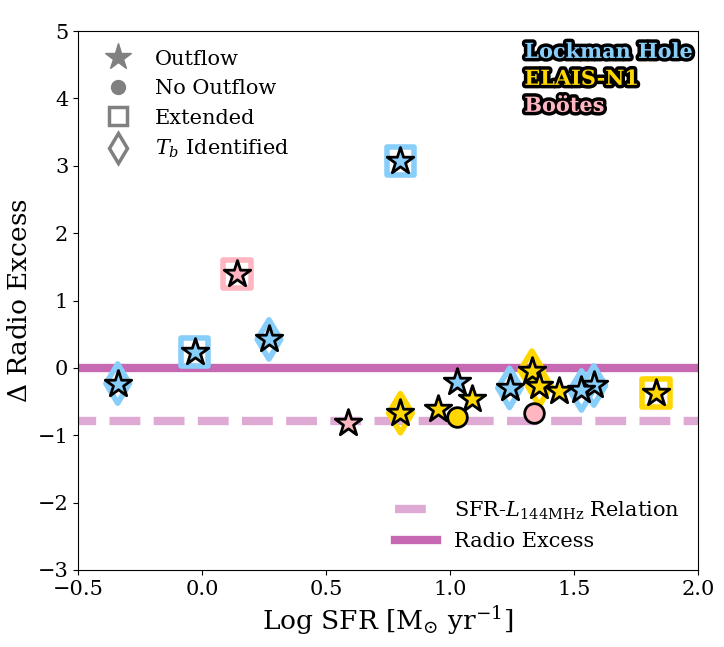}
    \caption{The relationship between $\Delta$ radio excess and log SFR. Only sources with a detection at 0.3\sarc\ contribute to this figure. The pink solid line is the radio excess divide which is centred at $y=0$, with the SFR and $L_{\mathrm{144MHz}}$ relation as the dashed pink line. We define these relations using the Bo\"{o}tes field (pink), and hence we scale sources within Lockman Hole (blue) and ELAIS-N1 (yellow), which have a different radio excess definition, to match the Bo\"{o}tes definition. Star markers represent AGN with [O~{\sc iii}] outflows and circular markers show AGN without [O~{\sc iii}] outflows. Markers surrounded by a square have an extended morphology and sources with a diamond around them are confirmed AGN cores using brightness temperature measurements.}
    \label{Brightness}
\end{figure}

Figure \ref{Brightness} shows the relationship between $\Delta$ radio excess and log star formation rates (SFR), where the SFR are from \cite{best_lofar_2023}, calculated by taking a consensus value from four different SED models. We only show detected sources at 0.3\sarc\ in this Figure and we note one detected source within Bo\"{o}tes is not shown due to lack of SFR information. We flatten the radio excess and SFR-$L_{\mathrm{144MHz}}$ relation to demonstrate the $\Delta$ radio excess as a function of SFR. We do this with respect to the Bo\"{o}tes radio excess definition as this field has a slight redshift adjustment \citep[see][for more details; here we use the maximum redshift of our sample, $z$=0.83, in this redshift dependent relation]{best_lofar_2023} in comparison to the other two fields. We scale sources within Lockman Hole (blue) and ELAIS-N1 (yellow) to match the radio excess definition in Bo\"{o}tes (pink). We note that the $L_{\mathrm{144MHz}}$ we use to obtain radio excess information is calculated using the 6\sarc\ flux density as the SFR-$L_{\mathrm{144MHz}}$ relation is derived using this resolution. Star markers show AGN with an [O~{\sc iii}] outflow and circles show AGN without an [O~{\sc iii}] outflow. Markers with a square around them are extended, and sources with a diamond are confirmed AGN from their $T_{b}$ values.

Figure \ref{Brightness} shows us that 3 of the 4 extended sources are radio excess, demonstrating that the radio emission in these sources are AGN-driven. Combining this with the brightness temperature results, we see that the majority of sources detected at 0.3\sarc\ are AGN dominated. We also note that all sources classified as an AGN via brightness temperature host an [O~{\sc iii}] outflow, thereby compact sources with no [O~{\sc iii}] outflow do not have an AGN identification.

\begin{figure*}
    \centering
    \includegraphics[width=\textwidth]{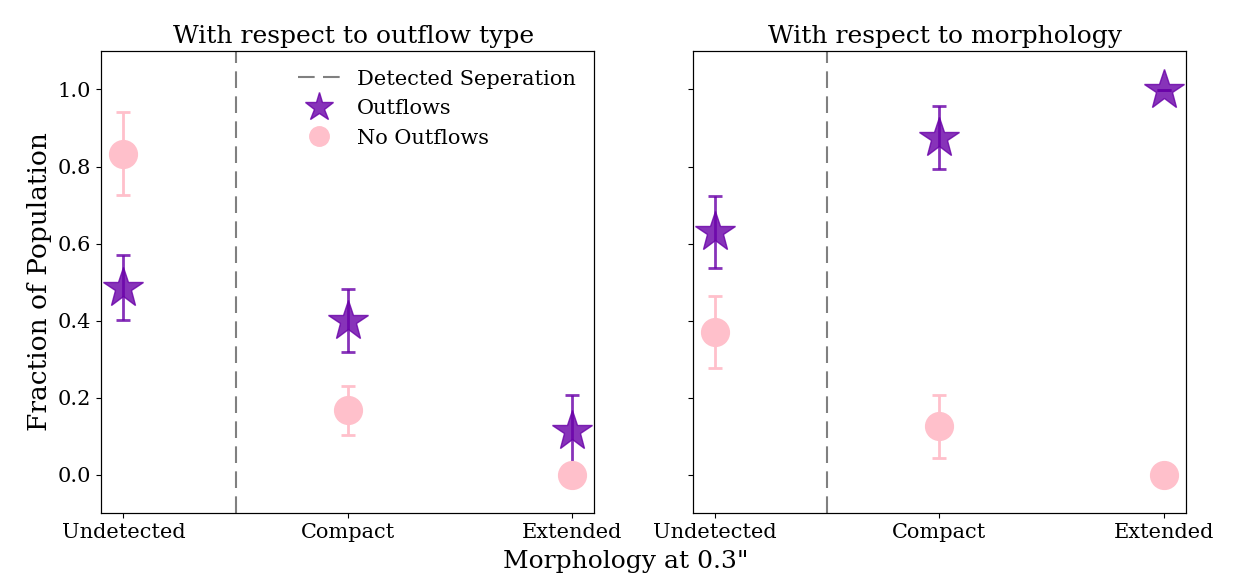}
    \caption{Fraction of population in relation to sub-arcsecond low-frequency radio morphologies, where a radio source is classified as either undetected, compact, or extended. \textit{Left:} Distribution of morphology with respect to [O~{\sc iii}] outflow population, i.e. for each [O~{\sc iii}] outflow population, the fractions of different morphology categories add up to unity. \textit{Right:} Distribution of [O~{\sc iii}] outflow population with respect to the morphology category, i.e. for each morphology class, the fractions of the [O~{\sc iii}] outflow population adds to unity. The pink circle markers demonstrate the no [O~{\sc iii}] outflow population and the purple stars represent the outflowing population. In cases where the fraction of the population is either zero or one, the uncertainty is omitted. The grey dashed line separates the undetected AGN from the detected AGN.}
    \label{morph_outflow}
\end{figure*}

\begin{table}
    \centering
    \textbf{(a) with respect to outflow type}\\[0.5em]
    \begin{tabular}{ccc}
    \hline
         \shortstack{Morphology} & \shortstack{Outflow}  &  \shortstack{No Outflow} \\
         \hline
         \hline
        Undetected & 48.6 $\pm$ 8 & 83.3 $\pm$ 11 \\
        Compact & 40.0 $\pm$ 8  & 17.0 $\pm$ 6 \\
        Extended &  11.4 $\pm$ 9 & 0 $\pm$ 0 \\

        \hline
    \end{tabular}
    \centering
    \vspace{1em}
    \begin{tabular}{cccc}
    \multicolumn{4}{c}{\textbf{(b) with respect to morphology}}\\[0.6em]
    \hline
         \shortstack{Population} & \shortstack{Undetected}  &  \shortstack{Compact} & \shortstack{Extended} \\
         \hline
         \hline
        [O~{\sc iii}] outflow & 63.0 $\pm$ 9 & 87.5 $\pm$ 8 & 100 $\pm$ 0 \\
        No [O~{\sc iii}] outflow & 33.8 $\pm$ 9 & 12.5 $\pm$ 8 & 0 $\pm$ 0 \\
        \hline
    \end{tabular}
    \caption{Tables summarising results shown in Figure \ref{morph_outflow} illustrating the connection between sub-arcsecond 144 MHz radio morphology and [O~{\sc iii}] outflows. Table (a), with respect to [O~{\sc iii}] outflow type. Table (b), with respect to morphology category. Values are percentages and we note that for Table (b), the uncertainties for the complementary categories are the same due to the binomial nature of the distribution.}
    \label{tab:morph}
\end{table}

\section{Discussion} \label{disscussion}

\subsection{ILT Wide-Field High-Resolution Image Comparison} \label{ilt}

Now that we have access to three sub-arcsecond Deep Field images, we will briefly discuss how the 0.3\sarc\ image of Bo\"{o}tes compares to the other 0.3\sarc\ images of Lockman Hole \citep{sweijen_deep_2022} and ELAIS-N1 \citep{de_jong_into_2024}. The source count varies substantially between the three fields, with Lockman Hole having least ($\sim$2500), then Bo\"{o}tes ($\sim$4000), and ELAIS-N1 having the largest source count ($\sim$9000). The increased source count in ELAIS-N1 can clearly be explained because this image combines four different 8-h observations to create an image with a 32-h integration time. This results in the image being significantly deeper than both the Lockman Hole image and the Bo\"{o}tes image, which use single 8-h observations. The increased depth means that ELAIS-N1 probes sources with lower flux densities than Bo\"{o}tes and Lockman Hole. The source count rapidly increases towards fainter radio luminosity and this hence explains the higher source count in this image \citep{shimwell_lofar_2025}.

The variation between source counts from Lockman Hole and Bo\"{o}tes can be understood in two ways. Firstly, significant progress has been made in the calibration strategies that we use to create these ILT wide-field images \citep[see][for more detail]{de_jong_into_2024}. However this is unlikely to be the only reason for the increased source count. In this paper we use a different cataloguing strategy than the previous Deep Fields. For both Lockman Hole and ELAIS-N1 only sources that are also detected in the LoTSS Deep Field images which have an sub-arcsecond resolution counterpart associated are included in these field's catalogues to ensure no false detections are present in them. For Bo\"{o}tes we take a different approach by keeping all sources detected by pyBDSF (which is on the order of 9000) and remove false detections by consulting the LoTSS noise map to see whether the high-resolution flux density recorded could be a noise fluctuation (see Section \ref{catalog} for more detail). This also ensures we do not remove high-resolution components from sources, which are unresolved at 6\sarc\, that have now become resolved at 0.3\sarc. Therefore by not removing these sources we ensure all real detections of extended sources are within the catalogues. 

We further note that the 0.3\sarc, Bo\"{o}tes wide-field image has a higher noise level compared to the Lockman Hole image, despite both using a single 8-h observation. This is likely due to the low declination of the Bo\"{o}tes field because at low declinations calibration becomes more difficult.

\subsection{Origin of Radio Emission from Radio-Quiet AGN} \label{origin}

\begin{figure}
    \centering
    \includegraphics[width=\linewidth]{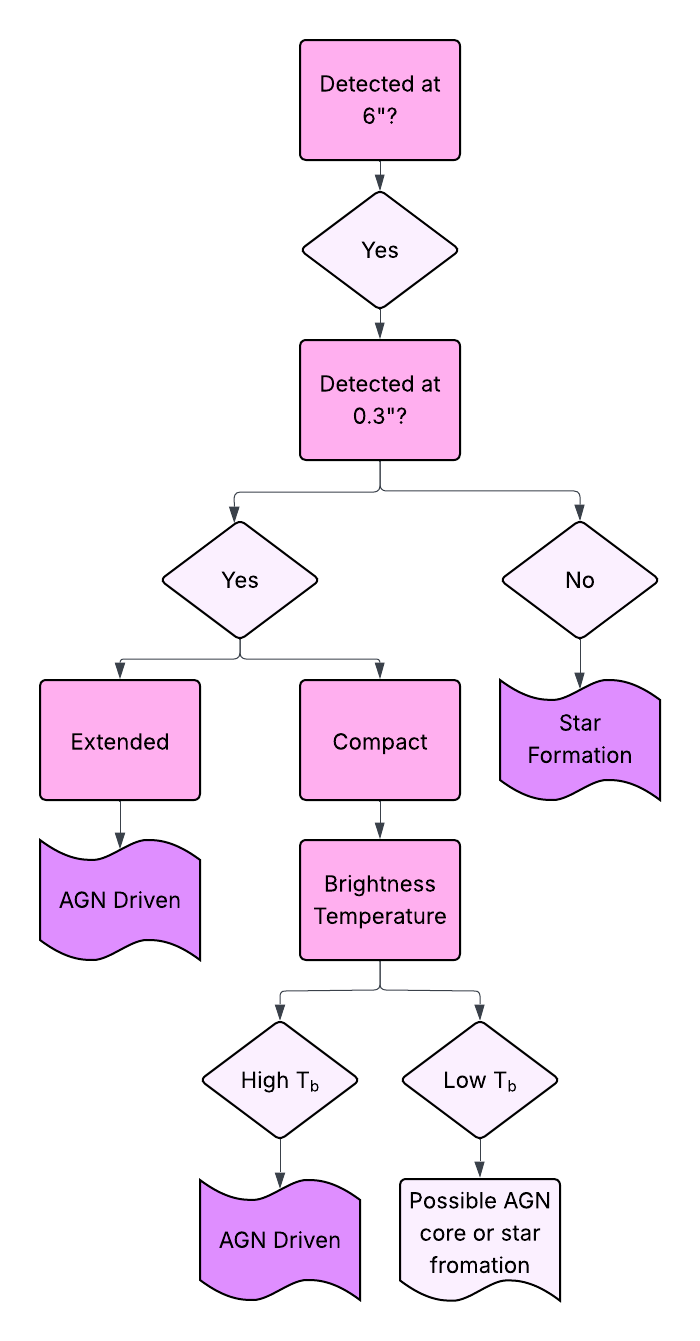}
    \caption{A decision tree demonstrating the procedure we use to determine the origin of radio emission from an AGN.}
    \label{fig:tree}
\end{figure}

Although the radio-loud and radio excess classes are defined in different ways, they both seek to identify where the radio emission is clearly dominated by AGN. The converse to these populations is considered to be `radio-quiet' AGN and it is unclear what is the dominant form of radio emission in these sources \citep[e.g.][]{laor_origin_2008,panessa_origin_2019, baldi_pg-rqs_2021}.

Figure \ref{Brightness} shows the relationship between SFR and radio luminosity at 144~MHz. A similar figure is also shown in \cite{escott_unveiling_2025}, however in this paper we add morphological information alongside brightness temperature identifications. We continue to see the majority of sources to be non-radio excess when a 0.3\sarc\ detection is present. Only four sources lie above the radio excess divide (ILTJ143445.82+332818.1, ILTJ104058.81+581703.4, ILTJ105141.05+591305.4, and ILTJ105421.20+572544.2), and all the AGN showing radio excess host an [O~{\sc iii}] outflow. This tells us that the radio mechanism produced by these radio AGN can not be produced by star formation alone and is likely to be produced by radio jets. Thus the origin of radio emission from 3 of the 4 extended sources is AGN-driven due to their radio excess nature. We would also assume that the radio emission from extended sources at high-resolution is AGN-driven as we can visibility see a radio jet in these sources such as in Figure \ref{fig:spec_montage}. The other radio excess source is compact as well as being a radio AGN core, as confirmed using brightness temperature (ILTJ105421.20+572544.2). As this is a radio AGN core, there will be a radio jet present \citep[e.g.][]{blandford_relativistic_1979} which at this resolution with this frequency is unresolved. So we can now confirm that for the radio excess AGN, the radio emission is produced by radio jets and therefore the dominant radio mechanism is AGN-driven.

For AGN which are not radio excess, we could previously only rule out that the dominant radio emission mechanism was high powered jets. Due to the new high-resolution images, we can now look at the morphologies and brightness temperatures of the sources to determine the dominant radio emission mechanism. In the non-radio excess sources, we see a mix of sub-arcsecond morphologies, containing AGN with compact morphologies, some being a confirmed radio AGN core and some not, and also a single extended AGN. We also note that all AGN which do not host an [O~{\sc iii}] outflow are non-radio excess, show a compact morphology, and are not a detected radio AGN core.

It is also important to highlight that for our 42 radio-quiet AGN, 64.3$\pm$7 per cent are detected at 6\sarc\ and are undetected at 0.3\sarc. As previously discussed, the responsible mechanism of this emission would likely be due to the large scale phenomenon of star formation. Therefore, the radio emission is not due to jet and/or hotspots or AGN core emission. For clarity, we present a decision tree in Figure \ref{fig:tree} which demonstrates the procedure in which we determine the origin of radio emission depending on both morphologies and brightness temperature.

We note that for sources detected at 6" but not at 0.3", the radio emission could originate from remnant emission associated with previous AGN activity rather than star formation. However, \cite{jurlin_multi-frequency_2021} shows that only 7 per cent of a sample of active, restarted and remnant candidates identified by \cite{brienza_search_2017} are confirmed remnant radio sources. Given the expected scarcity of these sources, we therefore classify the sources undetected at 0.3" as star-forming. To confirm whether the radio emission arises from remnant AGN activity rather than star formation requires spectral index measurements. If a significant fraction of these sources are dominated by remnant AGN emission, this would imply a long remnant phase and a correspondingly low AGN duty cycle, given the high fraction of sources undetected at 0.3".


Summing all this up, in this population of 42 non-radio excess AGN, i.e. radio-quiet AGN, there appears to be a combination of mechanisms driving the radio emission. Star formation appears to drive 64.3$\pm$7 per cent (27 out of 42) of these AGN, as the majority of sources are not detected at small scales (0.3\sarc). Small scale AGN-driven emission is also driving a considerable amount of the radio emission as 19.0$\pm$6 per cent (8 out of 42) show either an extended morphology, indicating small scale jets, or are a high $T_b$ core.

\subsection{Driving Radio Mechanism of [O~{\sc iii}] outflows} \label{OIII radio}

\begin{figure}
    \centering
    \includegraphics[width=\linewidth]{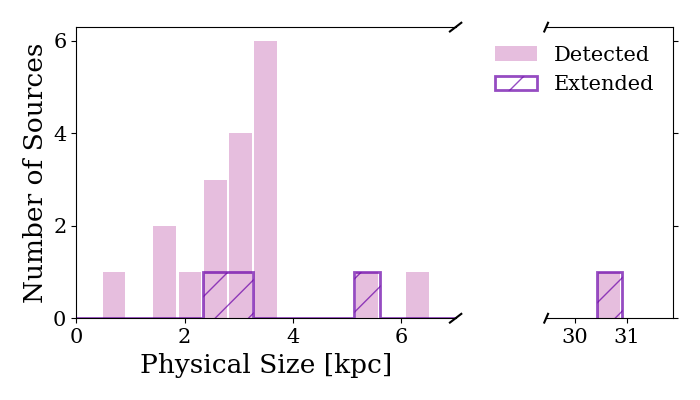}
    \caption{Histogram showing the distribution of the physical sizes in kpc of the detected sources at 0.3\sarc. The pink solid histogram shows all detected AGN within the $L_{\mathrm{6\muup\\ m}}$ and $\textit{z}$ matched population. The purple dashed histogram shows the sizes of the extended AGN. The figure includes a break in the x-axis to account for a single outlying source with a physical size of approximately 30.5~kpc.}
    \label{fig:sizes}
\end{figure}

We discuss how Figure \ref{morph_outflow} provides us with insight about the relationship between sub-arcsecond resolution morphology and the detection rates in the [O~{\sc iii}] outflow and no [O~{\sc iii}] outflow populations. We see that the majority of AGN without an [O~{\sc iii}] outflow are undetected at 0.3\sarc, but detected at 6\sarc. Therefore, the emission from undetected AGN is produced on large scales, as demonstrated in Figure \ref{fig:size_lum}, and hence is consistent with star formation. We show the physical size distribution of our detected sources in Figure \ref{fig:sizes} (see Section \ref{phy_size} for details about size calculation). The physical sizes are all under 7~kpc, aside from one extended source which is $>$30~kpc. Hence, emission on the scale of 0.3\sarc\ is small-scale.

Around half of the AGN with [O~{\sc iii}] outflows are undetected at 0.3\sarc, also indicating that this emission is driven by star formation. The remaining half of AGN with [O~{\sc iii}] outflows are detected at 0.3\sarc, with all extended AGN and all high $T_b$ cores hosting an [O~{\sc iii}] outflow. This implies that the radio emission is consistent with AGN-driven activities.
 
The differences we see between the outflowing population and non-outflowing population is not driven by an AGN luminosity bias as we have matched in $L_{\mathrm{6\muup\\ m}}$ and $\textit{z}$ and we note that the radio luminosity at 0.3\sarc\ for the extended sources, is not significantly higher than the radio luminosity for the compact sources ($<$2$\sigmaup$).

[O~{\sc iii}] traces recent nuclear activity as it is produced from a region about 1 kpc in size, however the radio emission produced by extended radio structure which can stretch 100s of kpcs occur on a much longer time scale of $10^{8}$ years. Therefore if the central engine has recently been shut down, the narrow line region, and hence [O~{\sc iii}], may no longer be detected in the AGN, although the radio lobes will still be present. Therefore, for our detected sources, where we are mostly probing emission on scales $<$10~kpc, the timescales of nuclear activity traced by [O~{\sc iii}] are likely comparable to those of the associated radio emission. In contrast, for our undetected sources the emission is on larger scales, which is unlikely to be physically associated with the same timescales as the nuclear [O~{\sc iii}] activity.

Putting all this together, we are seeing a physical link between small scale radio emission due to AGN activity, and ionised gas outflows, as traced by [O~{\sc iii}]. The majority of sources which do not host an [O~{\sc iii}] outflow are undetected at 0.3\sarc, whereas the majority of sources with a compact morphology host an [O~{\sc iii}] outflow. Furthermore, all the sources with a compact morphology that are also a $T_{b}$ core AGN, host an [O~{\sc iii}] outflow and all spatially resolved sources host an [O~{\sc iii}] outflow. This supports the idea that small scale emission produced by the AGN core, radio jets, disk winds, and/or hotspots plays a key role in driving these [O~{\sc iii}] outflows, while more diffuse, large scale emission from star formation is not a main driver. However, it appears overall that for AGN with [O~{\sc iii}] outflows there is around equal amounts of small and large scale emission contributing to the driving mechanism of these [O~{\sc iii}] outflows as $\sim$49$\pm$8 per cent of AGN with [O~{\sc iii}] outflows are undetected in the 0.3\sarc\ image.

In Appendix \ref{CDF} we investigate how [O~{\sc iii}] kinematics vary between the detected AGN population and the undetected population. This is similar analysis to \cite{escott_unveiling_2025}, although the authors investigate the radio emission at 6\sarc, whereas here we are probing emission at 0.3\sarc. We see similar enhanced integrated flux of the broad area as well as enhanced $W_{80}$ in the detected AGN compared to the undetected AGN. This suggests the enhancement we see is not driven by the radio emission which is resolved out between 6\sarc\ and 0.3\sarc. Due to the limited sample size and lack of uncertainties we do not draw conclusions from Figures \ref{fig:CDF_area} and \ref{fig:CDF_w80}.

\subsection{Comparing Kpc-Scale Radio Emission and [O~{\sc iii}] Studies}

When investigating the origin of radio emission it is important to compare the results we see in this work to other work conducted at other frequencies and resolutions. \cite{jarvis_prevalence_2019} follows up 10 low redshift radio quiet AGN which are originally selected in \cite{mullaney_narrow-line_2013} with images from the VLA and e-MERLIN. VLA images are between 0.3 and 1 arcseconds with a frequency range of 1.5-6~GHz and e-MERLIN images having a resolution around 0.25\sarc\ at 1.5~GHz for 9 targets and they discover that star formation only accounts for $\sim$10 per cent of the radio emission with around 90 per cent of these 9 targets showing extended radio structures on 1-25~kpc scales.

\cite{njeri_quasar_2025} presents similar work but at higher resolution with 42 e-MERLIN images at 6~GHz of low redshift quasars from the Quasar Feedback Survey. These images allow the authors to investigate sub-kiloparsec emission. Similarly to this study the authors use a combination of morphology and brightness temperature to classify their AGN. Combining these e-MERLIN and previous VLA images from \cite{jarvis_quasar_2021} they find that over 86 per cent of these quasars are classified as radio AGN and therefore this emission is likely to be dominated by jet driven lobes and outflow driven shocks.

The results from \cite{jarvis_prevalence_2019, jarvis_quasar_2021} and \cite{njeri_quasar_2025} appear to be contradictory to the results we present in this work, as we see that the majority of our matched AGN (27 out of 47) are undetected at 144~MHz and 0.3\sarc\ resolution and therefore the majority of the radio emission appears to be from star formation. A key reason for the difference could be due to the low redshift nature of the samples presented in these as the AGN from these studies are below 0.2 redshift, whereas our sample spans to around 0.85. We note that the four extended sources within our sample are at low redshift. The high angular resolution images from the VLA and e-MERLIN in \cite{jarvis_prevalence_2019, jarvis_quasar_2021} are at GHz frequencies whereas the high-resolution images for our sample are at MHz frequencies so we are probing a different regime of emission where the radio emission from the ILT is dominated by synchrotron emission. At GHz frequencies, there is an increased contribution from Bremsstrahlung emission. We also note that the selection of their targets is based on strong [O~{\sc iii}] emission whereas our sample is optically selected and independent of [O~{\sc iii}] properties until we split the population into sources with an [O~{\sc iii}] outflow and those without. We see a clear correlation with [O~{\sc iii}] outflows in our sample and resolved emission, therefore it may be the presence of powerful [O~{\sc iii}] emissions that could be driving the kpc-scale and sub-kpc-scale resolved emission that is seen in \cite{jarvis_prevalence_2019, jarvis_quasar_2021} and \cite{njeri_quasar_2025}. Finally, the Quasar Feedback Survey targets AGN with AGN luminosity above $10^{45}$ erg~$\mathrm{s^{-1}}$, whereas all our AGN have an AGN luminosity below $10^{45}$ erg~$\mathrm{s^{-1}}$. The radio luminosities of sources in this survey is higher than ours. Taking a spectral index of -0.7 our sample probes emission down to $L_{\mathrm{1.4GHz}}\sim\mathrm{10^{20}}$~W~$\mathrm{Hz^{-1}}$. Therefore, these studies and ours lie in a different regime of both radio and AGN luminosity.

This high-resolution resolved study of the connection between [O~{\sc iii}] and radio emission is consistent with previous unresolved analysis, which attribute the link to AGN activity \citep[e.g.,][]{molyneux_extreme_2019, kukreti_feedback_2025}. In addition, resolved Integral Field Unit (IFU) studies of individual sources, or small samples, have demonstrated that radio jets strongly interact with the ISM \citep[e.g.,][]{venturi_magnum_2021,girdhar_quasar_2022,speranza_multiphase_2024}, in agreement with predictions from simulations \citep{meenakshi_modelling_2022}. IFU follow-up for the resolved, extended sources in our sample will therefore help us to disentangle the spatial interplay between ionised outflows and the ISM.

\section{Conclusions} \label{conclusion}

We present sub-arcsecond, kpc-scale morphological results for 47 optically selected AGN from \cite{escott_unveiling_2025}, matched in $L_{\mathrm{6\muup\\ m}}$ and redshift, which lie within the ILT FoV of the LoTSS Deep Fields. Using wide-field VLBI techniques, we additionally release the first high-resolution images of the Bo\"{o}tes Deep Field at 144~MHz, presented at $\sim$0.3\sarc, $\sim$0.6\sarc, and $\sim$1.2\sarc. The $\sim$0.3\sarc\ image achieves a sensitivity of 33.8~$\muup$Jy~$\mathrm{beam^{-1}}$ with 4074 5$\sigmaup$ source detections.

For the 47 sources for which we have both [O~{\sc iii}] information and kpc-scale radio images, we define three morphological categories: undetected AGN, where these sources are detected at 6\sarc\ but are undetected at 0.3\sarc, compact AGN, which are unresolved at this resolution, and extended AGN, which show spatially resolved emission. We further separate our sample into two populations: AGN which host an [O~{\sc iii}] outflow (35 AGN), and another where the AGN do not host an [O~{\sc iii}] outflow (12 AGN). Below we present the key findings in this work:

\begin{itemize}[align=parleft, left=\leftmargin]
    \item Outflows traced by [O~{\sc iii}] appear to be AGN-driven for the following reasons:
    \begin{itemize}
        \item If AGN are detected at both small, kpc-scales (0.3\sarc) and large-scales (6\sarc), 90$\pm$7 per cent host an [O~{\sc iii}] outflow.
        \item All 9 detected compact sources which are a high $T_b$ radio core host an [O~{\sc iii}] outflow.
        \item All four detected, extended sources host an [O~{\sc iii}] outflow
        \item 87.5$\pm$8 per cent of sources with a compact morphology host an [O~{\sc iii}] outflow.
    \end{itemize}
    \item AGN without an [O~{\sc iii}] outflow are likely dominated by synchrotron star formation because:
    \begin{itemize}
        \item 83.3$\pm$11 per cent of sources without an [O~{\sc iii}] outflow are undetected on kpc-scales.
    \end{itemize}
    \item The origin of radio emission from the radio-quiet AGN is dominated by star formation as:
    \begin{itemize}
        \item 64.3$\pm$7 per cent of the non-radio excess AGN are undetected on small-scales.
        \item A considerable amount of the radio emission does appear to be AGN-driven as 19.0$\pm$6 per cent of the radio-quiet population are either extended or are a high $T_{b}$ AGN core.
    \end{itemize}
\end{itemize}

We acknowledge that our sample size is limited, with 35 AGN with [O~{\sc iii}] outflows and 12 without, primarily due to the limited number of sources with SDSS spectroscopy within the ILT coverage of the LoTSS Deep Fields. The upcoming deep tier of the WEAVE-LOFAR survey \citep{smith_weave-lofar_2016} is expected to produce an estimated 130,000 spectroscopic measurements of bright sources in these fields, many with sub-arcsecond morphologies. Furthermore, the first data release from the Dark Energy Spectroscopic Instrument \citep[DESI; ][]{desi_collaboration_data_2025} will also increase spectroscopic coverage. Together with the intermediate resolution images of Bo\"{o}tes and ELAIS-N1 \citep{de_jong_into_2024,ye_1-arcsecond_2024} at $\sim$0.6\sarc\ and $\sim$1.2\sarc, we can extend our high-resolution studies to investigate emission too diffuse to detect at the highest resolution.

\section*{Acknowledgements}

We thank Daniel Smith for his useful insights on this paper. ELE and LKM are grateful for support from the Medical Research Council [MR/T042842/1]. FS and LKM appreciate the support of STFC [ST/Y004159/1]. JMGHJdJ acknowledges support from project CORTEX (NWA.1160.18.316) of research programme NWA-ORC, which is (partly) financed by the Dutch Research Council (NWO). JMGHJdJ and RvW acknowledge support from the OSCARS project, which has received funding from the European Commission’s Horizon Europe Research and Innovation programme under grant agreement No. 101129751. CMH acknowledges funding from an United Kingdom Research and Innovation grant (code: MR/V022830/1). IP acknowledges support from INAF under the Large Grant 2022 funding scheme (project “MeerKAT and LOFAR Team up: a Unique Radio Window on Galaxy/AGN co-Evolution”).

This paper is based on data obtained with the International LOFAR Telescope (ILT). LOFAR \citep{haarlem_lofar_2013} is the low-frequency Array designed and constructed by ASTRON. It has observing, data processing, and data storage facilities in several countries, that are owned by various parties (each with their own funding sources), and that are collectively operated by the ILT foundation under a joint scientific policy. The ILT resources have benefited from the following recent major funding sources: CNRS-INSU, Observatoire de Paris and Université d'Orléans, France; BMBF, MIWF-NRW, MPG, Germany; Science Foundation Ireland (SFI), Department of Business, Enterprise and Innovation (DBEI), Ireland; NWO, The Netherlands; The Science and Technology Facilities Council, UK; Ministry of Science and Higher Education, Poland; The Istituto Nazionale di Astrofisica (INAF), Italy. \par

This research made use of the Dutch national e-infrastructure with support of the SURF Cooperative (e-infra 180169) and the LOFAR e-infra group. The Jülich LOFAR Long Term Archive and the German LOFAR network are both coordinated and operated by the Jülich Supercomputing Centre (JSC), and computing resources on the supercomputer JUWELS at JSC were provided by the Gauss Centre for Supercomputing e.V. (grant CHTB00) through the John von Neumann Institute for Computing (NIC). \par

This work used the DiRAC at Durham facility managed by the Institute for Computational Cosmology on behalf of the STFC DiRAC HPC Facility (www.dirac.ac.uk). The equipment was funded by BEIS capital funding via STFC capital grants ST/P002293/1, ST/R002371/1 and ST/S002502/1, Durham University and STFC operations grant ST/R000832/1. DiRAC is part of the National e-Infrastructure. \par

This work has been enabled by access to facilities and the scientific and technical support provided by the UK SKA Regional Centre (UKSRC). The UKSRC is a collaboration between the University of Cambridge, University of Edinburgh, Durham University, University of Hertfordshire, University of Manchester, University College London, and the UKRI STFC Scientific Computing (STFC) at RAL. The UKSRC is supported by funding from the UKRI STFC.

This work has made use of the University of Hertfordshire's high-performance computing facility. This work also made use of SciPy, NumPy, Matplotlib and Astropy.
\section*{Data Availability}

The complete catalogue of the 198 AGN presented in \cite{escott_unveiling_2025}, including the 76 AGN detected in the ILT which we discuss in this paper,  will be available on CDS on publication. This catalogue includes the [O~{\sc iii}] fitting properties.

The Bo\"{o}tes high-resolution images and associated catalogues will be publicly available at the time of publication at: \url{https://lofar-surveys.org/hd-bootes.html}. The Lockman Hole high-resolution image and associated catalogue are publicly available at: \url{https://lofar-surveys.org/hdfields.html}. The ELAIS-N1 high-resolution images and catalogues are publicly available at: \url{https://lofar-surveys.org/hd-en1.html}.

We use the SDSS Quasar catalogue from DR16 which can be accessed at \url{https://www.sdss4.org/dr17/algorithms/qso_catalog/} and the broad-line AGN catalogue from SDSS DR7 which can be found at \url{https://cdsarc.u-strasbg.fr/cgi-bin/ftp-index?/ftp/cats/J/ApJS/243/21}. The SDSS spectra used in the spectral fitting can be downloaded from \url{https://skyserver.sdss.org/dr19/SearchTools/SQS} by uploading the source’s location or plate information.



\bibliographystyle{mnras}
\bibliography{references} 

@article{rubeis_revealing_2026,
	title = {Revealing the intricacies of radio galaxies and filaments in the merging galaxy cluster {Abell} 2255. {II}. {Properties} of filaments using multi-frequency radio data},
	copyright = {https://www.edpsciences.org/en/authors/copyright-and-licensing},
	issn = {0004-6361, 1432-0746},
	url = {https://www.aanda.org/10.1051/0004-6361/202557651},
	doi = {10.1051/0004-6361/202557651},
	urldate = {2026-01-27},
	journal = {Astronomy \& Astrophysics},
	author = {Rubeis, E. and Bondi, M. and Botteon, A. and Van Weeren, R.J. and De Jong, J.M.G.H.J. and Brunetti, G. and Rudnick, L. and Brüggen, M. and Bruno, L. and Escott, E.L. and Gheller, C. and Morabito, L.K. and Rajpurohit, K. and Röttgering, H.J.A.},
	month = jan,
	year = {2026},
}

@article{jurlin_multi-frequency_2021,
	title = {Multi-frequency characterisation of remnant radio galaxies in the {Lockman} {Hole} field},
	volume = {653},
	copyright = {© ESO 2021},
	issn = {0004-6361, 1432-0746},
	url = {https://www.aanda.org/articles/aa/abs/2021/09/aa40102-20/aa40102-20.html},
	doi = {10.1051/0004-6361/202040102},
	language = {en},
	urldate = {2026-01-23},
	journal = {Astronomy \& Astrophysics},
	publisher = {EDP Sciences},
	author = {Jurlin, N. and Brienza, M. and Morganti, R. and Wadadekar, Y. and Ishwara-Chandra, C. H. and Maddox, N. and Mahatma, V.},
	month = sep,
	year = {2021},
	pages = {A110},
}

@article{brienza_search_2017,
	title = {Search and modelling of remnant radio galaxies in the {LOFAR} {Lockman} {Hole} field},
	volume = {606},
	issn = {0004-6361, 1432-0746},
	url = {http://arxiv.org/abs/1708.01904},
	doi = {10.1051/0004-6361/201730932},
	urldate = {2026-01-23},
	journal = {Astronomy \& Astrophysics},
	author = {Brienza, M. and Godfrey, L. and Morganti, R. and Prandoni, I. and Harwood, J. and Mahony, E. K. and Hardcastle, M. J. and Murgia, M. and Röttgering, H. J. A. and Shimwell, T. W. and Shulevski, A.},
	month = oct,
	year = {2017},
	note = {arXiv:1708.01904 [astro-ph]},
	pages = {A98},
}

@article{giroletti_low-power_2006,
	title = {Low-power compact radio galaxies at high angular resolution},
	url = {https://ui.adsabs.harvard.edu/abs/2006evn..confE..22G/abstract},
	doi = {10.22323/1.036.0022},
	language = {en},
	urldate = {2025-11-28},
	journal = {Proceedings of the 8th European VLBI Network Symposium},
	author = {Giroletti, Marcello and Giovannini, G. and Taylor, G. B.},
	year = {2006},
	pages = {22},
}

@article{baldi_lemmings_2020,
	title = {{LeMMINGs} – {II}. {The} \textit{e} -{MERLIN} legacy survey of nearby galaxies. {The} deepest radio view of the {Palomar} sample on parsec scale},
	volume = {500},
	copyright = {https://academic.oup.com/journals/pages/open\_access/funder\_policies/chorus/standard\_publication\_model},
	issn = {0035-8711, 1365-2966},
	url = {https://academic.oup.com/mnras/article/500/4/4749/5986635},
	doi = {10.1093/mnras/staa3519},
	language = {en},
	number = {4},
	urldate = {2025-12-02},
	journal = {Monthly Notices of the Royal Astronomical Society},
	author = {Baldi, R D and Williams, D R A and McHardy, I M and Beswick, R J and Brinks, E and Dullo, B T and Knapen, J H and Argo, M K and Aalto, S and Alberdi, A and Baan, W A and Bendo, G J and Corbel, S and Fenech, D M and Gallagher, J S and Green, D A and Kennicutt, R C and Klöckner, H-R and Körding, E and Maccarone, T J and Muxlow, T W B and Mundell, C G and Panessa, F and Peck, A B and Pérez-Torres, M A and Romero-Cañizales, C and Saikia, P and Shankar, F and Spencer, R E and Stevens, I R and Varenius, E and Ward, M J and Yates, J and Uttley, P},
	month = dec,
	year = {2020},
	pages = {4749--4767},
}

@article{baldi_pg-rqs_2021,
	title = {The {PG}-{RQS} survey. {Building} the radio spectral distribution of radio-quiet quasars. {I}. {The} 45-{GHz} data},
	volume = {510},
	copyright = {https://academic.oup.com/journals/pages/open\_access/funder\_policies/chorus/standard\_publication\_model},
	issn = {0035-8711, 1365-2966},
	url = {https://academic.oup.com/mnras/article/510/1/1043/6445035},
	doi = {10.1093/mnras/stab3445},
	language = {en},
	number = {1},
	urldate = {2025-11-28},
	journal = {Monthly Notices of the Royal Astronomical Society},
	author = {Baldi, R D and Laor, A and Behar, E and Horesh, A and Panessa, F and McHardy, I and Kimball, A},
	month = dec,
	year = {2021},
	pages = {1043--1058},
}

@article{krezinger_revealing_2024,
	title = {Revealing faint compact radio jets at redshifts above 5 with very long baseline interferometry},
	volume = {690},
	copyright = {© The Authors 2024},
	issn = {0004-6361, 1432-0746},
	url = {https://www.aanda.org/articles/aa/abs/2024/10/aa51025-24/aa51025-24.html},
	doi = {10.1051/0004-6361/202451025},
	language = {en},
	urldate = {2025-12-02},
	journal = {Astronomy \& Astrophysics},
	publisher = {EDP Sciences},
	author = {Krezinger, M. and Baldini, G. and Giroletti, M. and Sbarrato, T. and Ghisellini, G. and Giovannini, G. and An, T. and Gabányi, K. É and Frey, S.},
	month = oct,
	year = {2024},
	pages = {A321},
}

@article{garrett_agn_2001,
	title = {{AGN} and starbursts at high redshift: {High} resolution {EVN} radio observations of the {Hubble} {Deep} {Field}},
	volume = {366},
	copyright = {© ESO, 2001},
	issn = {0004-6361, 1432-0746},
	shorttitle = {{AGN} and starbursts at high redshift},
	url = {https://www.aanda.org/articles/aa/abs/2001/05/aacl101/aacl101.html},
	doi = {10.1051/0004-6361:20000537},
	language = {en},
	number = {2},
	urldate = {2025-12-02},
	journal = {Astronomy \& Astrophysics},
	publisher = {EDP Sciences},
	author = {Garrett, M. A. and Muxlow, T. W. B. and Garrington, S. T. and Alef, W. and Alberdi, A. and Langevelde, H. J. van and Venturi, T. and Polatidis, A. G. and Kellermann, K. I. and Baan, W. A. and Kus, A. and Wilkinson, P. N. and Richards, A. M. S.},
	month = feb,
	year = {2001},
	pages = {L5--L8},
}

@article{panessa_sub-parsec_2013,
	title = {Sub-parsec radio cores in nearby {Seyfert} galaxies},
	volume = {432},
	issn = {0035-8711},
	url = {https://doi.org/10.1093/mnras/stt547},
	doi = {10.1093/mnras/stt547},
	number = {2},
	urldate = {2025-12-02},
	journal = {Monthly Notices of the Royal Astronomical Society},
	author = {Panessa, Francesca and Giroletti, Marcello},
	month = jun,
	year = {2013},
	pages = {1138--1143},
}

@article{lister_mojave_2005,
	title = {{MOJAVE}: {Monitoring} of {Jets} in {Active} {Galactic} {Nuclei} with {VLBA} {Experiments}. {I}. {First}-{Epoch} 15 {GHz} {Linear} {Polarization} {Images}},
	volume = {130},
	issn = {1538-3881},
	shorttitle = {{MOJAVE}},
	url = {https://iopscience.iop.org/article/10.1086/432969/meta},
	doi = {10.1086/432969},
	language = {en},
	number = {4},
	urldate = {2025-12-02},
	journal = {The Astronomical Journal},
	publisher = {IOP Publishing},
	author = {Lister, M. L. and Homan, D. C.},
	month = oct,
	year = {2005},
	pages = {1389},
}

@article{baldi_lemmings_2018,
	title = {{LeMMINGs}. {I}. {The} {eMERLIN} legacy survey of nearby galaxies. 1.5-{GHz} parsec-scale radio structures and cores},
	volume = {476},
	issn = {0035-8711, 1365-2966},
	url = {http://arxiv.org/abs/1802.02162},
	doi = {10.1093/mnras/sty342},
	number = {3},
	urldate = {2025-12-02},
	journal = {Monthly Notices of the Royal Astronomical Society},
	author = {Baldi, R. D. and Williams, D. R. A. and McHardy, I. M. and Beswick, R. J. and Argo, M. K. and Dullo, B. T. and Knapen, J. H. and Brinks, E. and Muxlow, T. W. B. and Aalto, S. and Alberdi, A. and Bendo, G. J. and Corbel, S. and Evans, R. and Fenech, D. M. and Green, D. A. and Klöckner, H.-R. and Körding, E. and Kharb, P. and Maccarone, T. J. and Martí-Vidal, I. and Mundell, C. G. and Panessa, F. and Peck, A. B. and Pérez-Torres, M. A. and Saikia, D. J. and Saikia, P. and Shankar, F. and Spencer, R. E. and Stevens, I. R. and Uttley, P. and Westcott, J.},
	month = may,
	year = {2018},
	note = {arXiv:1802.02162 [astro-ph]},
	pages = {3478--3522},
}

@article{kharb_nature_2021,
	title = {The {Nature} of {Jets} in {Double}-peaked {Emission}-line {AGN} in the {KISSR} {Sample}},
	volume = {919},
	issn = {0004-637X},
	url = {https://doi.org/10.3847/1538-4357/ac0c82},
	doi = {10.3847/1538-4357/ac0c82},
	language = {en},
	number = {2},
	urldate = {2025-11-28},
	journal = {The Astrophysical Journal},
	publisher = {The American Astronomical Society},
	author = {Kharb, P. and Subramanian, S. and Das, M. and Vaddi, S. and Paragi, Z.},
	month = sep,
	year = {2021},
	pages = {108},
}

@article{gabuzda_survey_1992,
	title = {A survey of the milliarcsecond polarization properties of {BL} {Lacertae} objects at 5 {GHz}},
	volume = {388},
	issn = {0004-637X, 1538-4357},
	url = {http://adsabs.harvard.edu/doi/10.1086/171128},
	doi = {10.1086/171128},
	language = {en},
	urldate = {2025-11-28},
	journal = {The Astrophysical Journal},
	author = {Gabuzda, D. C. and Cawthorne, T. V. and Roberts, D. H. and Wardle, J. F. C.},
	month = mar,
	year = {1992},
	pages = {40},
}

@article{laor_origin_2008,
	title = {On the origin of radio emission in radio-quiet quasars},
	volume = {390},
	issn = {0035-8711},
	url = {https://doi.org/10.1111/j.1365-2966.2008.13806.x},
	doi = {10.1111/j.1365-2966.2008.13806.x},
	number = {2},
	urldate = {2025-11-28},
	journal = {Monthly Notices of the Royal Astronomical Society},
	author = {Laor, Ari and Behar, Ehud},
	month = oct,
	year = {2008},
	pages = {847--862},
}

@article{vries_star-formation_2007,
	title = {Star-{Formation} in {Low} {Radio} {Luminosity} {AGN} from the {Sloan} {Digital} {Sky} {Survey}},
	volume = {134},
	issn = {0004-6256, 1538-3881},
	url = {http://arxiv.org/abs/0704.2074},
	doi = {10.1086/518866},
	number = {2},
	urldate = {2025-11-28},
	journal = {The Astronomical Journal},
	author = {Vries, W. H. de and Hodge, J. A. and Becker, R. H. and White, R. L. and Helfand, D. J.},
	month = aug,
	year = {2007},
	note = {arXiv:0704.2074 [astro-ph]},
	pages = {457--465},
}

@article{maini_compact_2016,
	title = {Compact radio cores in radio-quiet active galactic nuclei},
	volume = {589},
	copyright = {© ESO, 2016},
	issn = {0004-6361, 1432-0746},
	url = {https://www.aanda.org/articles/aa/abs/2016/05/aa28305-16/aa28305-16.html},
	doi = {10.1051/0004-6361/201628305},
	language = {en},
	urldate = {2025-11-28},
	journal = {Astronomy \& Astrophysics},
	publisher = {EDP Sciences},
	author = {Maini, A. and Prandoni, I. and Norris, R. P. and Giovannini, G. and Spitler, L. R.},
	month = may,
	year = {2016},
	pages = {L3},
}

@misc{meenakshi_comparative_2024,
	title = {A comparative study of radio signatures from winds and jets: {Modelling} synchrotron emission and polarization},
	shorttitle = {A comparative study of radio signatures from winds and jets},
	url = {http://arxiv.org/abs/2408.00099},
	doi = {10.48550/arXiv.2408.00099},
	urldate = {2025-11-26},
	publisher = {arXiv},
	author = {Meenakshi, Moun and Mukherjee, Dipanjan and Bodo, Gianluigi and Rossi, Paola and Harrison, Chris M.},
	month = jul,
	year = {2024},
	note = {arXiv:2408.00099 [astro-ph]},
}

@article{tabatabaei_detailed_2013,
	title = {A {Detailed} {Study} of the {Radio}--{FIR} {Correlation} in {NGC6946} with {Herschel}-{PACS}/{SPIRE} from {KINGFISH}},
	volume = {552},
	issn = {0004-6361, 1432-0746},
	url = {http://arxiv.org/abs/1301.6884},
	doi = {10.1051/0004-6361/201220249},
	urldate = {2025-11-26},
	journal = {Astronomy \& Astrophysics},
	author = {Tabatabaei, F. S. and Schinnerer, E. and Murphy, E. J. and Beck, R. and Groves, B. and Meidt, S. and Krause, M. and Rix, H.-W. and Sandstrom, K. and Crocker, A. F. and Galametz, M. and Helou, G. and Wilson, C. D. and Kennicutt, R. and Calzetti, D. and Draine, B. and Aniano, G. and Dale, D. and Dumas, G. and Engelbracht, C. W. and Gordon, K. D. and Hinz, J. and Kreckel, K. and Montiel, E. and Roussel, H.},
	month = apr,
	year = {2013},
	note = {arXiv:1301.6884 [astro-ph]},
	pages = {A19},
}

@article{potzl_probing_2021,
	title = {Probing the innermost regions of {AGN} jets and their magnetic fields with {RadioAstron} - {IV}. {The} quasar {3C} 345 at 18 cm: {Magnetic} field structure and brightness temperature},
	volume = {648},
	copyright = {© F. M. Pötzl et al. 2021},
	issn = {0004-6361, 1432-0746},
	shorttitle = {Probing the innermost regions of {AGN} jets and their magnetic fields with {RadioAstron} - {IV}. {The} quasar {3C} 345 at 18 cm},
	url = {https://www.aanda.org/articles/aa/abs/2021/04/aa39493-20/aa39493-20.html},
	doi = {10.1051/0004-6361/202039493},
	language = {en},
	urldate = {2025-11-26},
	journal = {Astronomy \& Astrophysics},
	publisher = {EDP Sciences},
	author = {Pötzl, F. M. and Lobanov, A. P. and Ros, E. and Gómez, J. L. and Bruni, G. and Bach, U. and Fuentes, A. and Gurvits, L. I. and Jauncey, D. L. and Kovalev, Y. Y. and Kravchenko, E. V. and Lisakov, M. M. and Savolainen, T. and Sokolovsky, K. V. and Zensus, J. A.},
	month = apr,
	year = {2021},
	pages = {A82},
}

@article{kravchenko_mojave_2025,
	title = {{MOJAVE} -- {XXII}. {Brightness} temperature distributions and geometric profiles along parsec-scale {AGN} jets},
	volume = {538},
	issn = {0035-8711, 1365-2966},
	url = {http://arxiv.org/abs/2502.14516},
	doi = {10.1093/mnras/staf343},
	number = {3},
	urldate = {2025-11-26},
	journal = {Monthly Notices of the Royal Astronomical Society},
	author = {Kravchenko, E. V. and Pashchenko, I. N. and Homan, D. C. and Kovalev, Y. Y. and Lister, M. L. and Pushkarev, A. B. and Ros, E. and Savolainen, T.},
	month = mar,
	year = {2025},
	note = {arXiv:2502.14516 [astro-ph]},
	pages = {2008--2030},
}

@article{homan_intrinsic_2006,
	title = {Intrinsic {Brightness} {Temperatures} of {AGN} {Jets}},
	volume = {642},
	issn = {0004-637X, 1538-4357},
	url = {http://arxiv.org/abs/astro-ph/0603837},
	doi = {10.1086/504715},
	number = {2},
	urldate = {2025-11-26},
	journal = {The Astrophysical Journal},
	author = {Homan, D. C. and Kovalev, Y. Y. and Lister, M. L. and Ros, E. and Kellermann, K. I. and Cohen, M. H. and Vermeulen, R. C. and Zensus, J. A. and Kadler, M.},
	month = may,
	year = {2006},
	note = {arXiv:astro-ph/0603837},
	pages = {L115--L118},
}

@article{walter_things_2008,
	title = {{THINGS}: {THE} {H} i {NEARBY} {GALAXY} {SURVEY}},
	volume = {136},
	issn = {1538-3881},
	shorttitle = {{THINGS}},
	url = {https://doi.org/10.1088/0004-6256/136/6/2563},
	doi = {10.1088/0004-6256/136/6/2563},
	language = {en},
	number = {6},
	urldate = {2025-11-25},
	journal = {The Astronomical Journal},
	publisher = {The American Astronomical Society},
	author = {Walter, Fabian and Brinks, Elias and de Blok, W. J. G. and Bigiel, Frank and Kennicutt, Robert C. and Thornley, Michele D. and Leroy, Adam},
	month = nov,
	year = {2008},
	pages = {2563},
}

@article{heesen_calibrating_2019,
	title = {Calibrating the relation of low-frequency radio continuum to star formation rate at 1 kpc scale with {LOFAR}},
	volume = {622},
	issn = {0004-6361, 1432-0746},
	url = {http://arxiv.org/abs/1811.07968},
	doi = {10.1051/0004-6361/201833905},
	urldate = {2025-11-25},
	journal = {Astronomy \& Astrophysics},
	author = {Heesen, V. and Buie, E. and Huff, C. J. and Perez, L. A. and Woolsey, J. G. and Rafferty, D. A. and Basu, A. and Beck, R. and Brinks, E. and Horellou, C. and Scannapieco, E. and Brüggen, M. and Dettmar, R.-J. and Sendlinger, K. and Nikiel-Wroczyński, B. and Chyży, K. T. and Best, P. N. and Heald, G. H. and Paladino, R.},
	month = feb,
	year = {2019},
	note = {arXiv:1811.07968 [astro-ph]},
	pages = {A8},
}

@article{hwang_winds_2018,
	title = {Winds as the origin of radio emission in \$z=2.5\$ radio-quiet extremely red quasars},
	volume = {477},
	issn = {0035-8711, 1365-2966},
	url = {http://arxiv.org/abs/1803.02821},
	doi = {10.1093/mnras/sty742},
	number = {1},
	urldate = {2025-11-25},
	journal = {Monthly Notices of the Royal Astronomical Society},
	author = {Hwang, Hsiang-Chih and Zakamska, Nadia L. and Alexandroff, Rachael M. and Hamann, Fred and Greene, Jenny E. and Perrotta, Serena and Richards, Gordon T.},
	month = jun,
	year = {2018},
	note = {arXiv:1803.02821 [astro-ph]},
	pages = {830--844},
}

@misc{desi_collaboration_data_2025,
	title = {Data {Release} 1 of the {Dark} {Energy} {Spectroscopic} {Instrument}},
	url = {http://arxiv.org/abs/2503.14745},
	doi = {10.48550/arXiv.2503.14745},
	language = {en},
	urldate = {2025-06-26},
	publisher = {arXiv},
	author = {DESI Collaboration, DESI and Abdul-Karim, M. and Adame, A. G. and Aguado, D. and Aguilar, J. and Ahlen, S. and Alam, S. and Aldering, G. and Alexander, D. M. and Alfarsy, R. and Allen, L. and Prieto, C. Allende and Alves, O. and Anand, A. and Andrade, U. and Armengaud, E. and Avila, S. and Aviles, A. and Awan, H. and Bailey, S. and Lizancos, A. Baleato and Ballester, O. and Bault, A. and Bautista, J. and BenZvi, S. and Silva, L. Beraldo e and Bermejo-Climent, J. R. and Beutler, F. and Bianchi, D. and Blake, C. and Blum, R. and Bolton, A. S. and Bonici, M. and Brieden, S. and Brodzeller, A. and Brooks, D. and Buckley-Geer, E. and Burtin, E. and Canning, R. and Rosell, A. Carnero and Carr, A. and Carrilho, P. and Casas, L. and Castander, F. J. and Cereskaite, R. and Cervantes-Cota, J. L. and Chaussidon, E. and Chaves-Montero, J. and Chen, S. and Chen, X. and Claybaugh, T. and Cole, S. and Cooper, A. P. and Cousinou, M.-C. and Cuceu, A. and Davis, T. M. and Dawson, K. S. and Belsunce, R. de and Cruz, R. de la and Macorra, A. de la and Mattia, A. de and Deiosso, N. and Costa, J. Della and Demina, R. and Demirbozan, U. and DeRose, J. and Dey, A. and Dey, B. and Ding, J. and Ding, Z. and Doel, P. and Douglass, K. and Dowicz, M. and Ebina, H. and Edelstein, J. and Eisenstein, D. J. and Elbers, W. and Emas, N. and Escoffier, S. and Fagrelius, P. and Fan, X. and Fanning, K. and Fawcett, V. A. and Fernández-García, E. and Ferraro, S. and Findlay, N. and Font-Ribera, A. and Forero-Romero, J. E. and Forero-Sánchez, D. and Frenk, C. S. and Gänsicke, B. T. and Galbany, L. and García-Bellido, J. and Garcia-Quintero, C. and Garrison, L. H. and Gaztañaga, E. and Gil-Marín, H. and Gnedin, O. Y. and Gontcho, S. Gontcho A. and Gonzalez-Morales, A. X. and Gonzalez-Perez, V. and Gordon, C. and Graur, O. and Green, D. and Gruen, D. and Gsponer, R. and Guandalin, C. and Gutierrez, G. and Guy, J. and Hahn, C. and Han, J. J. and Han, J. and He, S. and Herrera-Alcantar, H. K. and Honscheid, K. and Hou, J. and Howlett, C. and Huterer, D. and Iršič, V. and Ishak, M. and Jacques, A. and Jimenez, J. and Jing, Y. P. and Joachimi, B. and Joudaki, S. and Joyce, R. and Jullo, E. and Juneau, S. and Karaçaylı, N. G. and Karim, T. and Kehoe, R. and Kent, S. and Khederlarian, A. and Kirkby, D. and Kisner, T. and Kitaura, F.-S. and Kizhuprakkat, N. and Kong, H. and Koposov, S. E. and Kremin, A. and Krolewski, A. and Lahav, O. and Lai, Y. and Lamman, C. and Lan, T.-W. and Landriau, M. and Lang, D. and Lange, J. U. and Lasker, J. and Goff, J. M. Le and Guillou, L. Le and Leauthaud, A. and Levi, M. E. and Li, S. and Li, T. S. and Lodha, K. and Lokken, M. and Luo, Y. and Magneville, C. and Manera, M. and Manser, C. J. and Margala, D. and Martini, P. and Maus, M. and McCullough, J. and McDonald, P. and Medina, G. E. and Medina-Varela, L. and Meisner, A. and Mena-Fernández, J. and Menegas, A. and Mezcua, M. and Miquel, R. and Montero-Camacho, P. and Moon, J. and Moustakas, J. and Muñoz-Gutiérrez, A. and Muñoz-Santos, D. and Myers, A. D. and Myles, J. and Nadathur, S. and Najita, J. and Napolitano, L. and Newman, J. A. and Nikakhtar, F. and Nikutta, R. and Niz, G. and Noriega, H. E. and Padmanabhan, N. and Paillas, E. and Palanque-Delabrouille, N. and Palmese, A. and Pan, J. and Pan, Z. and Parkinson, D. and Peacock, J. and Percival, W. J. and Pérez-Fernández, A. and Pérez-Ràfols, I. and Peterson, P. and Piat, J. and Pieri, M. M. and Pinon, M. and Poppett, C. and Porredon, A. and Prada, F. and Pucha, R. and Qin, F. and Rabinowitz, D. and Raichoor, A. and Ramírez-Pérez, C. and Ramirez-Solano, S. and Rashkovetskyi, M. and Ravoux, C. and Riley, A. H. and Rocher, A. and Rockosi, C. and Rohlf, J. and Ross, A. J. and Rossi, G. and Ruggeri, R. and Ruhlmann-Kleider, V. and Sabiu, C. G. and Said, K. and Saintonge, A. and Samushia, L. and Sanchez, E. and Sanders, N. and Saulder, C. and Schlafly, E. F. and Schlegel, D. and Scholte, D. and Schubnell, M. and Seo, H. and Shafieloo, A. and Sharples, R. and Silber, J. and Siudek, M. and Smith, A. and Sprayberry, D. and Suárez-Pérez, J. and Swanson, J. and Tan, T. and Tarlé, G. and Taylor, P. and Thomas, G. and Tojeiro, R. and Turner, R. J. and Turner, W. and Ureña-López, L. A. and Vaisakh, R. and Valluri, M. and Vargas-Magaña, M. and Verde, L. and Walther, M. and Wang, B. and Wang, M. S. and Wang, W. and Weaver, B. A. and Weaverdyck, N. and Wechsler, R. H. and White, M. and Wolfson, M. and Yang, J. and Yèche, C. and Youles, S. and Yu, J. and Yuan, S. and Zaborowski, E. A. and Zarrouk, P. and Zhang, H. and Zhao, C. and Zhao, R. and Zheng, Z. and Zhou, R. and Zou, H. and Zou, S. and Zu, Y.},
	month = mar,
	year = {2025},
	note = {arXiv:2503.14745 [astro-ph]},
}

@article{sweijen_deep_2022,
	title = {Deep sub-arcsecond wide-field imaging of the {Lockman} {Hole} field at 144 {MHz}},
	volume = {6},
	issn = {2397-3366},
	url = {https://ui.adsabs.harvard.edu/abs/2022NatAs...6..350S},
	doi = {10.1038/s41550-021-01573-z},
	urldate = {2025-09-01},
	journal = {Nature Astronomy},
	author = {Sweijen, F. and van Weeren, R. J. and Röttgering, H. J. A. and Morabito, L. K. and Jackson, N. and Offringa, A. R. and van der Tol, S. and Veenboer, B. and Oonk, J. B. R. and Best, P. N. and Bondi, M. and Shimwell, T. W. and Tasse, C. and Thomson, A. P.},
	month = jan,
	year = {2022},
	note = {ADS Bibcode: 2022NatAs...6..350S},
	pages = {350--356},
}

@article{shimwell_lofar_2025,
	title = {The {LOFAR} {Two}-metre {Sky} {Survey}: {Deep} {Fields} {Data} {Release} 2. {I}. {The} {ELAIS}-{N1} field},
	volume = {695},
	issn = {0004-6361, 1432-0746},
	shorttitle = {The {LOFAR} {Two}-metre {Sky} {Survey}},
	url = {http://arxiv.org/abs/2501.04093},
	doi = {10.1051/0004-6361/202452930},
	language = {en},
	urldate = {2025-04-13},
	journal = {Astronomy \& Astrophysics},
	author = {Shimwell, T. W. and Hale, C. L. and Best, P. N. and Botteon, A. and Drabent, A. and Hardcastle, M. J. and Jelić, V. and Jong, J. M. G. H. J. de and Kondapally, R. and Röttgering, H. J. A. and Tasse, C. and Weeren, R. J. van and Williams, W. L. and Bonafede, A. and Bondi, M. and Brüggen, M. and Brunetti, G. and Callingham, J. R. and Gasperin, F. De and Duncan, K. J. and Horellou, C. and Iyer, S. and Ruiter, I. de and Małek, K. and Nair, D. G. and Morabito, L. K. and Prandoni, I. and Rowlinson, A. and Sabater, J. and Shulevski, A. and Smith, D. J. B. and Sweijen, F.},
	month = mar,
	year = {2025},
	note = {arXiv:2501.04093 [astro-ph]},
	pages = {A80},
}

@article{bonzini_star_2015,
	title = {Star formation properties of sub-{mJy} radio sources},
	volume = {453},
	issn = {0035-8711, 1365-2966},
	url = {https://academic.oup.com/mnras/article-lookup/doi/10.1093/mnras/stv1675},
	doi = {10.1093/mnras/stv1675},
	language = {en},
	number = {1},
	urldate = {2023-11-08},
	journal = {Monthly Notices of the Royal Astronomical Society},
	author = {Bonzini, M. and Mainieri, V. and Padovani, P. and Andreani, P. and Berta, S. and Bethermin, M. and Lutz, D. and Rodighiero, G. and Rosario, D. and Tozzi, P. and Vattakunnel, S.},
	month = oct,
	year = {2015},
	pages = {1079--1094},
}

@article{ashby_spitzer_2009,
	title = {The {Spitzer} {Deep}, {Wide}-field {Survey}},
	volume = {701},
	issn = {0004-637X},
	url = {https://ui.adsabs.harvard.edu/abs/2009ApJ...701..428A},
	doi = {10.1088/0004-637X/701/1/428},
	urldate = {2025-09-01},
	journal = {The Astrophysical Journal},
	author = {Ashby, M. L. N. and Stern, D. and Brodwin, M. and Griffith, R. and Eisenhardt, P. and Kozłowski, S. and Kochanek, C. S. and Bock, J. J. and Borys, C. and Brand, K. and Brown, M. J. I. and Cool, R. and Cooray, A. and Croft, S. and Dey, A. and Eisenstein, D. and Gonzalez, A. H. and Gorjian, V. and Grogin, N. A. and Ivison, R. J. and Jacob, J. and Jannuzi, B. T. and Mainzer, A. and Moustakas, L. A. and Röttgering, H. J. A. and Seymour, N. and Smith, H. A. and Stanford, S. A. and Stauffer, J. R. and Sullivan, I. and van Breugel, W. and Willner, S. P. and Wright, E. L.},
	month = aug,
	year = {2009},
	note = {ADS Bibcode: 2009ApJ...701..428A},
	pages = {428--453},
}

@article{intema_gmrt_2017,
	title = {The {GMRT} 150 {MHz} all-sky radio survey. {First} alternative data release {TGSS} {ADR1}},
	volume = {598},
	issn = {0004-6361},
	url = {https://ui.adsabs.harvard.edu/abs/2017A%26A...598A..78I/abstract},
	doi = {10.1051/0004-6361/201628536},
	language = {en},
	urldate = {2022-04-27},
	journal = {Astronomy \& Astrophysics, Volume 598, id.A78},
	author = {Intema, H. T. and Jagannathan, P. and Mooley, K. P. and Frail, D. A.},
	month = feb,
	year = {2017},
	pages = {A78},
}

@article{fiore_agn_2017,
	title = {{AGN} wind scaling relations and the co-evolution of black holes and galaxies},
	volume = {601},
	issn = {0004-6361, 1432-0746},
	url = {http://www.aanda.org/10.1051/0004-6361/201629478},
	doi = {10.1051/0004-6361/201629478},
	language = {en},
	urldate = {2025-07-01},
	journal = {Astronomy \& Astrophysics},
	author = {Fiore, F. and Feruglio, C. and Shankar, F. and Bischetti, M. and Bongiorno, A. and Brusa, M. and Carniani, S. and Cicone, C. and Duras, F. and Lamastra, A. and Mainieri, V. and Marconi, A. and Menci, N. and Maiolino, R. and Piconcelli, E. and Vietri, G. and Zappacosta, L.},
	month = may,
	year = {2017},
	pages = {A143},
}

@article{magorrian_demography_1998,
	title = {The {Demography} of {Massive} {Dark} {Objects} in {Galaxy} {Centers}},
	volume = {115},
	issn = {0004-6256},
	url = {https://ui.adsabs.harvard.edu/abs/1998AJ....115.2285M},
	doi = {10.1086/300353},
	urldate = {2022-05-05},
	journal = {The Astronomical Journal},
	author = {Magorrian, John and Tremaine, Scott and Richstone, Douglas and Bender, Ralf and Bower, Gary and Dressler, Alan and Faber, S. M. and Gebhardt, Karl and Green, Richard and Grillmair, Carl and Kormendy, John and Lauer, Tod},
	month = jun,
	year = {1998},
	note = {ADS Bibcode: 1998AJ....115.2285M},
	pages = {2285--2305},
}

@article{speranza_multiphase_2024,
	title = {Multiphase characterization of {AGN} winds in five local type-2 quasars},
	volume = {681},
	issn = {0004-6361},
	url = {https://ui.adsabs.harvard.edu/abs/2024A&A...681A..63S},
	doi = {10.1051/0004-6361/202347715},
	urldate = {2025-08-22},
	journal = {Astronomy and Astrophysics},
	author = {Speranza, G. and Ramos Almeida, C. and Acosta-Pulido, J. A. and Audibert, A. and Holden, L. R. and Tadhunter, C. N. and Lapi, A. and González-Martín, O. and Brusa, M. and López, I. E. and Musiimenta, B. and Shankar, F.},
	month = jan,
	year = {2024},
	note = {ADS Bibcode: 2024A\&A...681A..63S},
	pages = {A63},
}

@article{venturi_magnum_2021,
	title = {{MAGNUM} survey: {Compact} jets causing large turmoil in galaxies. {Enhanced} line widths perpendicular to radio jets as tracers of jet-{ISM} interaction},
	volume = {648},
	issn = {0004-6361},
	shorttitle = {{MAGNUM} survey},
	url = {https://ui.adsabs.harvard.edu/abs/2021A&A...648A..17V},
	doi = {10.1051/0004-6361/202039869},
	urldate = {2025-08-22},
	journal = {Astronomy and Astrophysics},
	author = {Venturi, G. and Cresci, G. and Marconi, A. and Mingozzi, M. and Nardini, E. and Carniani, S. and Mannucci, F. and Marasco, A. and Maiolino, R. and Perna, M. and Treister, E. and Bland-Hawthorn, J. and Gallimore, J.},
	month = apr,
	year = {2021},
	note = {ADS Bibcode: 2021A\&A...648A..17V},
	pages = {A17},
}

@article{meenakshi_modelling_2022,
	title = {Modelling observable signatures of jet-{ISM} interaction: thermal emission and gas kinematics},
	volume = {516},
	issn = {0035-8711},
	shorttitle = {Modelling observable signatures of jet-{ISM} interaction},
	url = {https://ui.adsabs.harvard.edu/abs/2022MNRAS.516..766M},
	doi = {10.1093/mnras/stac2251},
	urldate = {2025-08-22},
	journal = {Monthly Notices of the Royal Astronomical Society},
	author = {Meenakshi, Moun and Mukherjee, Dipanjan and Wagner, Alexander Y. and Nesvadba, Nicole P. H. and Bicknell, Geoffrey V. and Morganti, Raffaella and Janssen, Reinier M. J. and Sutherland, Ralph S. and Mandal, Ankush},
	month = oct,
	year = {2022},
	note = {ADS Bibcode: 2022MNRAS.516..766M},
	pages = {766--786},
}

@article{kukreti_feedback_2025,
	title = {Feedback from low-to-moderate-luminosity radio-active galactic nuclei with {MaNGA}},
	volume = {698},
	issn = {0004-6361},
	url = {https://ui.adsabs.harvard.edu/abs/2025A&A...698A..99K},
	doi = {10.1051/0004-6361/202453307},
	urldate = {2025-08-22},
	journal = {Astronomy and Astrophysics},
	author = {Kukreti, Pranav and Wylezalek, Dominika and Albán, Marco and Dall'Agnol de Oliveira, Bruno},
	month = jun,
	year = {2025},
	note = {ADS Bibcode: 2025A\&A...698A..99K},
	pages = {A99},
}

@misc{jong_scalable_2025,
	title = {Scalable and robust wide-field facet calibration with {LOFAR}'s longest baselines},
	url = {http://arxiv.org/abs/2508.12115},
	doi = {10.48550/arXiv.2508.12115},
	urldate = {2025-08-21},
	publisher = {arXiv},
	author = {Jong, J. M. G. H. J. de and Veefkind, L. and Weeren, R. J. van and Oonk, J. B. R. and Schlimbach, R. J. and Kampert, D. N. G. and Wild, M. van der and Morabito, L. K. and Sweijen, F. and Offringa, A. R. and Röttgering, H. J. A.},
	month = aug,
	year = {2025},
	note = {arXiv:2508.12115 [astro-ph]},
}

@article{veilleux_cool_2020,
	title = {Cool outflows in galaxies and their implications},
	volume = {28},
	issn = {0935-4956},
	url = {https://ui.adsabs.harvard.edu/abs/2020A&ARv..28....2V},
	doi = {10.1007/s00159-019-0121-9},
	urldate = {2025-08-21},
	journal = {Astronomy and Astrophysics Review},
	author = {Veilleux, Sylvain and Maiolino, Roberto and Bolatto, Alberto D. and Aalto, Susanne},
	month = apr,
	year = {2020},
	note = {ADS Bibcode: 2020A\&ARv..28....2V},
	pages = {2},
}

@article{padovani_faint_2016,
	title = {The faint radio sky: radio astronomy becomes mainstream},
	volume = {24},
	issn = {0935-4956, 1432-0754},
	shorttitle = {The faint radio sky},
	url = {http://arxiv.org/abs/1609.00499},
	doi = {10.1007/s00159-016-0098-6},
	language = {en},
	number = {1},
	urldate = {2025-08-16},
	journal = {The Astronomy and Astrophysics Review},
	author = {Padovani, Paolo},
	month = dec,
	year = {2016},
	note = {arXiv:1609.00499 [astro-ph]},
	pages = {13},
}

@article{harrison_agn_2018,
	title = {{AGN} outflows and feedback twenty years on},
	volume = {2},
	copyright = {2018 The Author(s)},
	issn = {2397-3366},
	url = {https://www.nature.com/articles/s41550-018-0403-6},
	doi = {10.1038/s41550-018-0403-6},
	language = {en},
	number = {3},
	urldate = {2025-08-15},
	journal = {Nature Astronomy},
	publisher = {Nature Publishing Group},
	author = {Harrison, C. M. and Costa, T. and Tadhunter, C. N. and Flütsch, A. and Kakkad, D. and Perna, M. and Vietri, G.},
	month = mar,
	year = {2018},
	pages = {198--205},
}

@article{condon_radio_1992,
	title = {Radio emission from normal galaxies.},
	volume = {30},
	issn = {0066-4146},
	url = {https://ui.adsabs.harvard.edu/abs/1992ARA&A..30..575C},
	doi = {10.1146/annurev.aa.30.090192.003043},
	urldate = {2025-08-13},
	journal = {Annual Review of Astronomy and Astrophysics},
	author = {Condon, J. J.},
	month = jan,
	year = {1992},
	note = {ADS Bibcode: 1992ARA\&A..30..575C},
	pages = {575--611},
}

@article{blandford_relativistic_1979,
	title = {Relativistic jets as compact radio sources.},
	volume = {232},
	issn = {0004-637X},
	url = {https://ui.adsabs.harvard.edu/abs/1979ApJ...232...34B},
	doi = {10.1086/157262},
	urldate = {2025-08-05},
	journal = {The Astrophysical Journal},
	publisher = {IOP},
	author = {Blandford, R. D. and Königl, A.},
	month = aug,
	year = {1979},
	note = {ADS Bibcode: 1979ApJ...232...34B},
	pages = {34--48},
}

@article{charlot_third_2020,
	title = {The third realization of the {International} {Celestial} {Reference} {Frame} by very long baseline interferometry},
	volume = {644},
	copyright = {© P. Charlot et al. 2020},
	issn = {0004-6361, 1432-0746},
	url = {https://www.aanda.org/articles/aa/abs/2020/12/aa38368-20/aa38368-20.html},
	doi = {10.1051/0004-6361/202038368},
	language = {en},
	urldate = {2025-08-05},
	journal = {Astronomy \& Astrophysics},
	publisher = {EDP Sciences},
	author = {Charlot, P. and Jacobs, C. S. and Gordon, D. and Lambert, S. and Witt, A. de and Böhm, J. and Fey, A. L. and Heinkelmann, R. and Skurikhina, E. and Titov, O. and Arias, E. F. and Bolotin, S. and Bourda, G. and Ma, C. and Malkin, Z. and Nothnagel, A. and Mayer, D. and MacMillan, D. S. and Nilsson, T. and Gaume, R.},
	month = dec,
	year = {2020},
	pages = {A159},
}

@article{ye_1-arcsecond_2024,
	title = {1-arcsecond imaging of the {ELAIS}-{N1} field at {144MHz} using the {LoTSS} survey with the international {LOFAR} telescope},
	volume = {691},
	copyright = {© The Authors 2024},
	issn = {0004-6361, 1432-0746},
	url = {https://www.aanda.org/articles/aa/abs/2024/11/aa48103-23/aa48103-23.html},
	doi = {10.1051/0004-6361/202348103},
	language = {en},
	urldate = {2025-08-05},
	journal = {Astronomy \& Astrophysics},
	publisher = {EDP Sciences},
	author = {Ye, Haoyang and Sweijen, Frits and Weeren, Reinout J. van and Williams, Wendy and Jong, Jurjen de and Morabito, Leah K. and Rottgering, Huub and Shimwell, Timothy W. and Best, P. N. and Bondi, Marco and Brüggen, Marcus and Gasperin, Francesco de and Tasse, Cyril},
	month = nov,
	year = {2024},
	pages = {A347},
}

@article{van_weeren_lofar_2021,
	title = {{LOFAR} observations of galaxy clusters in {HETDEX}. {Extraction} and self-calibration of individual {LOFAR} targets},
	volume = {651},
	issn = {0004-6361},
	url = {https://ui.adsabs.harvard.edu/abs/2021A&A...651A.115V},
	doi = {10.1051/0004-6361/202039826},
	urldate = {2025-08-05},
	journal = {Astronomy and Astrophysics},
	publisher = {EDP},
	author = {van Weeren, R. J. and Shimwell, T. W. and Botteon, A. and Brunetti, G. and Brüggen, M. and Boxelaar, J. M. and Cassano, R. and Di Gennaro, G. and Andrade-Santos, F. and Bonnassieux, E. and Bonafede, A. and Cuciti, V. and Dallacasa, D. and de Gasperin, F. and Gastaldello, F. and Hardcastle, M. J. and Hoeft, M. and Kraft, R. P. and Mandal, S. and Rossetti, M. and Röttgering, H. J. A. and Tasse, C. and Wilber, A. G.},
	month = jul,
	year = {2021},
	note = {ADS Bibcode: 2021A\&A...651A.115V},
	pages = {A115},
}

@article{scaife_broad-band_2012,
	title = {A broad-band flux scale for low-frequency radio telescopes},
	volume = {423},
	issn = {1745-3925},
	url = {https://doi.org/10.1111/j.1745-3933.2012.01251.x},
	doi = {10.1111/j.1745-3933.2012.01251.x},
	number = {1},
	urldate = {2025-08-05},
	journal = {Monthly Notices of the Royal Astronomical Society: Letters},
	author = {Scaife, Anna M. M. and Heald, George H.},
	month = jan,
	year = {2012},
	pages = {L30--L34},
}

@article{offringa_aoflagger_2010,
	title = {{AOFlagger}: {RFI} {Software}},
	shorttitle = {{AOFlagger}},
	url = {https://ui.adsabs.harvard.edu/abs/2010ascl.soft10017O},
	urldate = {2025-08-05},
	journal = {Astrophysics Source Code Library},
	author = {Offringa, A. R.},
	month = oct,
	year = {2010},
	note = {ADS Bibcode: 2010ascl.soft10017O},
	pages = {ascl:1010.017},
}

@misc{harrison_observational_2024,
	title = {Observational {Tests} of {Active} {Galactic} {Nuclei} {Feedback}: {An} {Overview} of {Approaches} and {Interpretation}},
	shorttitle = {Observational {Tests} of {Active} {Galactic} {Nuclei} {Feedback}},
	url = {http://arxiv.org/abs/2404.08050},
	doi = {10.48550/arXiv.2404.08050},
	language = {en},
	urldate = {2025-08-04},
	publisher = {arXiv},
	author = {Harrison, Chris M. and Almeida, Cristina Ramos},
	month = apr,
	year = {2024},
	note = {arXiv:2404.08050 [astro-ph]},
}

@article{becker_first_1995,
	title = {The {FIRST} {Survey}: {Faint} {Images} of the {Radio} {Sky} at {Twenty} {Centimeters}},
	volume = {450},
	issn = {0004-637X},
	shorttitle = {The {FIRST} {Survey}},
	url = {https://ui.adsabs.harvard.edu/abs/1995ApJ...450..559B},
	doi = {10.1086/176166},
	urldate = {2025-08-04},
	journal = {The Astrophysical Journal},
	publisher = {IOP},
	author = {Becker, Robert H. and White, Richard L. and Helfand, David J.},
	month = sep,
	year = {1995},
	note = {ADS Bibcode: 1995ApJ...450..559B},
	pages = {559},
}

@misc{ward_agn-driven_2024,
	title = {{AGN}-driven outflows in clumpy media: multiphase structure and scaling relations},
	shorttitle = {{AGN}-driven outflows in clumpy media},
	url = {http://arxiv.org/abs/2407.17593},
	doi = {10.48550/arXiv.2407.17593},
	language = {en},
	urldate = {2025-07-28},
	publisher = {arXiv},
	author = {Ward, Samuel Ruthven and Costa, Tiago and Harrison, Chris M. and Mainieri, Vincenzo},
	month = aug,
	year = {2024},
	note = {arXiv:2407.17593 [astro-ph]},
}

@article{miley_structure_1980,
	title = {The {Structure} of {Extended} {Extragalactic} {Radio} {Sources}},
	volume = {18},
	issn = {0066-4146, 1545-4282},
	url = {https://www.annualreviews.org/doi/10.1146/annurev.aa.18.090180.001121},
	doi = {10.1146/annurev.aa.18.090180.001121},
	language = {en},
	number = {1},
	urldate = {2025-07-28},
	journal = {Annual Review of Astronomy and Astrophysics},
	author = {Miley, George},
	month = sep,
	year = {1980},
	pages = {165--218},
}

@inproceedings{bertin_terapix_2002,
	title = {The {TERAPIX} {Pipeline}},
	volume = {281},
	url = {https://ui.adsabs.harvard.edu/abs/2002ASPC..281..228B},
	urldate = {2025-07-16},
	author = {Bertin, Emmanuel and Mellier, Yannick and Radovich, Mario and Missonnier, Gilles and Didelon, Pierre and Morin, Bertrand},
	month = jan,
	year = {2002},
	note = {ADS Bibcode: 2002ASPC..281..228B},
	pages = {228},
}

@article{moldon_lofar_2015,
	title = {The {LOFAR} long baseline snapshot calibrator survey},
	volume = {574},
	copyright = {© ESO, 2015},
	issn = {0004-6361, 1432-0746},
	url = {https://www.aanda.org/articles/aa/abs/2015/02/aa25042-14/aa25042-14.html},
	doi = {10.1051/0004-6361/201425042},
	language = {en},
	urldate = {2025-07-16},
	journal = {Astronomy \& Astrophysics},
	publisher = {EDP Sciences},
	author = {Moldón, J. and Deller, A. T. and Wucknitz, O. and Jackson, N. and Drabent, A. and Carozzi, T. and Conway, J. and Kapińska, A. D. and McKean, J. P. and Morabito, L. and Varenius, E. and Zarka, P. and Anderson, J. and Asgekar, A. and Avruch, I. M. and Bell, M. E. and Bentum, M. J. and Bernardi, G. and Best, P. and Bîrzan, L. and Bregman, J. and Breitling, F. and Broderick, J. W. and Brüggen, M. and Butcher, H. R. and Carbone, D. and Ciardi, B. and Gasperin, F. de and Geus, E. de and Duscha, S. and Eislöffel, J. and Engels, D. and Falcke, H. and Fallows, R. A. and Fender, R. and Ferrari, C. and Frieswijk, W. and Garrett, M. A. and Grießmeier, J. and Gunst, A. W. and Hamaker, J. P. and Hassall, T. E. and Heald, G. and Hoeft, M. and Juette, E. and Karastergiou, A. and Kondratiev, V. I. and Kramer, M. and Kuniyoshi, M. and Kuper, G. and Maat, P. and Mann, G. and Markoff, S. and McFadden, R. and McKay-Bukowski, D. and Morganti, R. and Munk, H. and Norden, M. J. and Offringa, A. R. and Orru, E. and Paas, H. and Pandey-Pommier, M. and Pizzo, R. and Polatidis, A. G. and Reich, W. and Röttgering, H. and Rowlinson, A. and Scaife, A. M. M. and Schwarz, D. and Sluman, J. and Smirnov, O. and Stappers, B. W. and Steinmetz, M. and Tagger, M. and Tang, Y. and Tasse, C. and Thoudam, S. and Toribio, M. C. and Vermeulen, R. and Vocks, C. and Weeren, R. J. van and White, S. and Wise, M. W. and Yatawatta, S. and Zensus, A.},
	month = feb,
	year = {2015},
	pages = {A73},
}

@article{franzen_atlas_2015,
	title = {{ATLAS} – {I}. {Third} release of 1.4 {GHz} mosaics and component catalogues},
	volume = {453},
	issn = {0035-8711},
	url = {https://doi.org/10.1093/mnras/stv1866},
	doi = {10.1093/mnras/stv1866},
	number = {4},
	urldate = {2025-07-01},
	journal = {Monthly Notices of the Royal Astronomical Society},
	author = {Franzen, T. M. O. and Banfield, J. K. and Hales, C. A. and Hopkins, A. and Norris, R. P. and Seymour, N. and Chow, K. E. and Herzog, A. and Huynh, M. T. and Lenc, E. and Mao, M. Y. and Middelberg, E.},
	month = nov,
	year = {2015},
	pages = {4020--4036},
}

@article{jackson_sub-arcsecond_2022,
	title = {Sub-arcsecond imaging with the {International} {LOFAR} {Telescope} - {II}. {Completion} of the {LOFAR} {Long}-{Baseline} {Calibrator} {Survey}},
	volume = {658},
	copyright = {© ESO 2022},
	issn = {0004-6361, 1432-0746},
	url = {https://www.aanda.org/articles/aa/abs/2022/02/aa40756-21/aa40756-21.html},
	doi = {10.1051/0004-6361/202140756},
	language = {en},
	urldate = {2025-07-01},
	journal = {Astronomy \& Astrophysics},
	publisher = {EDP Sciences},
	author = {Jackson, N. and Badole, S. and Morgan, J. and Chhetri, R. and Prūsis, K. and Nikolajevs, A. and Morabito, L. and Brentjens, M. and Sweijen, F. and Iacobelli, M. and Orrù, E. and Sluman, J. and Blaauw, R. and Mulder, H. and Dijk, P. van and Mooney, S. and Deller, A. and Moldon, J. and Callingham, J. R. and Harwood, J. and Hardcastle, M. and Heald, G. and Drabent, A. and McKean, J. P. and Asgekar, A. and Avruch, I. M. and Bentum, M. J. and Bonafede, A. and Brouw, W. N. and Brüggen, M. and Butcher, H. R. and Ciardi, B. and Coolen, A. and Corstanje, A. and Damstra, S. and Duscha, S. and Eislöffel, J. and Falcke, H. and Garrett, M. and Gasperin, F. de and Griessmeier, J.-M. and Gunst, A. W. and Haarlem, M. P. van and Hoeft, M. and Horst, A. J. van der and Jütte, E. and Koopmans, L. V. E. and Krankowski, A. and Maat, P. and Mann, G. and Miley, G. K. and Nelles, A. and Norden, M. and Paas, M. and Pandey, V. N. and Pandey-Pommier, M. and Pizzo, R. F. and Reich, W. and Rothkaehl, H. and Rowlinson, A. and Ruiter, M. and Shulevski, A. and Schwarz, D. J. and Smirnov, O. and Tagger, M. and Vocks, C. and Weeren, R. J. van and Wijers, R. and Wucknitz, O. and Zarka, P. and Zensus, J. A. and Zucca, P.},
	month = feb,
	year = {2022},
	pages = {A2},
}

@article{morabito_hidden_2025,
	title = {A hidden active galactic nucleus population: the first radio luminosity functions constructed by physical process},
	volume = {536},
	issn = {1745-3925},
	shorttitle = {A hidden active galactic nucleus population},
	url = {https://doi.org/10.1093/mnrasl/slae104},
	doi = {10.1093/mnrasl/slae104},
	number = {1},
	urldate = {2025-07-01},
	journal = {Monthly Notices of the Royal Astronomical Society: Letters},
	author = {Morabito, Leah K and Kondapally, R and Best, P N and Yue, B -H and de Jong, J M G H J and Sweijen, F and Bondi, Marco and Schwarz, Dominik J and Smith, D J B and van Weeren, R J and Röttgering, H J A and Shimwell, T W and Prandoni, Isabella},
	month = jan,
	year = {2025},
	pages = {L32--L37},
}

@article{offringa_wsclean_2014,
	title = {wsclean: an implementation of a fast, generic wide-field imager for radio astronomy},
	volume = {444},
	issn = {0035-8711, 1365-2966},
	shorttitle = {wsclean},
	url = {http://academic.oup.com/mnras/article/444/1/606/1010067/wsclean-an-implementation-of-a-fast-generic},
	doi = {10.1093/mnras/stu1368},
	language = {en},
	number = {1},
	urldate = {2025-02-17},
	journal = {Monthly Notices of the Royal Astronomical Society},
	author = {Offringa, A. R. and McKinley, B. and Hurley-Walker, N. and Briggs, F. H. and Wayth, R. B. and Kaplan, D. L. and Bell, M. E. and Feng, L. and Neben, A. R. and Hughes, J. D. and Rhee, J. and Murphy, T. and Bhat, N. D. R. and Bernardi, G. and Bowman, J. D. and Cappallo, R. J. and Corey, B. E. and Deshpande, A. A. and Emrich, D. and Ewall-Wice, A. and Gaensler, B. M. and Goeke, R. and Greenhill, L. J. and Hazelton, B. J. and Hindson, L. and Johnston-Hollitt, M. and Jacobs, D. C. and Kasper, J. C. and Kratzenberg, E. and Lenc, E. and Lonsdale, C. J. and Lynch, M. J. and McWhirter, S. R. and Mitchell, D. A. and Morales, M. F. and Morgan, E. and Kudryavtseva, N. and Oberoi, D. and Ord, S. M. and Pindor, B. and Procopio, P. and Prabu, T. and Riding, J. and Roshi, D. A. and Shankar, N. Udaya and Srivani, K. S. and Subrahmanyan, R. and Tingay, S. J. and Waterson, M. and Webster, R. L. and Whitney, A. R. and Williams, A. and Williams, C. L.},
	month = oct,
	year = {2014},
	pages = {606--619},
}

@article{intema_ionospheric_2009,
	title = {Ionospheric calibration of low frequency radio interferometric observations using the peeling scheme - {I}. {Method} description and first results},
	volume = {501},
	copyright = {© ESO, 2009},
	issn = {0004-6361, 1432-0746},
	url = {https://www.aanda.org/articles/aa/abs/2009/27/aa11094-08/aa11094-08.html},
	doi = {10.1051/0004-6361/200811094},
	language = {en},
	number = {3},
	urldate = {2025-04-13},
	journal = {Astronomy \& Astrophysics},
	publisher = {EDP Sciences},
	author = {Intema, H. T. and Tol, S. van der and Cotton, W. D. and Cohen, A. S. and Bemmel, I. M. van and Röttgering, H. J. A.},
	month = jul,
	year = {2009},
	note = {Number: 3},
	pages = {1185--1205},
}

@article{delvecchio_vla-cosmos_2017,
	title = {The {VLA}-{COSMOS} 3 {GHz} {Large} {Project}: {AGN} and host-galaxy properties out to z ≲ 6},
	volume = {602},
	copyright = {© ESO, 2017},
	issn = {0004-6361, 1432-0746},
	shorttitle = {The {VLA}-{COSMOS} 3 {GHz} {Large} {Project}},
	url = {https://www.aanda.org/articles/aa/abs/2017/06/aa29367-16/aa29367-16.html},
	doi = {10.1051/0004-6361/201629367},
	language = {en},
	urldate = {2023-11-08},
	journal = {Astronomy \& Astrophysics},
	publisher = {EDP Sciences},
	author = {Delvecchio, I. and Smolčić, V. and Zamorani, G. and Lagos, C. Del P. and Berta, S. and Delhaize, J. and Baran, N. and Alexander, D. M. and Rosario, D. J. and Gonzalez-Perez, V. and Ilbert, O. and Lacey, C. G. and Fèvre, O. Le and Miettinen, O. and Aravena, M. and Bondi, M. and Carilli, C. and Ciliegi, P. and Mooley, K. and Novak, M. and Schinnerer, E. and Capak, P. and Civano, F. and Fanidakis, N. and Ruiz, N. Herrera and Karim, A. and Laigle, C. and Marchesi, S. and McCracken, H. J. and Middleberg, E. and Salvato, M. and Tasca, L.},
	month = jun,
	year = {2017},
	pages = {A3},
}

@article{intema_deep_2011,
	title = {Deep low-frequency radio observations of the {NOAO} {Boötes} field - {I}. {Data} reduction and catalog construction},
	volume = {535},
	copyright = {© ESO, 2011},
	issn = {0004-6361, 1432-0746},
	url = {https://www.aanda.org/articles/aa/abs/2011/11/aa14253-10/aa14253-10.html},
	doi = {10.1051/0004-6361/201014253},
	language = {en},
	urldate = {2025-06-13},
	journal = {Astronomy \& Astrophysics},
	publisher = {EDP Sciences},
	author = {Intema, H. T. and Weeren, R. J. van and Röttgering, H. J. A. and Lal, D. V.},
	month = nov,
	year = {2011},
	pages = {A38},
}

@misc{smith_weave-lofar_2016,
	title = {The {WEAVE}-{LOFAR} {Survey}},
	url = {http://arxiv.org/abs/1611.02706},
	doi = {10.48550/arXiv.1611.02706},
	language = {en},
	urldate = {2025-06-12},
	publisher = {arXiv},
	author = {Smith, D. J. B. and Best, P. N. and Duncan, K. J. and Hatch, N. A. and Jarvis, M. J. and Röttgering, H. J. A. and Simpson, C. J. and Stott, J. P. and Cochrane, R. K. and Coppin, K. E. and Dannerbauer, H. and Davis, T. A. and Geach, J. E. and Hale, C. L. and Hardcastle, M. J. and Hatfield, P. W. and Houghton, R. C. W. and Maddox, N. and McGee, S. L. and Morabito, L. and Nisbet, D. and Pandey-Pommier, M. and Prandoni, I. and Saxena, A. and Shimwell, T. W. and Tarr, M. and Bemmel, I. van and Verma, A. and White, G. J. and Williams, W. L.},
	month = nov,
	year = {2016},
	note = {arXiv:1611.02706 [astro-ph]},
}

@misc{rubeis_revealing_2025,
	title = {Revealing the intricacies of radio galaxies and filaments in the merging galaxy cluster {Abell} 2255. {I}. {Insights} from deep {LOFAR}-{VLBI} sub-arcsecond resolution images},
	url = {http://arxiv.org/abs/2505.13595},
	doi = {10.48550/arXiv.2505.13595},
	language = {en},
	urldate = {2025-05-23},
	publisher = {arXiv},
	author = {Rubeis, E. De and Bondi, M. and Botteon, A. and Weeren, R. J. van and Jong, J. M. G. H. J. de and Rudnick, L. and Brunetti, G. and Rajpurohit, K. and Gheller, C. and Röttgering, H. J. A.},
	month = may,
	year = {2025},
	note = {arXiv:2505.13595 [astro-ph]},
}

@article{morabito_decade_2025,
	title = {A decade of sub-arcsecond imaging with the {International} {LOFAR} {Telescope}},
	volume = {370},
	issn = {1572-946X},
	url = {https://doi.org/10.1007/s10509-025-04406-x},
	doi = {10.1007/s10509-025-04406-x},
	language = {en},
	number = {2},
	urldate = {2025-04-13},
	journal = {Astrophysics and Space Science},
	author = {Morabito, Leah K. and Jackson, Neal and de Jong, Jurjen and Escott, Emmy and Groeneveld, Christian and Mahatma, Vijay and Petley, James and Sweijen, Frits and Timmerman, Roland and van Weeren, Reinout J.},
	month = feb,
	year = {2025},
	pages = {19},
}

@article{weeren_lofar_2016,
	title = {{LOFAR} {FACET} {CALIBRATION}},
	volume = {223},
	issn = {0067-0049},
	url = {https://dx.doi.org/10.3847/0067-0049/223/1/2},
	doi = {10.3847/0067-0049/223/1/2},
	language = {en},
	number = {1},
	urldate = {2025-02-14},
	journal = {The Astrophysical Journal Supplement Series},
	publisher = {The American Astronomical Society},
	author = {Weeren, R. J. van and Williams, W. L. and Hardcastle, M. J. and Shimwell, T. W. and Rafferty, D. A. and Sabater, J. and Heald, G. and Sridhar, S. S. and Dijkema, T. J. and Brunetti, G. and Brüggen, M. and Andrade-Santos, F. and Ogrean, G. A. and Röttgering, H. J. A. and Dawson, W. A. and Forman, W. R. and Gasperin, F. de and Jones, C. and Miley, G. K. and Rudnick, L. and Sarazin, C. L. and Bonafede, A. and Best, P. N. and Bîrzan, L. and Cassano, R. and Chyży, K. T. and Croston, J. H. and Ensslin, T. and Ferrari, C. and Hoeft, M. and Horellou, C. and Jarvis, M. J. and Kraft, R. P. and Mevius, M. and Intema, H. T. and Murray, S. S. and Orrú, E. and Pizzo, R. and Simionescu, A. and Stroe, A. and Tol, S. van der and White, G. J.},
	month = mar,
	year = {2016},
	pages = {2},
}

@article{alban_mapping_2024,
	title = {Mapping {AGN} winds: {A} connection between radio-mode {AGNs} and the {AGN} feedback cycle},
	volume = {691},
	copyright = {https://creativecommons.org/licenses/by/4.0},
	issn = {0004-6361, 1432-0746},
	shorttitle = {Mapping {AGN} winds},
	url = {https://www.aanda.org/10.1051/0004-6361/202451738},
	doi = {10.1051/0004-6361/202451738},
	language = {en},
	urldate = {2025-01-20},
	journal = {Astronomy \& Astrophysics},
	author = {Albán, M. and Wylezalek, D. and Comerford, J. M. and Greene, J. E. and Riffel, R. A.},
	month = nov,
	year = {2024},
	pages = {A124},
}

@article{escott_unveiling_2025,
	title = {Unveiling {AGN} outflows: [{O} iii] outflow detection rates and correlation with low-frequency radio emission},
	volume = {536},
	issn = {0035-8711},
	shorttitle = {Unveiling {AGN} outflows},
	url = {https://doi.org/10.1093/mnras/stae2645},
	doi = {10.1093/mnras/stae2645},
	number = {2},
	urldate = {2025-01-15},
	journal = {Monthly Notices of the Royal Astronomical Society},
	author = {Escott, Emmy L and Morabito, Leah K and Scholtz, Jan and Hickox, Ryan C and Harrison, Chris M and Alexander, David M and Arnaudova, Marina I and Smith, Daniel J B and Duncan, Kenneth J and Petley, James and Kondapally, Rohit and Calistro Rivera, Gabriela and Kolwa, Sthabile},
	month = jan,
	year = {2025},
	pages = {1166--1179},
}

@misc{njeri_quasar_2025,
	title = {The {Quasar} {Feedback} {Survey}: zooming into the origin of radio emission with e-{MERLIN}},
	shorttitle = {The {Quasar} {Feedback} {Survey}},
	url = {http://arxiv.org/abs/2501.03433},
	doi = {10.1093/mnras/staf020},
	language = {en},
	urldate = {2025-01-08},
	author = {Njeri, Ann and Harrison, Chris M. and Kharb, Preeti and Beswick, Robert and Calistro-Rivera, Gabriela and Circosta, Chiara and Mainieri, Vincenzo and Molyneux, Stephen and Mullaney, James and Sasikumar, Silpa},
	month = jan,
	year = {2025},
	note = {arXiv:2501.03433 [astro-ph]},
}

@article{bondi_lofar_2024,
	title = {{LOFAR} {HBA} observations of the {Euclid} {Deep} {Field} {North} ({EDFN})},
	volume = {683},
	copyright = {https://creativecommons.org/licenses/by/4.0},
	issn = {0004-6361, 1432-0746},
	url = {https://www.aanda.org/10.1051/0004-6361/202348333},
	doi = {10.1051/0004-6361/202348333},
	language = {en},
	urldate = {2024-11-03},
	journal = {Astronomy \& Astrophysics},
	author = {Bondi, M. and Scaramella, R. and Zamorani, G. and Ciliegi, P. and Vitello, F. and Arias, M. and Best, P. N. and Bonato, M. and Botteon, A. and Brienza, M. and Brunetti, G. and Hardcastle, M. J. and Magliocchetti, M. and Massaro, F. and Morabito, L. K and Pentericci, L. and Prandoni, I. and Röttgering, H. J. A. and Shimwell, T. W. and Tasse, C. and Van Weeren, R. J. and White, G. J.},
	month = mar,
	year = {2024},
	pages = {A179},
}

@article{lyke_sloan_2020,
	title = {The {Sloan} {Digital} {Sky} {Survey} {Quasar} {Catalog}: {Sixteenth} {Data} {Release}},
	volume = {250},
	issn = {0067-0049, 1538-4365},
	shorttitle = {The {Sloan} {Digital} {Sky} {Survey} {Quasar} {Catalog}},
	url = {http://arxiv.org/abs/2007.09001},
	doi = {10.3847/1538-4365/aba623},
	language = {en},
	number = {1},
	urldate = {2024-08-16},
	journal = {The Astrophysical Journal Supplement Series},
	author = {Lyke, Brad W. and Higley, Alexandra N. and McLane, J. N. and Schurhammer, Danielle P. and Myers, Adam D. and Ross, Ashley J. and Dawson, Kyle and Chabanier, Solène and Martini, Paul and Busca, Nicolás G. and Bourboux, Hélion du Mas des and Salvato, Mara and Streblyanska, Alina and Zarrouk, Pauline and Burtin, Etienne and Anderson, Scott F. and Bautista, Julian and Bizyaev, Dmitry and Brandt, W. N. and Brinkmann, Jonathan and Brownstein, Joel R. and Comparat, Johan and Green, Paul and de la Macorra, Axel and Gutiérrez, Andrea Muñoz and Hou, Jiamin and Newman, Jeffrey A. and Palanque-Delabrouille, Nathalie and Pâris, Isabelle and Percival, Will J. and Petitjean, Patrick and Rich, James and Rossi, Graziano and Schneider, Donald P. and Smith, Alexander and Vivek, M. and Weaver, Benjamin Alan},
	month = sep,
	year = {2020},
	note = {arXiv:2007.09001 [astro-ph]},
	pages = {8},
}

@misc{de_jong_into_2024,
	title = {Into the depths: {Unveiling} {ELAIS}-{N1} with {LOFAR}'s deepest sub-arcsecond wide-field images},
	shorttitle = {Into the depths},
	url = {http://arxiv.org/abs/2407.13247},
	language = {en},
	urldate = {2024-08-15},
	publisher = {arXiv},
	author = {de Jong, J. M. G. H. J. and van Weeren, R. J. and Sweijen, F. and Oonk, J. B. R. and Shimwell, T. W. and Offringa, A. R. and Morabito, L. K. and Röttgering, H. J. A. and Kondapally, R. and Escott, E. L. and Best, P. N. and Bondi, M. and Ye, H. and Petley, J. W.},
	month = jul,
	year = {2024},
	note = {arXiv:2407.13247 [astro-ph]},
}

@misc{kukreti_connecting_2024,
	title = {Connecting the radio {AGN} life cycle to feedback: {Ionised} gas is more disturbed in young radio {AGN}},
	shorttitle = {Connecting the radio {AGN} life cycle to feedback},
	url = {http://arxiv.org/abs/2407.06265},
	language = {en},
	urldate = {2024-07-11},
	publisher = {arXiv},
	author = {Kukreti, Pranav and Morganti, Raffaella},
	month = jul,
	year = {2024},
	note = {arXiv:2407.06265 [astro-ph]},
}

@article{klein_radio_2018,
	title = {Radio synchrotron spectra of star-forming galaxies},
	volume = {611},
	copyright = {© ESO 2018},
	issn = {0004-6361, 1432-0746},
	url = {https://www.aanda.org/articles/aa/abs/2018/03/aa31673-17/aa31673-17.html},
	doi = {10.1051/0004-6361/201731673},
	language = {en},
	urldate = {2024-04-26},
	journal = {Astronomy \& Astrophysics},
	publisher = {EDP Sciences},
	author = {Klein, U. and Lisenfeld, U. and Verley, S.},
	month = mar,
	year = {2018},
	pages = {A55},
}

@article{sabater_lofar_2021,
	title = {The {LOFAR} {Two}-meter {Sky} {Survey}: {Deep} {Fields} {Data} {Release} 1 - {II}. {The} {ELAIS}-{N1} {LOFAR} deep field},
	volume = {648},
	copyright = {© ESO 2021},
	issn = {0004-6361, 1432-0746},
	shorttitle = {The {LOFAR} {Two}-meter {Sky} {Survey}},
	url = {https://www.aanda.org/articles/aa/abs/2021/04/aa38828-20/aa38828-20.html},
	doi = {10.1051/0004-6361/202038828},
	language = {en},
	urldate = {2024-01-18},
	journal = {Astronomy \& Astrophysics},
	publisher = {EDP Sciences},
	author = {Sabater, J. and Best, P. N. and Tasse, C. and Hardcastle, M. J. and Shimwell, T. W. and Nisbet, D. and Jelic, V. and Callingham, J. R. and Röttgering, H. J. A. and Bonato, M. and Bondi, M. and Ciardi, B. and Cochrane, R. K. and Jarvis, M. J. and Kondapally, R. and Koopmans, L. V. E. and O’Sullivan, S. P. and Prandoni, I. and Schwarz, D. J. and Smith, D. J. B. and Wang, L. and Williams, W. L. and Zaroubi, S.},
	month = apr,
	year = {2021},
	pages = {A2},
}

@article{lacy_karl_2020,
	title = {The {Karl} {G}. {Jansky} {Very} {Large} {Array} {Sky} {Survey} ({VLASS}). {Science} {Case} and {Survey} {Design}},
	volume = {132},
	issn = {0004-6280},
	url = {https://ui.adsabs.harvard.edu/abs/2020PASP..132c5001L},
	doi = {10.1088/1538-3873/ab63eb},
	urldate = {2024-01-12},
	journal = {Publications of the Astronomical Society of the Pacific},
	author = {Lacy, M. and Baum, S. A. and Chandler, C. J. and Chatterjee, S. and Clarke, T. E. and Deustua, S. and English, J. and Farnes, J. and Gaensler, B. M. and Gugliucci, N. and Hallinan, G. and Kent, B. R. and Kimball, A. and Law, C. J. and Lazio, T. J. W. and Marvil, J. and Mao, S. A. and Medlin, D. and Mooley, K. and Murphy, E. J. and Myers, S. and Osten, R. and Richards, G. T. and Rosolowsky, E. and Rudnick, L. and Schinzel, F. and Sivakoff, G. R. and Sjouwerman, L. O. and Taylor, R. and White, R. L. and Wrobel, J. and Andernach, H. and Beasley, A. J. and Berger, E. and Bhatnager, S. and Birkinshaw, M. and Bower, G. C. and Brandt, W. N. and Brown, S. and Burke-Spolaor, S. and Butler, B. J. and Comerford, J. and Demorest, P. B. and Fu, H. and Giacintucci, S. and Golap, K. and Güth, T. and Hales, C. A. and Hiriart, R. and Hodge, J. and Horesh, A. and Ivezić, Ž. and Jarvis, M. J. and Kamble, A. and Kassim, N. and Liu, X. and Loinard, L. and Lyons, D. K. and Masters, J. and Mezcua, M. and Moellenbrock, G. A. and Mroczkowski, T. and Nyland, K. and O'Dea, C. P. and O'Sullivan, S. P. and Peters, W. M. and Radford, K. and Rao, U. and Robnett, J. and Salcido, J. and Shen, Y. and Sobotka, A. and Witz, S. and Vaccari, M. and van Weeren, R. J. and Vargas, A. and Williams, P. K. G. and Yoon, I.},
	month = mar,
	year = {2020},
	note = {ADS Bibcode: 2020PASP..132c5001L},
	pages = {035001},
}

@article{haarlem_lofar_2013,
	title = {{LOFAR}: {The} {LOw}-{Frequency} {ARray}},
	volume = {556},
	copyright = {© ESO, 2013},
	issn = {0004-6361, 1432-0746},
	shorttitle = {{LOFAR}},
	url = {https://www.aanda.org/articles/aa/abs/2013/08/aa20873-12/aa20873-12.html},
	doi = {10.1051/0004-6361/201220873},
	language = {en},
	urldate = {2023-12-30},
	journal = {Astronomy \& Astrophysics},
	publisher = {EDP Sciences},
	author = {Haarlem, M. P. van and Wise, M. W. and Gunst, A. W. and Heald, G. and McKean, J. P. and Hessels, J. W. T. and Bruyn, A. G. de and Nijboer, R. and Swinbank, J. and Fallows, R. and Brentjens, M. and Nelles, A. and Beck, R. and Falcke, H. and Fender, R. and Hörandel, J. and Koopmans, L. V. E. and Mann, G. and Miley, G. and Röttgering, H. and Stappers, B. W. and Wijers, R. a. M. J. and Zaroubi, S. and Akker, M. van den and Alexov, A. and Anderson, J. and Anderson, K. and Ardenne, A. van and Arts, M. and Asgekar, A. and Avruch, I. M. and Batejat, F. and Bähren, L. and Bell, M. E. and Bell, M. R. and Bemmel, I. van and Bennema, P. and Bentum, M. J. and Bernardi, G. and Best, P. and Bîrzan, L. and Bonafede, A. and Boonstra, A.-J. and Braun, R. and Bregman, J. and Breitling, F. and Brink, R. H. van de and Broderick, J. and Broekema, P. C. and Brouw, W. N. and Brüggen, M. and Butcher, H. R. and Cappellen, W. van and Ciardi, B. and Coenen, T. and Conway, J. and Coolen, A. and Corstanje, A. and Damstra, S. and Davies, O. and Deller, A. T. and Dettmar, R.-J. and Diepen, G. van and Dijkstra, K. and Donker, P. and Doorduin, A. and Dromer, J. and Drost, M. and Duin, A. van and Eislöffel, J. and Enst, J. van and Ferrari, C. and Frieswijk, W. and Gankema, H. and Garrett, M. A. and Gasperin, F. de and Gerbers, M. and Geus, E. de and Grießmeier, J.-M. and Grit, T. and Gruppen, P. and Hamaker, J. P. and Hassall, T. and Hoeft, M. and Holties, H. A. and Horneffer, A. and Horst, A. van der and Houwelingen, A. van and Huijgen, A. and Iacobelli, M. and Intema, H. and Jackson, N. and Jelic, V. and Jong, A. de and Juette, E. and Kant, D. and Karastergiou, A. and Koers, A. and Kollen, H. and Kondratiev, V. I. and Kooistra, E. and Koopman, Y. and Koster, A. and Kuniyoshi, M. and Kramer, M. and Kuper, G. and Lambropoulos, P. and Law, C. and Leeuwen, J. van and Lemaitre, J. and Loose, M. and Maat, P. and Macario, G. and Markoff, S. and Masters, J. and McFadden, R. A. and McKay-Bukowski, D. and Meijering, H. and Meulman, H. and Mevius, M. and Middelberg, E. and Millenaar, R. and Miller-Jones, J. C. A. and Mohan, R. N. and Mol, J. D. and Morawietz, J. and Morganti, R. and Mulcahy, D. D. and Mulder, E. and Munk, H. and Nieuwenhuis, L. and Nieuwpoort, R. van and Noordam, J. E. and Norden, M. and Noutsos, A. and Offringa, A. R. and Olofsson, H. and Omar, A. and Orrú, E. and Overeem, R. and Paas, H. and Pandey-Pommier, M. and Pandey, V. N. and Pizzo, R. and Polatidis, A. and Rafferty, D. and Rawlings, S. and Reich, W. and Reijer, J.-P. de and Reitsma, J. and Renting, G. A. and Riemers, P. and Rol, E. and Romein, J. W. and Roosjen, J. and Ruiter, M. and Scaife, A. and Schaaf, K. van der and Scheers, B. and Schellart, P. and Schoenmakers, A. and Schoonderbeek, G. and Serylak, M. and Shulevski, A. and Sluman, J. and Smirnov, O. and Sobey, C. and Spreeuw, H. and Steinmetz, M. and Sterks, C. G. M. and Stiepel, H.-J. and Stuurwold, K. and Tagger, M. and Tang, Y. and Tasse, C. and Thomas, I. and Thoudam, S. and Toribio, M. C. and Tol, B. van der and Usov, O. and Veelen, M. van and Veen, A.-J. van der and Veen, S. ter and Verbiest, J. P. W. and Vermeulen, R. and Vermaas, N. and Vocks, C. and Vogt, C. and Vos, M. de and Wal, E. van der and Weeren, R. van and Weggemans, H. and Weltevrede, P. and White, S. and Wijnholds, S. J. and Wilhelmsson, T. and Wucknitz, O. and Yatawatta, S. and Zarka, P. and Zensus, A. and Zwieten, J. van},
	month = aug,
	year = {2013},
	pages = {A2},
}

@article{petley_connecting_2022,
	title = {Connecting radio emission to {AGN} wind properties with broad absorption line quasars},
	volume = {515},
	issn = {0035-8711, 1365-2966},
	url = {https://academic.oup.com/mnras/article/515/4/5159/6649341},
	doi = {10.1093/mnras/stac2067},
	language = {en},
	number = {4},
	urldate = {2023-12-30},
	journal = {Monthly Notices of the Royal Astronomical Society},
	author = {Petley, J W and Morabito, L K and Alexander, D M and Rankine, A L and Fawcett, V A and Rosario, D J and Matthews, J H and Shimwell, T M and Drabent, A},
	month = aug,
	year = {2022},
	pages = {5159--5174},
}

@article{kukreti_ionised_2023,
	title = {Ionised gas outflows over the radio {AGN} life cycle},
	volume = {674},
	copyright = {© The Authors 2023},
	issn = {0004-6361, 1432-0746},
	url = {https://www.aanda.org/articles/aa/abs/2023/06/aa45691-22/aa45691-22.html},
	doi = {10.1051/0004-6361/202245691},
	language = {en},
	urldate = {2023-12-30},
	journal = {Astronomy \& Astrophysics},
	publisher = {EDP Sciences},
	author = {Kukreti, Pranav and Morganti, Raffaella and Tadhunter, Clive and Santoro, Francesco},
	month = jun,
	year = {2023},
	pages = {A198},
}

@article{zakamska_quasar_2014,
	title = {Quasar feedback and the origin of radio emission in radio-quiet quasars},
	volume = {442},
	issn = {0035-8711, 1365-2966},
	url = {https://academic.oup.com/mnras/article-lookup/doi/10.1093/mnras/stu842},
	doi = {10.1093/mnras/stu842},
	language = {en},
	number = {1},
	urldate = {2023-11-08},
	journal = {Monthly Notices of the Royal Astronomical Society},
	author = {Zakamska, N. L. and Greene, J. E.},
	month = jul,
	year = {2014},
	pages = {784--804},
}

@article{kormendy_inward_1995,
	title = {Inward {Bound}---{The} {Search} {For} {Supermassive} {Black} {Holes} {In} {Galactic} {Nuclei}},
	volume = {33},
	issn = {0066-4146},
	url = {https://ui.adsabs.harvard.edu/abs/1995ARA&A..33..581K},
	doi = {10.1146/annurev.aa.33.090195.003053},
	urldate = {2023-11-08},
	journal = {Annual Review of Astronomy and Astrophysics},
	author = {Kormendy, John and Richstone, Douglas},
	month = jan,
	year = {1995},
	note = {ADS Bibcode: 1995ARA\&A..33..581K},
	pages = {581},
}

@article{kormendy_evidence_1992,
	title = {Evidence for a {Supermassive} {Black} {Hole} in {NGC} 3115},
	volume = {393},
	issn = {0004-637X},
	url = {https://ui.adsabs.harvard.edu/abs/1992ApJ...393..559K},
	doi = {10.1086/171528},
	urldate = {2023-11-08},
	journal = {The Astrophysical Journal},
	author = {Kormendy, John and Richstone, Douglas},
	month = jul,
	year = {1992},
	note = {ADS Bibcode: 1992ApJ...393..559K},
	pages = {559},
}

@article{jarvis_prevalence_2019,
	title = {Prevalence of radio jets associated with galactic outflows and feedback from quasars},
	volume = {485},
	issn = {0035-8711},
	url = {https://ui.adsabs.harvard.edu/abs/2019MNRAS.485.2710J},
	doi = {10.1093/mnras/stz556},
	urldate = {2023-11-02},
	journal = {Monthly Notices of the Royal Astronomical Society},
	author = {Jarvis, M. E. and Harrison, C. M. and Thomson, A. P. and Circosta, C. and Mainieri, V. and Alexander, D. M. and Edge, A. C. and Lansbury, G. B. and Molyneux, S. J. and Mullaney, J. R.},
	month = may,
	year = {2019},
	note = {ADS Bibcode: 2019MNRAS.485.2710J},
	pages = {2710--2730},
}

@article{nesvadba_gas_2017,
	title = {Gas kinematics in powerful radio galaxies at z {\textasciitilde} 2: {Energy} supply from star formation, {AGN}, and radio jets},
	volume = {600},
	copyright = {© ESO, 2017},
	issn = {0004-6361, 1432-0746},
	shorttitle = {Gas kinematics in powerful radio galaxies at z {\textasciitilde} 2},
	url = {https://www.aanda.org/articles/aa/abs/2017/04/aa29357-16/aa29357-16.html},
	doi = {10.1051/0004-6361/201629357},
	language = {en},
	urldate = {2023-10-03},
	journal = {Astronomy \& Astrophysics},
	publisher = {EDP Sciences},
	author = {Nesvadba, N. P. H. and Drouart, G. and Breuck, C. De and Best, P. and Seymour, N. and Vernet, J.},
	month = apr,
	year = {2017},
	pages = {A121},
}

@article{zakamska_discovery_2016,
	title = {Discovery of extreme [{OIII}]{5007A} outflows in high-redshift red quasars},
	volume = {459},
	issn = {0035-8711, 1365-2966},
	url = {http://arxiv.org/abs/1512.02642},
	doi = {10.1093/mnras/stw718},
	number = {3},
	urldate = {2023-10-02},
	journal = {Monthly Notices of the Royal Astronomical Society},
	author = {Zakamska, Nadia L. and Hamann, Fred and Pâris, Isabelle and Brandt, W. N. and Greene, Jenny E. and Strauss, Michael A. and Villforth, Carolin and Wylezalek, Dominika and Alexandroff, Rachael M. and Ross, Nicholas P.},
	month = jul,
	year = {2016},
	note = {arXiv:1512.02642 [astro-ph]},
	pages = {3144--3160},
}

@misc{lonsdale_swire_2003,
	title = {{SWIRE}: {The} {SIRTF} {Wide}-{Area} {Infrared} {Extragalactic} {Survey}},
	language = {en},
	author = {Lonsdale, Carol J and Smith, Harding E and Rowan-Robinson, Michael and Surace, Jason and Shupe, David and Xu, Cong and Oliver, Sebastian and Padgett, Deborah and Fang, Fan and Conrow, Tim and Franceschini, Alberto and Gautier, Nick and Grifﬁn, Matt and Hacking, Perry and Masci, Frank and Morrison, Glenn and O’Linger, Joanne and Owen, Frazer and Perez-Fournon, Ismael and Pierre, Marguerite and Puetter, Rick and Stacey, Gordon and Castro, Sandra and Polletta, Maria Del Carmen and Farrah, Duncan and Jarrett, Tom and Frayer, Dave and Siana, Brian and Babbedge, Tom and Dye, Simon and Fox, Matt and Gonzalez-Solares, Eduardo and Salaman, Malcolm and Berta, Stefano and Condon, Jim J and Dole, Herve and Serjeant, Steve},
	year = {2003},
}

@misc{jannuzi_noao_1999,
	title = {The {NOAO} {Deep} {Wide}-{Field} {Survey}},
	url = {https://ui.adsabs.harvard.edu/abs/1999ASPC..193..258J},
	urldate = {2023-09-29},
	author = {Jannuzi, B. T. and Dey, A.},
	month = jan,
	year = {1999},
	note = {Conference Name: The Hy-Redshift Universe: Galaxy Formation and Evolution at High Redshift
ADS Bibcode: 1999ASPC..193..258J},
}

@article{shimwell_lofar_2017,
	title = {The {LOFAR} {Two}-metre {Sky} {Survey} - {I}. {Survey} {Description} and {Preliminary} {Data} {Release}},
	volume = {598},
	issn = {0004-6361, 1432-0746},
	url = {http://arxiv.org/abs/1611.02700},
	doi = {10.1051/0004-6361/201629313},
	urldate = {2022-04-22},
	journal = {Astronomy \& Astrophysics},
	author = {Shimwell, T. W. and Röttgering, H. J. A. and Best, P. N. and Williams, W. L. and Dijkema, T. J. and de Gasperin, F. and Hardcastle, M. J. and Heald, G. H. and Hoang, D. N. and Horneffer, A. and Intema, H. and Mahony, E. K. and Mandal, S. and Mechev, A. P. and Morabito, L. and Oonk, J. B. R. and Rafferty, D. and Retana-Montenegro, E. and Sabater, J. and Tasse, C. and van Weeren, R. J. and Brüggen, M. and Brunetti, G. and Chyży, K. T. and Conway, J. E. and Haverkorn, M. and Jackson, N. and Jarvis, M. J. and McKean, J. P. and Miley, G. K. and Morganti, R. and White, G. J. and Wise, M. W. and van Bemmel, I. M. and Beck, R. and Brienza, M. and Bonafede, A. and Rivera, G. Calistro and Cassano, R. and Clarke, A. O. and Cseh, D. and Deller, A. and Drabent, A. and van Driel, W. and Engels, D. and Falcke, H. and Ferrari, C. and Fröhlich, S. and Garrett, M. A. and Harwood, J. J. and Heesen, V. and Hoeft, M. and Horellou, C. and Israel, F. P. and Kapińska, A. D. and Kunert-Bajraszewska, M. and McKay, D. J. and Mohan, N. R. and Orrú, E. and Pizzo, R. F. and Prandoni, I. and Schwarz, D. J. and Shulevski, A. and Sipior, M. and Smith, D. J. B. and Sridhar, S. S. and Steinmetz, M. and Stroe, A. and Varenius, E. and van der Werf, P. P. and Zensus, J. A. and Zwart, J. T. L.},
	month = feb,
	year = {2017},
	note = {arXiv: 1611.02700},
	pages = {A104},
}

@article{hinshaw_nine-year_2013,
	title = {{NINE}-{YEAR} \textit{{WILKINSON} {MICROWAVE} {ANISOTROPY} {PROBE}} ( \textit{{WMAP}} ) {OBSERVATIONS}: {COSMOLOGICAL} {PARAMETER} {RESULTS}},
	volume = {208},
	issn = {0067-0049, 1538-4365},
	shorttitle = {{NINE}-{YEAR} \textit{{WILKINSON} {MICROWAVE} {ANISOTROPY} {PROBE}} ( \textit{{WMAP}} ) {OBSERVATIONS}},
	url = {https://iopscience.iop.org/article/10.1088/0067-0049/208/2/19},
	doi = {10.1088/0067-0049/208/2/19},
	language = {en},
	number = {2},
	urldate = {2023-08-23},
	journal = {The Astrophysical Journal Supplement Series},
	author = {Hinshaw, G. and Larson, D. and Komatsu, E. and Spergel, D. N. and Bennett, C. L. and Dunkley, J. and Nolta, M. R. and Halpern, M. and Hill, R. S. and Odegard, N. and Page, L. and Smith, K. M. and Weiland, J. L. and Gold, B. and Jarosik, N. and Kogut, A. and Limon, M. and Meyer, S. S. and Tucker, G. S. and Wollack, E. and Wright, E. L.},
	month = sep,
	year = {2013},
	pages = {19},
}

@article{girdhar_quasar_2022,
	title = {Quasar {Feedback} {Survey}: {Multi}-phase outflows, turbulence and evidence for feedback caused by low power radio jets inclined into the galaxy disk},
	volume = {512},
	issn = {0035-8711, 1365-2966},
	shorttitle = {Quasar {Feedback} {Survey}},
	url = {http://arxiv.org/abs/2201.02208},
	doi = {10.1093/mnras/stac073},
	language = {en},
	number = {2},
	urldate = {2023-08-23},
	journal = {Monthly Notices of the Royal Astronomical Society},
	author = {Girdhar, A. and Harrison, C. M. and Mainieri, V. and Bittner, A. and Costa, T. and Kharb, P. and Mukherjee, D. and Battaia, F. Arrigoni and Alexander, D. M. and Rivera, G. Calistro and Circosta, C. and De Breuck, C. and Edge, A. C. and Farina, E. P. and Kakkad, D. and Lansbury, G. B. and Molyneux, S. J. and Mullaney, J. R. and S., Silpa and Thomson, A. P. and Ward, S. R.},
	month = mar,
	year = {2022},
	note = {arXiv:2201.02208 [astro-ph]},
	pages = {1608--1628},
}

@article{scholtz_impact_2021,
	title = {The impact of ionized outflows from \textit{z} ∼ 2.5 quasars is not through instantaneous \textit{in situ} quenching: the evidence from {ALMA} and {VLT}/{SINFONI}},
	volume = {505},
	issn = {0035-8711, 1365-2966},
	shorttitle = {The impact of ionized outflows from \textit{z} ∼ 2.5 quasars is not through instantaneous \textit{in situ} quenching},
	url = {https://academic.oup.com/mnras/article/505/4/5469/6295321},
	doi = {10.1093/mnras/stab1631},
	language = {en},
	number = {4},
	urldate = {2023-08-22},
	journal = {Monthly Notices of the Royal Astronomical Society},
	author = {Scholtz, J and Harrison, C M and Rosario, D J and Alexander, D M and Knudsen, K K and Stanley, F and Chen, Chian-Chou and Kakkad, D and Mainieri, V and Mullaney, J},
	month = jul,
	year = {2021},
	pages = {5469--5487},
}

@article{best_lofar_2023,
	title = {The {LOFAR} {Two}-metre {Sky} {Survey}: {Deep} {Fields} data release 1. {V}. {Survey} description, source classifications, and host galaxy properties},
	volume = {523},
	issn = {0035-8711, 1365-2966},
	shorttitle = {The {LOFAR} {Two}-metre {Sky} {Survey}},
	url = {https://academic.oup.com/mnras/article/523/2/1729/7147321},
	doi = {10.1093/mnras/stad1308},
	language = {en},
	number = {2},
	urldate = {2023-07-25},
	journal = {Monthly Notices of the Royal Astronomical Society},
	author = {Best, P N and Kondapally, R and Williams, W L and Cochrane, R K and Duncan, K J and Hale, C L and Haskell, P and Małek, K and McCheyne, I and Smith, D J B and Wang, L and Botteon, A and Bonato, M and Bondi, M and Rivera, G Calistro and Gao, F and Gürkan, G and Hardcastle, M J and Jarvis, M J and Mingo, B and Miraghaei, H and Morabito, L K and Nisbet, D and Prandoni, I and Röttgering, H J A and Sabater, J and Shimwell, T and Tasse, C and van Weeren, R},
	month = may,
	year = {2023},
	pages = {1729--1755},
}

@article{mullaney_narrow-line_2013,
	title = {Narrow-line region gas kinematics of 24,264 optically-selected {AGN}: the radio connection},
	volume = {433},
	issn = {0035-8711, 1365-2966},
	shorttitle = {Narrow-line region gas kinematics of 24,264 optically-selected {AGN}},
	url = {http://arxiv.org/abs/1305.0263},
	doi = {10.1093/mnras/stt751},
	language = {en},
	number = {1},
	urldate = {2023-07-25},
	journal = {Monthly Notices of the Royal Astronomical Society},
	author = {Mullaney, J. R. and Alexander, D. M. and Fine, S. and Goulding, A. D. and Harrison, C. M. and Hickox, R. C.},
	month = jul,
	year = {2013},
	note = {arXiv:1305.0263 [astro-ph]},
	pages = {622--638},
}

@article{shimwell_lofar_2022,
	title = {The {LOFAR} {Two}-metre {Sky} {Survey} - {V}. {Second} data release},
	volume = {659},
	copyright = {© ESO 2022},
	issn = {0004-6361, 1432-0746},
	url = {https://www.aanda.org/articles/aa/abs/2022/03/aa42484-21/aa42484-21.html},
	doi = {10.1051/0004-6361/202142484},
	language = {en},
	urldate = {2023-04-21},
	journal = {Astronomy \& Astrophysics},
	publisher = {EDP Sciences},
	author = {Shimwell, T. W. and Hardcastle, M. J. and Tasse, C. and Best, P. N. and Röttgering, H. J. A. and Williams, W. L. and Botteon, A. and Drabent, A. and Mechev, A. and Shulevski, A. and Weeren, R. J. van and Bester, L. and Brüggen, M. and Brunetti, G. and Callingham, J. R. and Chyży, K. T. and Conway, J. E. and Dijkema, T. J. and Duncan, K. and Gasperin, F. de and Hale, C. L. and Haverkorn, M. and Hugo, B. and Jackson, N. and Mevius, M. and Miley, G. K. and Morabito, L. K. and Morganti, R. and Offringa, A. and Oonk, J. B. R. and Rafferty, D. and Sabater, J. and Smith, D. J. B. and Schwarz, D. J. and Smirnov, O. and O’Sullivan, S. P. and Vedantham, H. and White, G. J. and Albert, J. G. and Alegre, L. and Asabere, B. and Bacon, D. J. and Bonafede, A. and Bonnassieux, E. and Brienza, M. and Bilicki, M. and Bonato, M. and Rivera, G. Calistro and Cassano, R. and Cochrane, R. and Croston, J. H. and Cuciti, V. and Dallacasa, D. and Danezi, A. and Dettmar, R. J. and Gennaro, G. Di and Edler, H. W. and Enßlin, T. A. and Emig, K. L. and Franzen, T. M. O. and García-Vergara, C. and Grange, Y. G. and Gürkan, G. and Hajduk, M. and Heald, G. and Heesen, V. and Hoang, D. N. and Hoeft, M. and Horellou, C. and Iacobelli, M. and Jamrozy, M. and Jelić, V. and Kondapally, R. and Kukreti, P. and Kunert-Bajraszewska, M. and Magliocchetti, M. and Mahatma, V. and Małek, K. and Mandal, S. and Massaro, F. and Meyer-Zhao, Z. and Mingo, B. and Mostert, R. I. J. and Nair, D. G. and Nakoneczny, S. J. and Nikiel-Wroczyński, B. and Orrú, E. and Pajdosz-Śmierciak, U. and Pasini, T. and Prandoni, I. and Piggelen, H. E. van and Rajpurohit, K. and Retana-Montenegro, E. and Riseley, C. J. and Rowlinson, A. and Saxena, A. and Schrijvers, C. and Sweijen, F. and Siewert, T. M. and Timmerman, R. and Vaccari, M. and Vink, J. and West, J. L. and Wołowska, A. and Zhang, X. and Zheng, J.},
	month = mar,
	year = {2022},
	pages = {A1},
}

@article{molyneux_extreme_2019,
	title = {Extreme ionised outflows are more common when the radio emission is compact in {AGN} host galaxies},
	volume = {631},
	issn = {0004-6361, 1432-0746},
	url = {https://www.aanda.org/10.1051/0004-6361/201936408},
	doi = {10.1051/0004-6361/201936408},
	language = {en},
	urldate = {2023-03-23},
	journal = {Astronomy \& Astrophysics},
	author = {Molyneux, S. J. and Harrison, C. M. and Jarvis, M. E.},
	month = nov,
	year = {2019},
	pages = {A132},
}

@article{richards_spectral_2006,
	title = {Spectral {Energy} {Distributions} and {Multiwavelength} {Selection} of {Type} 1 {Quasars}},
	volume = {166},
	issn = {0067-0049, 1538-4365},
	url = {https://iopscience.iop.org/article/10.1086/506525},
	doi = {10.1086/506525},
	language = {en},
	number = {2},
	urldate = {2022-10-21},
	journal = {The Astrophysical Journal Supplement Series},
	author = {Richards, Gordon T. and Lacy, Mark and Storrie‐Lombardi, Lisa J. and Hall, Patrick B. and Gallagher, S. C. and Hines, Dean C. and Fan, Xiaohui and Papovich, Casey and Vanden Berk, Daniel E. and Trammell, George B. and Schneider, Donald P. and Vestergaard, Marianne and York, Donald G. and Jester, Sebastian and Anderson, Scott F. and Budavari, Tamas and Szalay, Alexander S.},
	month = oct,
	year = {2006},
	pages = {470--497},
}

@misc{morabito_identifying_2022,
	title = {Identifying active galactic nuclei via brightness temperature with sub-arcsecond {International} {LOFAR} {Telescope} observations},
	url = {http://arxiv.org/abs/2207.13096},
	doi = {10.48550/arXiv.2207.13096},
	urldate = {2022-07-28},
	publisher = {arXiv},
	author = {Morabito, Leah K. and Sweijen, F. and Radcliffe, J. F. and Best, P. N. and Kondapally, Rohit and Bondi, Marco and Bonato, Matteo and Duncan, K. J. and Prandoni, Isabella and Shimwell, T. W. and Williams, W. L. and van Weeren, R. J. and Conway, J. E. and Rivera, G. Calistro},
	month = jul,
	year = {2022},
	note = {arXiv:2207.13096 [astro-ph]},
}

@article{jarvis_quasar_2021,
	title = {The quasar feedback survey: discovering hidden {Radio}-{AGN} and their connection to the host galaxy ionized gas},
	volume = {503},
	issn = {0035-8711},
	shorttitle = {The quasar feedback survey},
	url = {https://ui.adsabs.harvard.edu/abs/2021MNRAS.503.1780J},
	doi = {10.1093/mnras/stab549},
	urldate = {2022-06-06},
	journal = {Monthly Notices of the Royal Astronomical Society},
	author = {Jarvis, M. E. and Harrison, C. M. and Mainieri, V. and Alexander, D. M. and Arrigoni Battaia, F. and Calistro Rivera, G. and Circosta, C. and Costa, T. and De Breuck, C. and Edge, A. C. and Girdhar, A. and Kakkad, D. and Kharb, P. and Lansbury, G. B. and Molyneux, S. J. and Mukherjee, D. and Mullaney, J. R. and Farina, E. P. and Silpa, S. and Thomson, A. P. and Ward, S. R.},
	month = may,
	year = {2021},
	note = {ADS Bibcode: 2021MNRAS.503.1780J},
	pages = {1780--1797},
}

@article{croton_erratum_2006,
	title = {Erratum: {The} many lives of active galactic nuclei: cooling flows, black holes and the luminosities and colours of galaxies},
	volume = {367},
	issn = {0035-8711},
	shorttitle = {Erratum},
	url = {https://doi.org/10.1111/j.1365-2966.2006.09994.x},
	doi = {10.1111/j.1365-2966.2006.09994.x},
	number = {2},
	urldate = {2022-05-13},
	journal = {Monthly Notices of the Royal Astronomical Society},
	author = {Croton, Darren J. and Springel, Volker and White, Simon D. M. and De Lucia, G. and Frenk, C. S. and Gao, L. and Jenkins, A. and Kauffmann, G. and Navarro, J. F. and Yoshida, N.},
	month = apr,
	year = {2006},
	pages = {864},
}

@article{gebhardt_relationship_2000,
	title = {A {Relationship} between {Nuclear} {Black} {Hole} {Mass} and {Galaxy} {Velocity} {Dispersion}},
	volume = {539},
	issn = {0004-637X},
	url = {https://ui.adsabs.harvard.edu/abs/2000ApJ...539L..13G},
	doi = {10.1086/312840},
	urldate = {2022-05-13},
	journal = {The Astrophysical Journal},
	author = {Gebhardt, Karl and Bender, Ralf and Bower, Gary and Dressler, Alan and Faber, S. M. and Filippenko, Alexei V. and Green, Richard and Grillmair, Carl and Ho, Luis C. and Kormendy, John and Lauer, Tod R. and Magorrian, John and Pinkney, Jason and Richstone, Douglas and Tremaine, Scott},
	month = aug,
	year = {2000},
	note = {ADS Bibcode: 2000ApJ...539L..13G},
	pages = {L13--L16},
}

@article{merritt_mbh-sigma_2001,
	title = {The {M}({BH})-{Sigma} {Relation} for {Supermassive} {Black} {Holes}},
	volume = {547},
	issn = {0004-637X, 1538-4357},
	url = {http://arxiv.org/abs/astro-ph/0008310},
	doi = {10.1086/318372},
	number = {1},
	urldate = {2022-05-13},
	journal = {The Astrophysical Journal},
	author = {Merritt, David and Ferrarese, Laura},
	month = jan,
	year = {2001},
	note = {arXiv:astro-ph/0008310},
	pages = {140--145},
}

@article{shimwell_lofar_2019,
	title = {The {LOFAR} {Two}-metre {Sky} {Survey} - {II}. {First} data release},
	volume = {622},
	copyright = {© ESO 2019},
	issn = {0004-6361, 1432-0746},
	url = {https://www.aanda.org/articles/aa/abs/2019/02/aa33559-18/aa33559-18.html},
	doi = {10.1051/0004-6361/201833559},
	language = {en},
	urldate = {2022-05-06},
	journal = {Astronomy \& Astrophysics},
	publisher = {EDP Sciences},
	author = {Shimwell, T. W. and Tasse, C. and Hardcastle, M. J. and Mechev, A. P. and Williams, W. L. and Best, P. N. and Röttgering, H. J. A. and Callingham, J. R. and Dijkema, T. J. and Gasperin, F. de and Hoang, D. N. and Hugo, B. and Mirmont, M. and Oonk, J. B. R. and Prandoni, I. and Rafferty, D. and Sabater, J. and Smirnov, O. and Weeren, R. J. van and White, G. J. and Atemkeng, M. and Bester, L. and Bonnassieux, E. and Brüggen, M. and Brunetti, G. and Chyży, K. T. and Cochrane, R. and Conway, J. E. and Croston, J. H. and Danezi, A. and Duncan, K. and Haverkorn, M. and Heald, G. H. and Iacobelli, M. and Intema, H. T. and Jackson, N. and Jamrozy, M. and Jarvis, M. J. and Lakhoo, R. and Mevius, M. and Miley, G. K. and Morabito, L. and Morganti, R. and Nisbet, D. and Orrú, E. and Perkins, S. and Pizzo, R. F. and Schrijvers, C. and Smith, D. J. B. and Vermeulen, R. and Wise, M. W. and Alegre, L. and Bacon, D. J. and Bemmel, I. M. van and Beswick, R. J. and Bonafede, A. and Botteon, A. and Bourke, S. and Brienza, M. and Rivera, G. Calistro and Cassano, R. and Clarke, A. O. and Conselice, C. J. and Dettmar, R. J. and Drabent, A. and Dumba, C. and Emig, K. L. and Enßlin, T. A. and Ferrari, C. and Garrett, M. A. and Génova-Santos, R. T. and Goyal, A. and Gürkan, G. and Hale, C. and Harwood, J. J. and Heesen, V. and Hoeft, M. and Horellou, C. and Jackson, C. and Kokotanekov, G. and Kondapally, R. and Kunert-Bajraszewska, M. and Mahatma, V. and Mahony, E. K. and Mandal, S. and McKean, J. P. and Merloni, A. and Mingo, B. and Miskolczi, A. and Mooney, S. and Nikiel-Wroczyński, B. and O’Sullivan, S. P. and Quinn, J. and Reich, W. and Roskowiński, C. and Rowlinson, A. and Savini, F. and Saxena, A. and Schwarz, D. J. and Shulevski, A. and Sridhar, S. S. and Stacey, H. R. and Urquhart, S. and Wiel, M. H. D. van der and Varenius, E. and Webster, B. and Wilber, A.},
	month = feb,
	year = {2019},
	pages = {A1},
}

@article{panessa_origin_2019,
	title = {The origin of radio emission from radio-quiet active galactic nuclei},
	volume = {3},
	copyright = {2019 Springer Nature Limited},
	issn = {2397-3366},
	url = {https://www.nature.com/articles/s41550-019-0765-4},
	doi = {10.1038/s41550-019-0765-4},
	language = {en},
	number = {5},
	urldate = {2022-05-06},
	journal = {Nature Astronomy},
	publisher = {Nature Publishing Group},
	author = {Panessa, Francesca and Baldi, Ranieri Diego and Laor, Ari and Padovani, Paolo and Behar, Ehud and McHardy, Ian},
	month = may,
	year = {2019},
	note = {Number: 5},
	pages = {387--396},
}

@article{rawlings_relations_1989,
	title = {The relations between radio and forbidden [{O} {III}]-emission line luminosities in {FRII} radiogalaxies},
	volume = {240},
	issn = {0035-8711},
	url = {https://doi.org/10.1093/mnras/240.3.701},
	doi = {10.1093/mnras/240.3.701},
	number = {3},
	urldate = {2022-05-06},
	journal = {Monthly Notices of the Royal Astronomical Society},
	author = {Rawlings, Steve and Saunders, Richard and Eales, S. A. and Mackay, C. D.},
	month = oct,
	year = {1989},
	pages = {701--722},
}

@article{bower_breaking_2006,
	title = {Breaking the hierarchy of galaxy formation},
	volume = {370},
	issn = {0035-8711},
	url = {https://ui.adsabs.harvard.edu/abs/2006MNRAS.370..645B},
	doi = {10.1111/j.1365-2966.2006.10519.x},
	urldate = {2022-05-05},
	journal = {Monthly Notices of the Royal Astronomical Society},
	author = {Bower, R. G. and Benson, A. J. and Malbon, R. and Helly, J. C. and Frenk, C. S. and Baugh, C. M. and Cole, S. and Lacey, C. G.},
	month = aug,
	year = {2006},
	note = {ADS Bibcode: 2006MNRAS.370..645B},
	pages = {645--655},
}

@article{jackson_lbcs_2016,
	title = {{LBCS}: {The} {LOFAR} {Long}-{Baseline} {Calibrator} {Survey}},
	volume = {595},
	copyright = {© ESO 2016},
	issn = {0004-6361, 1432-0746},
	shorttitle = {{LBCS}},
	url = {https://www.aanda.org/articles/aa/abs/2016/11/aa29016-16/aa29016-16.html},
	doi = {10.1051/0004-6361/201629016},
	language = {en},
	urldate = {2022-05-05},
	journal = {Astronomy \& Astrophysics},
	publisher = {EDP Sciences},
	author = {Jackson, N. and Tagore, A. and Deller, A. and Moldón, J. and Varenius, E. and Morabito, L. and Wucknitz, O. and Carozzi, T. and Conway, J. and Drabent, A. and Kapinska, A. and Orrù, E. and Brentjens, M. and Blaauw, R. and Kuper, G. and Sluman, J. and Schaap, J. and Vermaas, N. and Iacobelli, M. and Cerrigone, L. and Shulevski, A. and Veen, S. ter and Fallows, R. and Pizzo, R. and Sipior, M. and Anderson, J. and Avruch, I. M. and Bell, M. E. and Bemmel, I. van and Bentum, M. J. and Best, P. and Bonafede, A. and Breitling, F. and Broderick, J. W. and Brouw, W. N. and Brüggen, M. and Ciardi, B. and Corstanje, A. and Gasperin, F. de and Geus, E. de and Eislöffel, J. and Engels, D. and Falcke, H. and Garrett, M. A. and Grießmeier, J. M. and Gunst, A. W. and Haarlem, M. P. van and Heald, G. and Hoeft, M. and Hörandel, J. and Horneffer, A. and Intema, H. and Juette, E. and Kuniyoshi, M. and Leeuwen, J. van and Loose, G. M. and Maat, P. and McFadden, R. and McKay-Bukowski, D. and McKean, J. P. and Mulcahy, D. D. and Munk, H. and Pandey-Pommier, M. and Polatidis, A. G. and Reich, W. and Röttgering, H. J. A. and Rowlinson, A. and Scaife, A. M. M. and Schwarz, D. J. and Steinmetz, M. and Swinbank, J. and Thoudam, S. and Toribio, M. C. and Vermeulen, R. and Vocks, C. and Weeren, R. J. van and Wise, M. W. and Yatawatta, S. and Zarka, P.},
	month = nov,
	year = {2016},
	pages = {A86},
}

@article{van_diepen_dppp_2018,
	title = {{DPPP}: {Default} {Pre}-{Processing} {Pipeline}},
	shorttitle = {{DPPP}},
	url = {https://ui.adsabs.harvard.edu/abs/2018ascl.soft04003V},
	urldate = {2022-04-28},
	journal = {Astrophysics Source Code Library},
	author = {van Diepen, Ger and Dijkema, Tammo Jan and Offringa, André},
	month = apr,
	year = {2018},
	note = {ADS Bibcode: 2018ascl.soft04003V},
	pages = {ascl:1804.003},
}

@article{morabito_sub-arcsecond_2022,
	title = {Sub-arcsecond imaging with the {International} {LOFAR} {Telescope}: {I}. {Foundational} calibration strategy and pipeline},
	volume = {658},
	issn = {0004-6361, 1432-0746},
	shorttitle = {Sub-arcsecond imaging with the {International} {LOFAR} {Telescope}},
	url = {https://www.aanda.org/10.1051/0004-6361/202140649},
	doi = {10.1051/0004-6361/202140649},
	urldate = {2022-04-22},
	journal = {Astronomy \& Astrophysics},
	author = {Morabito, L. K. and Jackson, N. J. and Mooney, S. and Sweijen, F. and Badole, S. and Kukreti, P. and Venkattu, D. and Groeneveld, C. and Kappes, A. and Bonnassieux, E. and Drabent, A. and Iacobelli, M. and Croston, J. H. and Best, P. N. and Bondi, M. and Callingham, J. R. and Conway, J. E. and Deller, A. T. and Hardcastle, M. J. and McKean, J. P. and Miley, G. K. and Moldon, J. and Röttgering, H. J. A. and Tasse, C. and Shimwell, T. W. and van Weeren, R. J. and Anderson, J. M. and Asgekar, A. and Avruch, I. M. and van Bemmel, I. M. and Bentum, M. J. and Bonafede, A. and Brouw, W. N. and Butcher, H. R. and Ciardi, B. and Corstanje, A. and Coolen, A. and Damstra, S. and de Gasperin, F. and Duscha, S. and Eislöffel, J. and Engels, D. and Falcke, H. and Garrett, M. A. and Griessmeier, J. and Gunst, A. W. and van Haarlem, M. P. and Hoeft, M. and van der Horst, A. J. and Jütte, E. and Kadler, M. and Koopmans, L. V. E. and Krankowski, A. and Mann, G. and Nelles, A. and Oonk, J. B. R. and Orru, E. and Paas, H. and Pandey, V. N. and Pizzo, R. F. and Pandey-Pommier, M. and Reich, W. and Rothkaehl, H. and Ruiter, M. and Schwarz, D. J. and Shulevski, A. and Soida, M. and Tagger, M. and Vocks, C. and Wijers, R. A. M. J. and Wijnholds, S. J. and Wucknitz, O. and Zarka, P. and Zucca, P.},
	month = feb,
	year = {2022},
	pages = {A1},
}

@article{de_gasperin_systematic_2019,
	title = {Systematic effects in {LOFAR} data: {A} unified calibration strategy},
	volume = {622},
	issn = {0004-6361, 1432-0746},
	shorttitle = {Systematic effects in {LOFAR} data},
	url = {https://www.aanda.org/10.1051/0004-6361/201833867},
	doi = {10.1051/0004-6361/201833867},
	urldate = {2022-04-22},
	journal = {Astronomy \& Astrophysics},
	author = {de Gasperin, F. and Dijkema, T. J. and Drabent, A. and Mevius, M. and Rafferty, D. and van Weeren, R. and Brüggen, M. and Callingham, J. R. and Emig, K. L. and Heald, G. and Intema, H. T. and Morabito, L. K. and Offringa, A. R. and Oonk, R. and Orrù, E. and Röttgering, H. and Sabater, J. and Shimwell, T. and Shulevski, A. and Williams, W.},
	month = feb,
	year = {2019},
	pages = {A5},
}

@article{goodman_ensemble_2010,
	title = {Ensemble samplers with affine invariance},
	volume = {5},
	issn = {2157-5452, 1559-3940},
	url = {http://msp.org/camcos/2010/5-1/p04.xhtml},
	doi = {10.2140/camcos.2010.5.65},
	language = {en},
	number = {1},
	urldate = {2022-04-22},
	journal = {Communications in Applied Mathematics and Computational Science},
	author = {Goodman, Jonathan and Weare, Jonathan},
	month = jan,
	year = {2010},
	pages = {65--80},
}

@article{liu_comprehensive_2019,
	title = {A {Comprehensive} and {Uniform} {Sample} of {Broad}-line {Active} {Galactic} {Nuclei} from the {SDSS} {DR7}},
	volume = {243},
	issn = {1538-4365},
	url = {https://iopscience.iop.org/article/10.3847/1538-4365/ab298b},
	doi = {10.3847/1538-4365/ab298b},
	language = {en},
	number = {2},
	urldate = {2022-04-22},
	journal = {The Astrophysical Journal Supplement Series},
	author = {Liu, He-Yang and Liu, Wen-Juan and Dong, Xiao-Bo and Zhou, Hongyan and Wang, Tinggui and Lu, Honglin and Yuan, Weimin},
	month = jul,
	year = {2019},
	pages = {21},
}

@article{tasse_lofar_2021,
	title = {The {LOFAR} {Two} {Meter} {Sky} {Survey}: {Deep} {Fields}, {I} -- {Direction}-dependent calibration and imaging},
	volume = {648},
	issn = {0004-6361, 1432-0746},
	shorttitle = {The {LOFAR} {Two} {Meter} {Sky} {Survey}},
	url = {http://arxiv.org/abs/2011.08328},
	doi = {10.1051/0004-6361/202038804},
	urldate = {2022-04-21},
	journal = {Astronomy \& Astrophysics},
	author = {Tasse, C. and Shimwell, T. and Hardcastle, M. J. and O'Sullivan, S. P. and van Weeren, R. and Best, P. N. and Bester, L. and Hugo, B. and Smirnov, O. and Sabater, J. and Calistro-Rivera, G. and de Gasperin, F. and Morabito, L. K. and Röttgering, H. and Williams, W. L. and Bonato, M. and Bondi, M. and Botteon, A. and Brüggen, M. and Brunetti, G. and Chyży, K. T. and Garrett, M. A. and Gürkan, G. and Jarvis, M. J. and Kondapally, R. and Mandal, S. and Prandoni, I. and Repetti, A. and Retana-Montenegro, E. and Schwarz, D. J. and Shulevski, A. and Wiaux, Y.},
	month = apr,
	year = {2021},
	note = {arXiv: 2011.08328},
	pages = {A1},
}

@article{kondapally_lofar_2021,
	title = {The {LOFAR} {Two}-meter {Sky} {Survey}: {Deep} {Fields} {Data} {Release} 1: {III}. {Host}-galaxy identifications and value added catalogues},
	volume = {648},
	issn = {0004-6361, 1432-0746},
	shorttitle = {The {LOFAR} {Two}-meter {Sky} {Survey}},
	url = {https://www.aanda.org/10.1051/0004-6361/202038813},
	doi = {10.1051/0004-6361/202038813},
	urldate = {2022-04-11},
	journal = {Astronomy \& Astrophysics},
	author = {Kondapally, R. and Best, P. N. and Hardcastle, M. J. and Nisbet, D. and Bonato, M. and Sabater, J. and Duncan, K. J. and McCheyne, I. and Cochrane, R. K. and Bowler, R. A. A. and Williams, W. L. and Shimwell, T. W. and Tasse, C. and Croston, J. H. and Goyal, A. and Jamrozy, M. and Jarvis, M. J. and Mahatma, V. H. and Röttgering, H. J. A. and Smith, D. J. B. and Wołowska, A. and Bondi, M. and Brienza, M. and Brown, M. J. I. and Brüggen, M. and Chambers, K. and Garrett, M. A. and Gürkan, G. and Huber, M. and Kunert-Bajraszewska, M. and Magnier, E. and Mingo, B. and Mostert, R. and Nikiel-Wroczyński, B. and O’Sullivan, S. P. and Paladino, R. and Ploeckinger, T. and Prandoni, I. and Rosenthal, M. J. and Schwarz, D. J. and Shulevski, A. and Wagenveld, J. D. and Wang, L.},
	month = apr,
	year = {2021},
	pages = {A3},
}




\appendix

\section{ILT Wide-Field Data Reduction Details} \label{data reduction appen}

In this appendix, we discuss the data reduction techniques we implement to obtain the Bo\"{o}tes high-resolution, wide-field images. All the scripts we discuss in the following sections are integrated into the publicly available pipeline (van der Wild et al. in prep) \footnote{VLBI Github repository: \url{https://git.astron.nl/RD/VLBI-cwl}}.

Before the observation becomes available on the LTA, some pre-processing is required. Any time periods or frequencies affected by Radio Frequency Interference (RFI) are flagged using \texttt{AOFlagger}\footnote{\texttt{AOFlagger}: \url{https://aoflagger.readthedocs.io/en/latest/}} \citep{offringa_aoflagger_2010}. The first and last channels were flagged, as their edges contain various imperfections that can degrade data quality. The data was then averaged to 16 channels per subband with a frequency of 12.21 kHz per channel in order to reduce data volume and this averaged data is then uploaded to the LTA.

We note that the data reduction steps below are now being automated in the publicly available pipeline \citep[van der Wild et al. in prep.,][]{jong_scalable_2025}. We conducted these steps manually as the automated pipeline development was in early stages at the start of data processing for Bo\"{o}tes.

The second stage of reducing our Bo\"{o}tes LTA pointing uses the \texttt{PREFACTOR}\footnote{The \texttt{PREFACTOR} pipeline is publicly available at: \url{https://github.com/lofar-astron/prefactor}} pipeline which is split into two parts, \texttt{Pre-Facet-Calibrator} and \texttt{Pre-Facet-Target}. \texttt{PREFACTOR} is now deprecated and users should now turn to \texttt{LINC}\footnote{\texttt{LINC} is publicly available at: \url{https://git.astron.nl/RD/LINC.}}, but this was not fully operational at the time of processing.

\subsubsection{Pre-Facet-Calibrator} \label{calibrator}

The first step of running the \texttt{PREFACTOR} pipeline is running \texttt{Pre-Facet-Calibrator} which considers all stations, both Dutch and international. Here, we derive corrections for the DIEs, namely polarisation offset, bandpass, and clock offset.

This section of the pipeline uses a bright 3C source that was specifically specified for the particular observation and uses this 3C source as a flux calibrator. These sources have a known radio spectrum, morphology, and are unpolarised at 144~MHz \citep{scaife_broad-band_2012}. For this 8-h observation of Bo\"{o}tes the flux calibrator source was 3C 196.

\subsubsection{Pre-Facet-Target} \label{target}

After we run \texttt{Pre-Facet-Calibrator}, the solutions from the flux calibrator are applied to the target observation of the Dutch stations only and hence dramatically reduces the data volume. We then apply the calibrator solutions to the target data for the polarisation alignment, bandpass, clock offset, and finally, a beam correction. If required, bright off-axis sources are removed from the target observation during the A-Team clipping step, where the four brightest radio sources in the northern hemisphere (the supernova remnants Cassiopeia A and Taurus A, and the radio galaxies Cygnus A and Virgo A) are flagged if their emission exceeds a certain threshold and they are in close proximity to the centre of the observation. Then, we further average the data to 8 seconds and 97.64 KHz. We also apply a coarse rotation measure correction based on Global Position System (GPS) information alongside the flux calibrator solutions. The final step of \texttt{Pre-Facet-Target} is to perform a direction independent phase-only calibration, only on the Dutch stations, using a skymodel from TGSS \citep[Tata Institute of Fundamental Research (TIFR) Giant Metrewave Radio Telescope (GMRT) Sky Survey;][]{intema_gmrt_2017}. This provides an initial direction independent correction for ionospheric effects. Finally, these corrections are applied to the measurement sets, producing DIE corrected data sets.


\subsubsection{Source Subtraction Outside the International FoV} \label{sub}

When incorporating the international stations into the data reduction process, we significantly reduce our FoV. This is because the international stations of the ILT are physically larger in size than the Dutch core and remote stations, therefore the beams of the international stations are smaller in comparison to the beam of the Dutch stations. We therefore subtract sources from the $uv$ data that are outside of the international station FoV.

We run the \texttt{DDF} pipeline~\footnote{The \texttt{DDF} pipeline is publicly available at: \url{https://github.com/mhardcastle/ddf-pipeline}}, which provides both DDE and DIE solutions for the Dutch stations. We took the 6\sarc\ resolution image which we produce during our run of the \texttt{DDF} pipeline and we ran \texttt{sub-sources-outside-region.py} \footnote{\texttt{sub-sources-outside-region.py} script is publicly available: \url{https://github.com/rvweeren/lofar_facet_selfcal/blob/main/submods/sub_sources_outside_region.py}}. We subtracted sources outside of a 2.5~$\times$~2.5~$\mathrm{deg^{2}}$ box region which represents the international FoV. This step is taken to prevent areas of the sky visible only to the Dutch stations, and not to the international ones, from affecting calibration and imaging, especially in the presence of bright sources.

\subsubsection{In-field Calibration} \label{delay calibration}

The next stage of data processing is using the LOFAR-VLBI pipeline. As a first step it performs a bulk DIE correction for the international stations, using the best in-field delay calibrator within the FoV of the international stations. We apply the solutions we produce using this in-field calibrator to both the Dutch and international stations. This is a critical part of reducing any ILT pointing because once the solutions from the in-field calibrator have been applied, calibration errors introduced at this stage are difficult to correct for, due to the use of small time and frequency solution intervals. Before we select a calibrator, we first apply the DI solutions from both \texttt{Pre-Facet-Calibrator} (Section \ref{calibrator}) and \texttt{Pre-Facet-Target} (Section \ref{target}). Then we separate the data into 24 manageable measurements sets, each with $\sim$2 MHz bandwidth. During this process, we produce a list of potential LBCS \citep[LOFAR Long-Baseline Calibrator Survey; ][]{moldon_lofar_2015,jackson_lbcs_2016, jackson_sub-arcsecond_2022} sources. This catalogue contains potential in-field calibrator sources, taking into account the brightness of these sources at 6\sarc\ and the position relative to the pointing centre. The ideal in-field calibrator source is compact and bright at 0.3\sarc\ while being within the international station FWHM. At this stage of the data reduction process, the source's morphology at 0.3\sarc is not known. Therefore, the source ranked as the most likely best in-field calibrator in the LBCS catalogue may not be the best one, and hence we verify the LBCS selection. The previous pipeline version selected the LBCS calibrator with compact flux and was closest to the pointing centre, however, these two criteria alone are insufficient for identifying a suitable in-field calibrator. Consequently, we process multiple candidate calibrators during data reduction. We then split out these potential in-field calibrators from the field and calibrate them individually to determine which source is the best calibrator within the FoV. \par

To determine whether the sources which we select using LBCS are suitable in-field calibrators, we conduct self-calibration of these sources. The aim of self-calibration is to find the complex gains which we require to reproduce the sky intensity model given the visibilities observed. To do this, an initial model is created and the complex gains are then solved for and the corrected visibility is calculated. Using this corrected visibility, a new and improved model is formed. We repeat this process until the model created is representative of the observed source. We implement \texttt{facetselfcal} \citep{van_weeren_lofar_2021} to perform all self-calibration procedures to create these Bo\"{o}tes images. This software uses the default preprocessing pipeline (\texttt{DP3}; \cite{van_diepen_dppp_2018}) to calculate complex gains and \texttt{WSClean} \citep{offringa_wsclean_2014} for imaging. For our potential calibrator sources we set the number of self-calibration cycles to 10, as this generally leads to convergence \citep[e.g.][]{sweijen_deep_2022, ye_1-arcsecond_2024}. 

For this Bo\"{o}tes pointing, we reduce two LBCS sources to sub-arcsecond resolution, both of which are double sources at 0.3\sarc. Of the two, we chose ILTJ142905.10+342641.0 as the in-field calibrator because the calibration solutions are cleaner compared to the first source, and this source is closer to the pointing centre of the observation. This highlights the importance of considering multiple sources as potential in-field calibrators. \par

We applied the calibrator solutions which we produced from the self-calibration of the in-field calibrator to the subtracted measurement set. For this step, we phase-shift toward our in-field calibrator and form a "super station" (ST001) where we combine all the core stations to reduce the data volume, increase SNR, and suppress signals from other sources due to the reduced FoV.

\subsection{Direction Dependent Calibrators}

The solutions we apply from the in-field calibrator only provide corrections in the direction of the in-field calibrator, leaving residual DDEs that still require correcting. For wide-field imaging using the ILT, we require multiple calibrator sources across our FoV referred to as DDCs (Directional Dependent Calibrators) to correct for the residual DDEs.

The ideal candidates for these DDCs are the same as our selection of an in-field calibrator where high brightness and compactness are desirable. Unfortunately, at such high-resolution, these sources are rare. Therefore, we must loosen our requirements to obtain the best understanding of ionospheric conditions across our FoV. We do not have to be as strict with our selection of DDCs compared to the selection of our in-field calibrator, as our in-field calibrator has corrected for the bulk of distortions. We consequently include fainter and extended sources within our selection of these DDCs, as long as these objects' calibration solutions converge through the self-calibration cycles.

For this field we select 26 DDCs and hence have 27 directions in total within our field, including the in-field calibrator. To begin this selection, we consider sources identified as potential in-field calibrators, but ultimately do not adopt them as the in-field calibrator. We also consider the next 100 brightest objects in the 6\sarc\ image that lie within the international FoV. Therefore, we choose our DDCs from 112 sources. We split each of these sources from the field and then conduct self-calibration on each of them individually to determine if they are a suitable DDC. We then visually inspect the resulting images and solutions to determine their suitability.

Now that we have our DDCs, we split the field into facets using Voronoi tessellation \citep[e.g.][]{weeren_lofar_2016}, which sets the dimensions of each facet by assigning every point in the plane to its nearest calibrator source such that each facet contains all points closer to that calibrator than to any other. This results in the non-symmetric facets and from this step forward we consider each facet individually rather than the entire FoV. We demonstrate the facet layout for the sub-arcsecond resolution Bo\"{o}tes images in Figure \ref{fig:facet}, and we note that facet 20 has higher noise in comparison to other directions which is further described in Section \ref{postpro}.

\begin{figure}
    \centering
    \includegraphics[width=\linewidth]{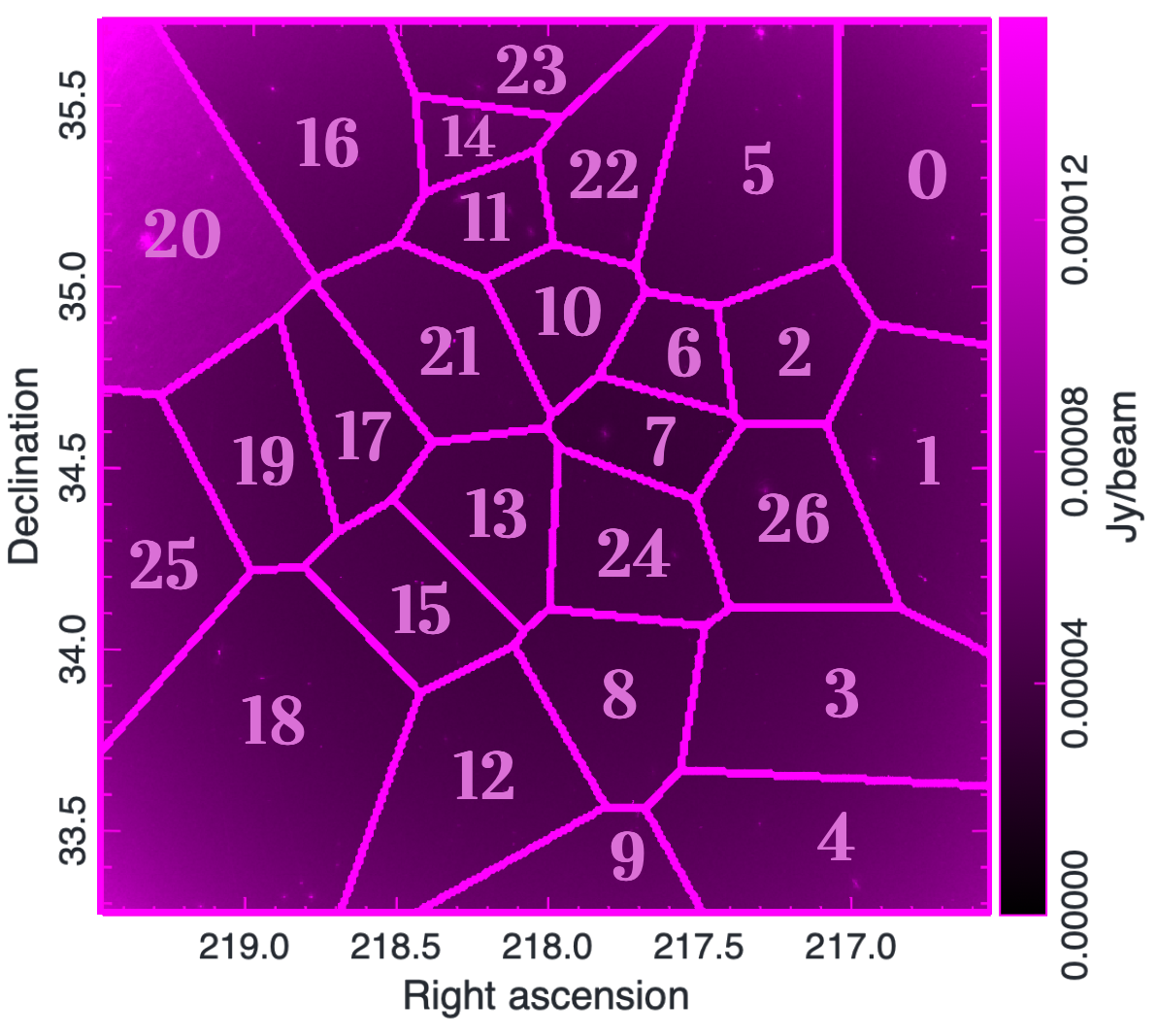}
    \caption{The layout of the 27 facets we use to produce the sub-arcsecond resolution, wide-field, images of the Bo\"{o}tes Deep Field. The background is the $\sim$0.3\sarc\ RMS image and facet 20 has a higher level of noise due to an extra flux scaling correction required for this facet.}
    \label{fig:facet}
\end{figure}

\subsection{High-Resolution Imaging}

The first image we create of Bo\"{o}tes (using \texttt{WSClean}) is at $\sim$1.2\sarc\ resolution. We produce this image first to check the quality of the calibration of the image because of its comparatively low computational cost. This 1.2\sarc\ image was imaged in one go using \texttt{WSClean}'s facet mode, however this is not possible for the 0.6\sarc\ and 0.3\sarc\ images. Instead, the 1.2\sarc\ model images were used to create individual datasets for each facet (refer to Figure \ref{fig:facet} for facet layout) which could then be imaged independently, allowing for faster and less computationally expensive sub-arcsecond imaging.

To create the individual datasets for each of these facets, we first subtract all sources within the field and using the \texttt{--predict} functionality of \texttt{WSClean} and the 1.2\sarc\ model images. We then add back the sources within the facet using the solutions of the corresponding DDCs, hence giving calibrated datasets for each facet. We complete this for each of the 27 facets, and are left with 27 facet images. We do this independently for the 0.3\sarc\ and 0.6\sarc\ images. 

The final step is to mosaic these facets together using \texttt{SWarp}\footnote{\texttt{SWarp} is publicly available at: \url{https://ascl.net/1010.068}} \citep{bertin_terapix_2002} to create the final image at 0.3\sarc\ (which we show in Figure \ref{fig:0.3}), as well as the final 0.6\sarc image.

\subsubsection{Flux Density Scale and Astrometry Corrections} \label{postpro}

The self-calibration which we implement to create these images is agnostic to the position of sources when there is not a reference source, therefore, the resulting images contain both an astrometry and flux density offset. These offsets are introduced by the delay calibrator rather than our DDCs, as during self-calibration of our DDCs we normalise the amplitudes to minimise any drift in the flux scale. A position shift may also be present because for Bo\"{o}tes, we do not have an astrometric reference for the delay calibrator selected. Similarly, although the amplitude scale was tied to archival data, residual offsets from the desired flux density scale may remain, so the source flux densities require scaling to match to known flux density measurements within the literature. To derive these corrections, we locate compact, high SNR, sources using a compactness measurement from the peak intensity and total flux intensity from \cite{kondapally_lofar_2021} as well as the respective high-resolution images from \texttt{pyBDSF}\footnote{\url{https://pybdsf.readthedocs.io/en/latest/}}. For flux scaling, we apply a universal correction across all facets, by taking the median flux scaling of all facets combined (due to the substantial uncertainties in the flux scaling for the ILT) and we perform the astrometry correction for each facet individually. We apply the correction calculated for the 0.3\sarc\ image for each resolution as both the astrometry and flux scaling offset is not dependent on the resolution as we create each resolution image using the same dataset.

To determine the flux scaling we require, we use the total flux density measurements at 6\sarc\ and compare this to the measurement of flux density at 0.3\sarc. For this correction we consider sources with $S_{\text{peak}}$/$S_{\text{total}}$~$>$~0.7 and from these sources, we use a SNR cut off of 10 using the peak intensity measurement and the RMS (root mean squared) island noise from the 0.3\sarc\ image. This leaves 343 sources at 0.3\sarc. We incorporate an initial universal flux scaling correction of 1.111 to the high-resolution images. We investigate the scatter in the flux scaling correction between facets to determine the uncertainties in this flux scale correction. We fit a Gaussian to the flux scaling correction distribution and adopt the standard deviation of the fit, 24.8 per cent, as the uncertainty in the flux scaling. For Bo\"{o}tes we add a further flux scale to correct for flux measurement deviations between the LoTSS Deep 6\sarc\ image and a 6\sarc\ image produced with this data set and calibration methods. To do this, we locate sources in LoTSS with SNR greater than 20 with $S_{\text{peak}}$/$S_{\text{total}}$~$>$~0.9, leaving 990 sources. Using the flux density measurements from LoTSS Deep and the new 6\sarc\ image, we find a flux scaling correction of 0.92. Combining this with our previous flux scaling correction, for the whole field we incorporate a flux scale of 1.02.

To correct for the astrometric offset, we use the RA and DEC of the optical counterparts from \cite{kondapally_lofar_2021} to calculate the offsets between the LoTSS Deep 6\sarc\ catalogue and the 0.3\sarc\ resolution image. Our definition for compact sources is more lenient for the position offset as we correct per facet for this offset rather than a universal correction and therefore prioritise count statistics per facet. We reduce our criteria to $S_{\text{peak}}$/$S_{\text{total}}$~$>$~0.6 followed by a SNR cut off of 7. This leaves 954 sources, with the largest number of compact sources in a single facet of 80 and a smallest of 3. We correct the astrometry per facet by taking the median RA and DEC offset in each facet. The RA correction ranges from 0.0982\sarc\ to 0.576\sarc\ with an overall median of 0.319" and the DEC correction ranges from -0.425\sarc\ to 0.289\sarc\ with a median of 0.0367\sarc.

The astrometry correction in the images for Bo\"{o}tes is larger than that of the ELAIS-N1 field. The key reason for this is that the in-field calibrator source for ELAIS-N1 has a high frequency VLBI position based on \cite{charlot_third_2020}. Therefore for this field the authors could achieve millisecond accuracy for the positioning of this source. For future fields that are to be reduced at sub-arcsecond scale, it is important to consider the astrometry early on and potentially select calibrators that have a highly accurate position available.

We note that facet 20 had relatively poor calibration in comparison to other facets within this image. Therefore, we apply a further flux scaling correction to this facet using the same method as above. In Figure \ref{fig:facet} we can see this has higher noise in comparison to the other facets. 

\subsubsection{Catalogue Construction and Component Association} \label{catalog}

\begin{figure}
    \centering
    \includegraphics[width=0.5\textwidth]{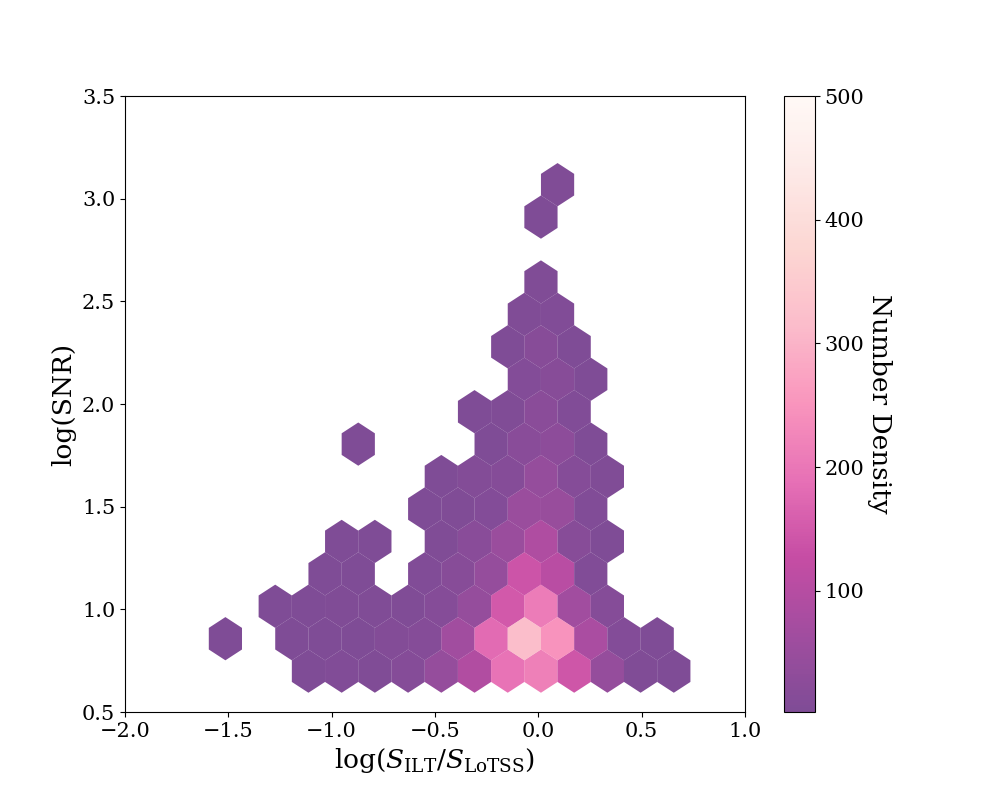}
    \caption{The SNR of the $\sim$0.3\sarc\ resolution image as a function of the ratio between the flux density at 0.3\sarc\ $(S_{\mathrm{ILT}})$ and flux density at 6\sarc\ $(S_{\mathrm{LoTSS}})$, where we trace the number density with the colour bar. For a hexagon to be plotted, at least 2 sources must lie within the bin.}    
    \label{fig:snr_res} 
\end{figure}

We produce the final catalogues, corrected for astrometry and flux scale, for each of the three resolution images using \texttt{pyBDSF} with a pixel detection threshold of 5$\sigmaup$ and a detection island threshold of 3$\sigmaup$, an RMS box of (120,15), and for bright-source RMS box of (40,15) following \cite{de_jong_into_2024}. For the 0.3\sarc, 0.6\sarc, and 1.2\sarc\ catalogues we detect 8048, 5851, and 2596 components respectively at 5$\sigmaup$. 

In the published source catalogues, the components have been associated in several stages. If a single component at the relevant high-resolution image is isolated within the 6\sarc\ beam of the LoTSS Deep Field image, we class it as a single component source. At 0.3\sarc\, if multiple components are present within a separation 30\sarc\ of each other, we perform visual inspection to check if components should be associated to a common source, by creating a 60\sarc\ $\times$ 60\sarc\ cutout. These cutouts also have information at other resolutions to allow simultaneous component association at each resolution. For instance, 6\sarc\ resolution images allow us to de-blend sources, which could be confused as multiple components for a single source rather than multiple sources. For a handful of cases, we require further information to determine whether components are blended sources rather than a single source. In these cases we consult the NOAO Deep Wide Field Survey \citep[NDWFS;][]{jannuzi_noao_1999} optical image of Bo\"{o}tes at I-band to either keep components separate or associate these components. Using a separation of 30\sarc\ misses giant radio galaxies within our images, therefore we also use the LOFAR Galaxy Zoo (LGZ) within \cite{kondapally_lofar_2021} to visually inspect sources with a LGZ size greater than 30\sarc. 

These methods result in a total of 585 components being merged into 223 sources at 0.3\sarc. At 0.6\sarc\, 434 components are merged into 152 sources, and finally at 1.2\sarc\, 166 components are merged to 65 sources.

When merging sources, we sum the total flux density of all components and perform standard error propagation on their errors. We take the mean of the island RMS noise levels as a proxy for the local RMS noise of the associated source. We take the highest peak intensity value alongside the associated error. The RA and DEC for these sources is the flux density weighted position. Finally we calculate the Largest Angular Size (LAS) for these sources, by considering each component's major and minor axes, as well as their respective position angle. We set the major and minor axes and the position angle of these sources to zero in the catalogue. We obtain the uncertainties for the LAS by using the uncertainty extremities of the major and minor axis, as well as the position angle. 

During the visual inspection of sources, several of the components within a certain separation appeared to be artifacts. We remove these from our final catalogues. For 0.3\sarc\ we remove 69 components, 45 at 0.6\sarc\, and 4 at 1.2\sarc.

Many false detections are present within the initial \texttt{pyBDSF} catalogues. The 6\sarc\ image of Bo\"{o}tes from LoTSS is deeper than the images presented in this paper and therefore can be used to help identify false detections. To remove these false detections, firstly we located sources which do not lie within 3\sarc\ of a LoTSS source, as sources within this radius would be associated to the LoTSS 6\sarc\ source. We only consider sources with a flux density below 500~$\muup$Jy measured at the given high-resolution image because sources below this threshold are where we expect to start to see false detections. We then measure the median 5$\sigmaup$ RMS noise within a 3\sarc\ radius region of the LoTSS position in the RMS map \footnote{The RMS map is available at: \url{https://lofar-surveys.org/deepfields_public_bootes.html}}. If the flux density in the high-resolution image of the source is greater than the 5$\sigmaup$ median RMS at 6\sarc\, then we class this as a false detection and remove this from our final catalogue. However, if the flux density is less than the 5$\sigmaup$ median RMS then this is a possible real detection that was not detected at 6\sarc. We remove 3541 false detections at 0.3\sarc\ with 278 potential new detections, at 0.6\sarc\ we remove 1070 false detections and have 13 new detections, and at 1.2\sarc\ we remove 10 detections and have no new detections at this resolution.

Finally, we conduct one final round of visual inspection of the whole field to ensure the brightest sources have been correctly associated. In the 0.3\sarc\ image we associate four components into two sources and in the 1.2\sarc\ we associate three components into one source. This leaves us with the final catalogue at each resolution. The 0.3\sarc\ catalogue contains 4074 sources, 4455 at 0.6\sarc, and 2480 at 1.2\sarc. We note that the highest source count of the three resolution images released in this paper is in the $\sim$0.6\sarc\ resolution image. This is because at the highest resolution, 0.3\sarc, we are probing the smallest scale emission at this frequency, therefore as we decrease resolution we increase sensitivity to lower surface brightness emission and we are more likely to detect diffuse emission on larger scales, and therefore at 0.6\sarc\ we can detected sources which cannot be detected at 0.3\sarc\ as these would be too diffuse. We do not however see the highest source count at the $\sim$1.2\sarc\ image despite this being the lowest resolution as this image has the lowest sensitivity due to gaps in the $uv$ coverage.

Figure \ref{fig:snr_res} shows us how the ratio of flux density at 0.3\sarc\ $(S_{\mathrm{ILT}})$ and flux density at 6\sarc\ $(S_{\mathrm{LoTSS}})$ varies with SNR which we measure from the 0.3\sarc\ image by taking the ratio of peak intensity and flux densities as measured by pyBDSF. The sources contributing to this figure are detected within 1\sarc\ of their LoTSS position. As expected, we see the majority of sources lie at low SNR and have a logarithmic ratio of $(S_{\mathrm{ILT}})$ and $(S_{\mathrm{LoTSS}})$ slightly below zero, as we expect some emission to be resolved out when we increase the length of our baselines. We do see some sources which lie above zero, and this appears to occur at a lower SNR ratio. This happens as at lower SNR it becomes increasingly difficult for pyBDSF to measure flux density and hence the fluxes maybe overestimated, because not all the flux detected is associated with the source, and is likely to be a noise contribution.

\section{[O~{\sc iii}] kinematics} \label{CDF}

\begin{figure}
    \centering
    \includegraphics[width=\linewidth]{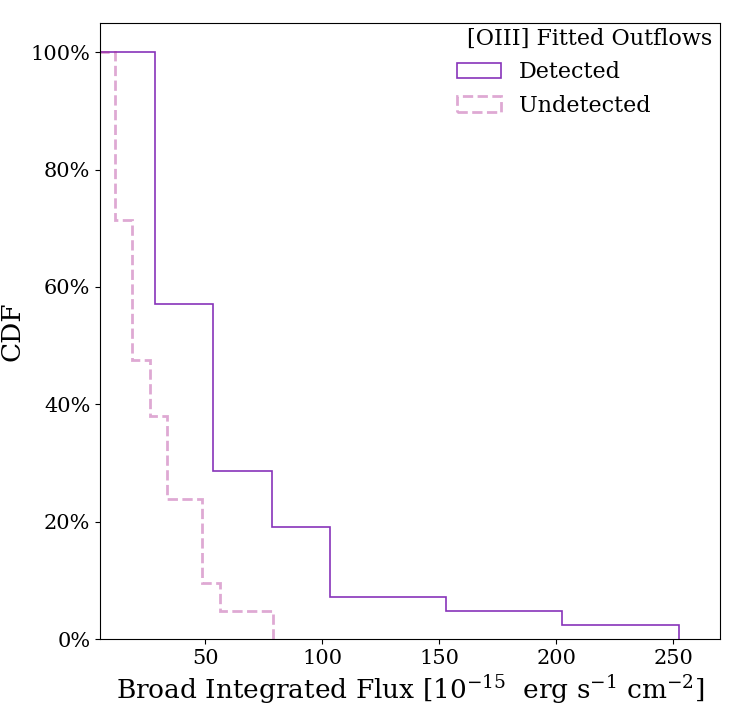}
    \caption{Cumulative Distribution Function AGN with an [O~{\sc iii}] fitted outflow comparing the integrated flux of the broad component fitted to [O~{\sc iii}] of detected and undetected AGN at 0.3\sarc. The solid purple histograms show the results for the detected AGN and the dashed pink histogram portrays the undetected AGN.}
    \label{fig:CDF_area}
\end{figure}

\begin{figure}
    \centering
    \includegraphics[width=\linewidth]{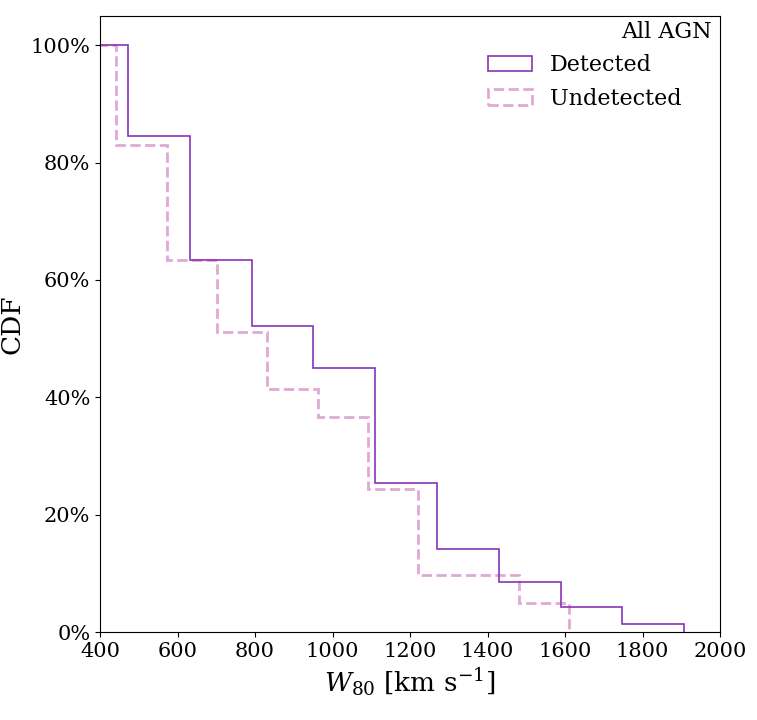}
    \caption{Cumulative Distribution Function all AGN, regardless of [O~{\sc iii}] outflow type comparing the $W_{80}$ of detected and undetected AGN at 0.3\sarc. The solid purple histograms show the results for the detected AGN and the dashed pink histogram portrays the undetected AGN.}
    \label{fig:CDF_w80}
\end{figure}

To further investigate the relationship between [O~{\sc iii}] and high-resolution radio morphologies, we investigate how [O~{\sc iii}] kinematics \citep[derived from][]{escott_unveiling_2025} vary depending on whether a source is detected or undetected at 0.3\sarc\ using cumulative distribution functions (CDFs). Here, we are not investigating the variation between AGN with [O~{\sc iii}] outflows and AGN with no [O~{\sc iii}] outflows, and therefore we do not use the $L_{\mathrm{6\muup\\ m}}$ and redshift matched population. We check that the 2D KS test for $L_{\mathrm{6\muup\\ m}}$ and redshift between the detected and undetected populations outputs a p value above 0.05, and therefore these two populations can be compared without matching being necessary. We note that we do not draw any significant conclusions from these CDFs due to our limited sample size as the detected population has 35 AGN and the undetected have 41 AGN.

Figure \ref{fig:CDF_area} and Figure \ref{fig:CDF_w80} illustrates the relationship between two kinematics properties of [O~{\sc iii}] and how they vary depending on if they are detected at sub-arcsecond resolution or not. Detected AGN are depicted with purple solid lines, and we show undetected AGN with dashed pink lines. Figure \ref{fig:CDF_area} shows the results for the integrated flux of the broad component of [O~{\sc iii}] which we define as the area of the component which we calculate using the peak and FWHM of the broad Gaussian component. Only AGN which are within the [O~{\sc iii}] fitted outflow category are included as we require a broad component to be present. We see a higher integrated flux of the broad component when the AGN are detected compared to AGN which are undetected at 0.3\sarc.

Figure \ref{fig:CDF_w80} compares how the $W_{80}$ varies between the two populations. Again, here we see that compact AGN have a larger $W_{80}$ (which is a non-parametric proxy) compared to the undetected AGN, however this is at a lesser extent in comparison to the increase we see in the integrated flux of the broad component.

Both these CDFs show similar results to the kinematic analysis in \cite{escott_unveiling_2025} where the authors compare these kinematics between AGN detected at 6\sarc\ and AGN not detected at 6\sarc, whereas in this work, we are comparing AGN which are all detected at 6\sarc, but the detected AGN are still detected at 0.3\sarc\ and the undetected population are not detected at 0.3\sarc. As we still see the enhanced kinematics of [O~{\sc iii}] in the detected population at 0.3\sarc, this suggests that the relation is not driven by the emission which is resolved out in-between 6\sarc\ and 0.3\sarc. Due to our limited sample size and lack of uncertainties, we require larger samples in further work to probe the concreteness of this result.


\bsp	
\label{lastpage}
\end{document}